\documentclass[a4j,11pt,onecolumn,oneside,notitlepage,final]{article}

\usepackage{url}
\usepackage{amsmath}
\usepackage{amsfonts}
\usepackage{amssymb}
\usepackage{latexsym}
\usepackage{color}
\usepackage{graphicx}
\usepackage{cite}
\usepackage{subfigure}

\def\be{\begin{equation}}
\def\ee{\end{equation}}
\def\ben{\begin{eqnarray}}
\def\een{\end{eqnarray}}

\begin{document}
\begin{center}
\vskip .5in

\begin{flushright}
YITP-18-04 \\
KUNS-2714 \\
RUP-18-2
\end{flushright}

{\Large \bf
Primordial Black Holes \\
- Perspectives in Gravitational Wave Astronomy -
}
\vskip .45in

{
Misao Sasaki$^{a}$,
Teruaki Suyama$^{b}$,
Takahiro Tanaka$^{c,a}$,
and Shuichiro Yokoyama$^{d,e}$
}

{\em
$^a$
   Center for Gravitational Physics, Yukawa Institute for Theoretical Physics,
Kyoto University, Kyoto 606-8502, Japan
}\\

{\em
$^b$
   Research Center for the Early Universe (RESCEU), Graduate School
  of Science,\\ The University of Tokyo, Tokyo 113-0033, Japan
}\\

{\em
$^c$
  Department of Physics, Kyoto University, Kyoto 606-8502, Japan
}\\

{\em
$^d$
  Department of Physics, Rikkyo University, Tokyo 171-8501, Japan
}\\

{\em
$^e$
Kavli IPMU (WPI), UTIAS, The University of Tokyo,
Kashiwa, Chiba 277-8583, Japan
}

\abstract{
This is a review article on the primordial black holes (PBHs), with particular focus
on the massive ones ($\gtrsim 10^{15}{\rm g}$) which have not evaporated by the
present epoch by the Hawking radiation.
By the detections of gravitational waves by LIGO, 
we have gained a completely novel tool to observationally
search for PBHs complementary to the electromagnetic waves.
Based on the perspective that gravitational-wave astronomy will make a significant progress
in the next decades, 
a purpose of this article is to give a comprehensive review covering a wide range of topics on PBHs.
After discussing PBH formation as well as several inflation models leading to PBH production, 
we summarize various existing and future observational constraints.
We then present topics on formation of PBH binaries, gravitational waves from PBH binaries, 
various observational tests of PBHs by using gravitational waves.
}

\end{center}

\newpage
\tableofcontents
\newpage

\vskip .4in

\section{Introduction}
History of primordial black holes (PBHs) dates back to sixties when Zeldovich and Novikov pointed out 
that BHs in the early Universe may grow catastrophically by accreting the surrounding radiation \cite{Zeldovich:1967ei}. 
In 1971, Hawking proposed \cite{Hawking:1971ei} that highly overdense region of inhomogeneities in the primordial Universe 
can directly undergo gravitational collapse to form BHs, 
which initiated the {\it modern} mechanism of the PBH formation. 
Contrary to the astrophysical processes (i.e. collapse of stars) for which only BHs heavier than a particular mass (around 3 solar mass \cite{Rhoades:1974fn}) are possible to form, 
extremely strong gravitational force inside the highly compressed radiation/matter 
that can be realized in the early Universe 
allows formation of not only stellar/super-massive BHs but also small BHs that could be in principle as light as Planck mass $\sim 10^{-5} {\rm g}$ (see {\it e.g.} \cite{Carr:2005zd} and references therein). 
After the advent of the inflationary cosmology, 
formation of PBHs and their properties such as mass and abundance 
had been studied in tight connection with inflation models. 
Conversely, knowledge of observational information about PBHs provides important 
clues to build inflation models. 
In particular, it is worth mentioning that even the non-detection of PBHs 
gives us useful information of the early Universe \cite{Carr:2005zd}.

Observational searches of PBHs have been conducted intensively and continuously over several decades. Depending on the mass, 
PBHs trigger different observational signals. 
PBHs lighter than a certain mass $M_c$ given by \cite{Page:1976df}
\be
M_c \simeq {\left( \frac{3\hbar c^4 \alpha_0}{G^2} t_0 \right)}^{\frac{1}{3}} \sim 10^{15}~{\rm g}
~{\left( \frac{\alpha_0}{4\times 10^{-4}} \right)}^{\frac{1}{3}}
{\left( \frac{t_0}{13.8~{\rm Gyr}} \right)}^{\frac{1}{3}},
\ee
have already evaporated by the cosmic age $t_0$ due to the Hawking radiation.
Thus, PBHs lighter than $\simeq 10^{15}~{\rm g}$ do not exist in the present Universe.
Nevertheless, they leave some traces from which we can investigate how many PBHs could have existed in the early Universe. 
For instance, PBHs in the mass range $10^9 \sim 10^{13}~{\rm g}$ change abundance of light elements produced by the Big Bang nucleosynthesis due to high energy particles emitted by the evaporating PBHs \cite{Miyama:1978mp}. 
Comparison between the observed light elements and the theoretical prediction tightly constrains the abundance of such PBHs (see {\it e.g.} \cite{Carr:2009jm} and references therein). 

PBHs heavier than $10^{15}~{\rm g}$ have not yet lost their mass significantly
by the evaporation and remain in the present Universe. 
They not only imprint observational traces in the early Universe (such as by accretion, 
and indirect effects by the primordial density perturbations that seed PBHs) but also produce various distinct signals at present time
such as gravitational lensing, dynamical effects on baryonic matter, and radiation emanating from the accreting matter into PBHs etc. 
What physical process among them becomes the most prominent to show us the existence of PBHs depends on the PBH mass. 
For instance, gravitational lensing of background stars is the most powerful method to search sub-solar PBHs (see Sec.~\ref{Observational-constraint}). 
Accretion and dynamical effects on baryonic matter become more important for heavier PBHs. 
One of the important questions regarding non-evaporating PBHs is whether they comprise all the dark matter or not. 
Thanks to achievements of many different types of cosmological and astrophysical observations over decades, 
stringent upper limit on the PBH abundance has been obtained for a vast PBH mass range \cite{Carr:2016drx}. 
Currently, it appears that PBHs do not explain all the dark matter and at most constitute fraction of dark matter \cite{Carr:2017jsz}. 
Reviewing the existing observational constraints on the non-evaporating PBHs is one of the main purpose of this article.

The LIGO discovery of the merger event (GW150914) of binary BHs \cite{Abbott:2016blz} triggered a renewed interest of PBHs, 
especially in the stellar mass range. 
Unexpected largeness of the detected BHs (around $30~M_\odot$) brought us a new mystery about the component of the Universe.
After the LIGO event, elucidating the origin of the BHs and binary formation has emerged as an important topic
in cosmology and astrophysics ({\it e.g.} see \cite{TheLIGOScientific:2016htt}).
Soon after the LIGO's announcement on the first detection of the BH merger,
several research groups \cite{Bird:2016dcv, Clesse:2016vqa, Sasaki:2016jop} independently pointed out that the inferred merger rate can be explained by the merger of PBHs
without violating the trivial bound that the PBH abundance is equal to or less than the total dark matter abundance.
In \cite{Bird:2016dcv, Clesse:2016vqa}, binary formation by the accidental encounters of PBHs in dense environment,
which works in the low-redshift Universe, has been considered,
while different mechanism of the binary formation by the tidal perturbation caused by the distant PBHs,
which works in the radiation dominated epoch in the early Universe and was originally proposed earlier in \cite{Nakamura:1997sm},
has been investigated in \cite{Sasaki:2016jop}.
These studies demonstrate that gravitational waves (GWs), brand new observable,
provide a powerful and useful tool to probe parameter region of PBHs (mass, abundance etc) 
that have not been possible only by the electromagnetic waves.
In other words, roles of GWs are complementary to electromagnetic waves.

In addition to the PBH scenario, several astrophysical processes such as the field binary scenario
and the dynamical formation in the dense stellar environment have also been proposed to 
explain the observed properties of the binary black holes (for instance, see \cite{TheLIGOScientific:2016htt} and references therein).
At the time of writing this article (autumn 2017), LIGO and Virgo detected five BH merger events
\cite{Abbott:2016blz, Abbott:2016nmj, Abbott:2017vtc, Abbott:2017oio, Abbott:2017gyy}.
For the moment, both the PBH scenario and the astrophysical scenarios are allowed as a possible explanation 
of these events because of a limited number of the detected events.

We are at the dawn of the golden age of the GW astronomy.
In the future, the ongoing experiments such as LIGO and Virgo (and soon KAGRA) gain better sensitivity,
and further upgraded and new type of experiments such as the Einstein Telescope \cite{Punturo:2010zz}, Cosmic Explorer \cite{Evans:2016mbw}, LISA \cite{Seoane:2013qna}, 
and DECIGO \cite{Seto:2001qf} covering 
different frequency bands will follow.
Very likely, much more merger events will be detected and the statistical information of BHs and BH binaries such as mass, spin,
eccentricity, redshift, spatial inhomogeneities, etc. will become available.
Those information will enable us to test the individual scenarios and possibly to pin down the best one,
or it is also conceivable that the truth is a mixture of multiple scenarios, {\it e.g.} some events due to PBHs
and the others due to astrophysical BHs.
In order to reveal the true nature of the BHs and their binaries, it is indispensable to go through three stages;
to devise possible scenarios as many as we can (first), to theoretically understand each scenario and derive 
distinct features predicted in each scenario (second), and then to test the predictions by real observations
and identify the answer (third).
In addition to the main purpose as mentioned above, another purpose of this article is to review the PBH scenario 
in connection with the LIGO events
and various proposals to test this scenario by using the future GW observations, {\it i.e.} the first and 
the second stages of the above classification.

Organization of this article is as follows.
In the next section, we will review the basics of the PBH formation and related topics. 
Inflation models leading to the production of PBHs are also introduced.
In Sec.~\ref{Observational-constraint}, we cover various observational constraints on the abundance
of non-evaporating PBHs obtained by the electromagnetic observations.
Expected constraints in the future are also briefly presented.
We do not put much focus on smaller PBHs that have already evaporated 
but briefly mention the constraints because they are less relevant to GWs.
Regarding the constraints on such small PBHs, Ref.~\cite{Carr:2009jm} provides 
a detailed analysis.
In Sec.~\ref{GW-PBH}, we review the PBH scenario as an explanation of the LIGO events
and future prospects of its observational test by GWs.
Final section is a summary.

Throughout this article, we use a natural unit in which $c=\hbar=1$.

\section{Formation of PBHs}

Until now, several mechanisms  to form PBHs in the early Universe have been proposed.  As examples, Ref.~\cite{Garriga:2015fdk,Deng:2016vzb} recently discussed the possibility of PBH formation
by domain walls and
also Ref.~\cite{Garriga:2015fdk,Deng:2017uwc} proposed the PBH formation scenario by
vacuum bubbles which nucleate during the inflation. There are also several works about the PBH formation from the cosmic string loops~\cite{Hawking:1987bn,Polnarev:1988dh,Garriga:1992nm}. However, the most frequently studied PBH formation scenario must be a gravitational collapse of the overdense region in the early Universe. Here, we briefly review the formation process of the PBHs and also the inflationary models which could produce such an overdense region.

\subsection{Basis of primordial black holes formation in the early Universe}
\subsubsection{PBH formation based on the simple physical picture}

In the early radiation-dominated Universe,
a highly overdense region would gravitationally collapse into a black hole, directly.
Such a black hole formed in the early Universe is called primordial black hole (PBH).
Details of the PBH formation from the overdense region in the early Universe has been extensively
investigated numerically and analytically \cite{Niemeyer:1997mt,Shibata:1999zs,Musco:2004ak,Polnarev:2006aa,Musco:2008hv,Nakama:2013ica, Harada:2013epa}.
Here, let us give a brief review of a rough sketch of the PBH formation
which unifies the traditional view based on the density 
contrast~\cite{Carr:1975qj,Yokoyama:1995ex} with the recent view 
based on the curvature perturbation~\cite{Shibata:1999zs,Young:2014ana}.

First, in the early Universe after inflation, the background spacetime can be
 well-described by
the spatially-flat Friedmann-Lemaitre-Robertson-Walker (FLRW) metric
 (homogeneous and isotropic space):
\begin{equation}
ds^2 = -dt^2 + a(t)^2 \delta_{ij} dx^i dx^j,
\end{equation}
where $a(t)$ is a scale factor. From the Einstein equation, 
we can derive a background Friedmann equation as
\begin{equation}
\left( {\dot{a} \over a} \right)^2 = {8 \pi G \over 3} \bar{\rho} (t),
\end{equation}
where a dot denotes the derivative in terms of $t$ and $\bar{\rho}$ is 
the background energy density.

On this background we consider a locally perturbed region that would eventually
collapse to a black hole. Such a region will be a very rare region in the
space. Hence, it may be approximated by a spherically symmetric region of positive
curvature. Since the comoving size of such a region is initially much larger than the
Hubble horizon size, one may apply the separate universe approach
or the leading order spatial gradient expansion to it,
that is, we may assume the metric of the form,
\begin{align}
ds^2= - dt^2 + a(t)^2e^{2\psi(r)}\delta_{ij}dx^idx^j\,,
\end{align}
where $\psi>0$ and is assumed to be 
monotonically decreasing to zero as $r\to\infty$.
It is known that the above form agrees with the
metric on comoving slices on superhorizon scales, where
$\psi$ corresponds to the nonlinear version of
the conserved comoving curvature perturbation,
that is $\psi={\cal R}_c$~\cite{Lyth:2004gb}.
The above metric can be cast into a more familiar form of a locally
closed universe with the metric,
\begin{align}
ds^2=-dt^2+ a(t)^2 \left[ {dR^2 \over 1 - K(R) R^2} 
+ R^2 (d\theta^2 + \sin^2 \theta d \varphi^2) \right]
\label{eq:closedU}
\end{align}
where the coordinates
$r$ and $R$ are related to each other as $R=re^{\psi(r)}$, and
$K$ is given by
\begin{align}
K=-\frac{\psi'(r)}{r}\frac{2+r\psi'(r)}{e^{2\psi(r)}}\,.
\end{align}
We note that the 3-curvature of the $t=$const. hypersurface is given by
\begin{align}
R^{(3)}=-\frac{e^{-2\psi}}{3a^2}\delta^{ij}
\left[2\partial_i\partial_j\psi+\partial_i\psi\partial_j\psi
\right]
=\frac{K}{a^2}\left(1+\frac{d\ln K(R)}{3d\ln R}\right)\,.
\label{eq:3-curvature}
\end{align}

Ignoring the spatial derivative of $K$ in the spirit of
leading order gradient expansion,
the time-time component of the Einstein equations (the
Hamiltonian constraint) gives
\begin{equation}
H^2 +{K(r) \over a^2}= {8 \pi G \over 3} \rho\,,
\label{eq:localFriedmann}
\end{equation}
where $H=\dot a/a$. This is equivalent to the Friedmann equation except for
a small inhomogeneity induced by the curvature term. One could regard
this as the Hamiltonian constraint on the comoving hypersurface, or
that on the uniform Hubble hypersurface on which the expansion
rate is spatially homogeneous and isotropic. 

The above equation naturally leads us to define the density contrast on
the comoving hypersurface by
\begin{equation}
\Delta := {\rho - \bar{\rho}\over \bar{\rho}}
= \frac{3K}{8 \pi G \bar{\rho}a^2}=\frac{K}{H^2a^2}\,.
\label{eq:delta}
\end{equation}
From the fact that $\bar\rho(t) \propto a^{-4}$ during the radiation-dominated
 universe, this is vanishingly small initially, being consistent with the
picture that it is the curvature perturbation that induces the density perturbation.

As the universe evolves $\Delta$ grows to become of order unity.
If we would ignore the spatial dependence of $K$, the universe
with $K>0$ would eventually stop expanding and recollapse.
This happens when $3K/a^2=8\pi G\rho$, 
namely when the comoving scale of this positively curved region becomes of the order of the Hubble
horizon scale, at which our separate universe approximation precisely
breaks down. Also the equivalence between the comoving and uniform
Hubble slices no longer holds. 
Nevertheless, we may expect that Eq. (\ref{eq:localFriedmann}) 
will still be used in obtaining a qualitatively acceptable criterion for the black hole formation,
which has been actually shown to be valid in fully nonlinear numerical studies.

Since $\Delta=1$ is the time when the universe stops expanding
if it were homogeneous and isotropic, let us assume this epoch
to be the time of black hole formation, $t=t_c$.
Since a perturbation on scales smaller than the Jeans length cannot collapse,
we set this to happen at $c_s^2k^2/a^2=H^2$ or $k^2/a^2=3H^2$ for $c_s^2=1/3$.
Namely, we have
\begin{align}
1=\Delta(t_c)=\frac{K}{k^2}\frac{k^2}{H^2a^2}=\frac{K}{c_s^2k^2}.
\end{align}
This implies we should identify $K$ with $c_s^2k^2$.
It is then straightforward to find the criterion for
the black hole formation. The condition is that the comoving slice density contrast
at the time when the scale of interest re-enters the Hubble horizon is
greater than $\Delta_c=c_s^2$,
\begin{align}
\Delta(t_k)=\frac{K}{H^2(t_k)a^2(t_k)}
=\frac{c_s^2k^2}{H^2(t_k)a^2(t_k)}\geq\Delta_c=c_s^2=\frac{1}{3}\,, \label{PBH-threshold}
\end{align}
where $t_k$ is the time at which $k/a=H$\footnote{Originally, in Ref. \cite{Carr:1975qj} there was an upper 
bound on $\Delta(t_k)$ in order to avoid the formation of the separate 
closed universe. However, recently Ref. \cite{Kopp:2010sh} pointed out that 
this is not the case. Actually from the point of view
of the curvature perturbation on comoving slices, it is by definition
the fact that $\Delta$ can never exceeds unity on superhorizon scales. 
}.

Within the scope of the present level of approximation, it is not meaningful to distinguish between
the Jeans length, $R=c_sH^{-1}$, and the Hubble horizon.
Crudely speaking, the mass of the formed PBH is equal to the horizon mass at the time of formation.

In the above, we have presented a basis of the PBH formation out of the primordial perturbation
based on the simple physical picture.
Although such analysis captures the essence of the PBH formation,
it is also important to clarify the impact of the various effects that have been ignored in the above discussion,
which are addressed below.

\subsubsection{Precise value of the threshold}
The equation (\ref{PBH-threshold}) only tells us that the PBH formation occurs when the density perturbation
becomes comparable to $1/3$.
A precise value of the threshold for PBH formation has been extensively
investigated both numerically and analytically \cite{Niemeyer:1997mt,Shibata:1999zs,Musco:2004ak,Polnarev:2006aa,Musco:2008hv,Nakama:2013ica, Harada:2013epa}.
For instance, Ref.~\cite{Harada:2013epa} has derived a new analytic formula for the threshold as 
$\delta^{\rm UH}_{Hc} = \sin^2 [\pi \sqrt{w}/(1+3w)]$ for the uniform overdense profile surrounded
by the underdense layer,
where $\delta^{\rm UH}_{Hc}$ is the amplitude of the density perturbation at the horizon crossing time 
in the uniform Hubble slice and $w$ is the equation of state of the dominant component in the Universe at the formation.
Although the analytic formula given above shows good agreement with the results of the numerical simulations,
numerical simulations also demonstrate that there is no unique value of the threshold. 
Different density/curvature perturbation profiles collape to BH above the different threshold,
which is quite natural from the physical point of view.
In terms of the comoving density perturbation, the spread of the threshold was found to be $0.3 - 0.66$.
Interestingly, the value $1/3$, which have been obtained in the crude approximation, lies in this range.
Notice that the spread changes if we use perturbation variable defined in the different time slicing.

\subsubsection{Effects of long wavelength modes}
The physical picture presented above shows that PBHs can form when the size of the 
overdense region becomes equal to the Hubble horizon.
Because of causality, any additional large-scale perturbations longer than the Hubble horizon at the
time of the PBH formation must not affect the PBH formation.
However, this is not explicitly visible for some perturbation variables.

For example, if we use the curvature variable $\psi$ on the uniform Hubble slicing,
which is identical to the comoving curvature perturbation $\mathcal{R}_c$
on super-horizon scales (and is also equal to the curvature perturbation on 
uniform energy density slicing $\zeta$),
the threshold depends on how much the longer wavelength modes are.
The reason for this is that the curvature variable is not the local quantity but
the quantity which requires information of the distant distribution of matter
for it to be determined \cite{Harada:2015yda}.
Although one can in principle use the curvature variable (or any other perturbation
variables as long as it is well-defined) for computing any observables such as
the abundance of the PBHs, appropriate prescription is needed to obtain the correct results
if the considered perturbation variable is not local.

In Ref. \cite{Young:2014ana}, the formation criterion is 
discussed in terms of the spatial 3-curvature and the density contrast
both of which naturally suppress the super-Hubble modes.
Actually the picture we presented in the above is based on this approach.
As clear from Eq.~(\ref{eq:3-curvature}), it is the spatial derivative
of $\psi$ that determines the 3-curvature, and when normalized with
the background expansion term $H^2$, it approximately equal to
the density contrast on the comoving hypersurface $\Delta$, as
defined in Eq.~(\ref{eq:delta}). 
Use of these quantities would be especially convenient when one studies the PBH formation out
of the inhomogeneities which are randomly distributed and whose overdensity regions are non-spherical.

\subsubsection{Near critical collapse}
It is known that when the overdensity $\delta$ is very close to the threshold 
there is a simple scaling-law for the mass of the formed black holes as \cite{Choptuik:1992jv,Evans:1994pj,Koike:1995jm}
\begin{equation}
M_{\rm BH} = C (\delta - \delta_{\rm th})^\gamma.
\end{equation}
In Refs.~\cite{Niemeyer:1997mt,Yokoyama:1998xd},
based on the above scaling-law, the mass function of the PBHs has been calculated.
The results show that the typical mass of the PBHs is still about the horizon mass evaluated
at the time of formation.
Yet, since the critical collapse produces smaller mass PBHs, mass function has a power-law tail
for the smaller mass range.
This could be important when one tries to precisely adopt the 
observational constraint for the abundance of PBHs with extended 
mass spectrum to constrain inflation models \cite{Carr:2016drx,Kuhnel:2015vtw}. 

\subsubsection{Non-sphericity of the overdense region}
In the above discussion, the criterion for the PBH formation is considered 
for an isolated spherically symmetric perturbation where
the universe approaches a perfectly homogeneous and isotropic
spatially flat universe in the large scale limit.
In reality, any overdense region for the random density perturbation is generically not completely spherically symmetric. 
Yet, compared to the studies of the PBH formation out of the spherically symmetric over density, 
studies on how the non-sphericity affects the PBH formation are short. 
In Ref.~\cite{Kuhnel:2016exn}, non-spherical overdense region was approximated as ellipsoid, 
and it was demanded that spherical region enclosed by the shortest axis of the ellipsoid satisfies 
the collapse criterion for the spherical overdensity as the collapse condition. 
Then, combining the above condition with the Carr’s argument \cite{Carr:1975qj} that relates the density contrast 
at the time of horizon crossing to the mass of the overdense region, 
it was shown that the threshold density contrast of the ellipsoid over density $\delta_{\rm ec}$ is given by
\be
\frac{\delta_{\rm ec}}{\delta_c} \approx 1+3e, \label{NSOR}
\ee
where $e$ is the ellipticity of the ellipsoid and $\delta_c$ is the threshold for the spherical overdensity. 
This shows that larger density contrast than the spherical one is needed for the non-spherical overdense region to turn into the PBH.
Although the numerical value on the right hand side of Eq.~(\ref{NSOR}) is obtained by the crude approximation in \cite{Kuhnel:2016exn},
this result is qualitatively natural since more overdensity is needed than the spherical case
to pull the overdense region along the longest axis by gravity. 
Thus, for primordial density perturbations obeying the probability distribution function for which 
higher-sigma peaks are more suppressed, 
which are realized in most inflation models, 
the overdense regions for the PBH formation are nearly spherically symmetric. 
More quantitative argument is possible if we further assume the Gaussianity for the density perturbations. 
In this case, typical value of the ellipticity for the peak amplitue $\delta$ much
greater than the square root of the variance $\sigma$ ($\delta \gg \sigma$) is given by 
\be
e \sim \frac{3\sigma}{\sqrt{10 \pi}\delta}.
\ee
Thus, if the PBHs are formed out of the $n$-sigma peaks,
the degree of the non-sphericity for the overdense regions is suppressed by ${\cal O}(n^{-1})$ 
\cite{1970Afz.....6..581D,Bardeen:1985tr,Kuhnel:2016exn}.

\subsubsection{PBH formation during a matter-dominated-like era}
Finally let us mention the case of PBH formation during a matter-dominated
era. Of course, primordial black holes may also form in
the matter-dominated era, and it has been also discussed in the 
literature, and it has been emphasized that taking account of
deviations from spherical configurations is 
essential~\cite{Khlopov:1980mg,Harada:2015ewt,Harada:2016mhb}. 
Besides, in reality the matter cannot be exactly dust with $P=0$ but
it behaves either as a fluid with small pressure or
as a collisionless fluid. These effects should be very important
but they have not been fully explored yet.
The massive scalar field is known to behave like a non-relativistic matter 
when it oscillates around the potential minimum. Thus, for models of inflation
in which the inflaton or another scalar field oscillates after inflation
and dominates the universe, the PBH formation during that era may become
important~\cite{Lyth:2005ze,Alabidi:2013wtp,Torres-Lomas:2014bua}.

\subsection{Clustering of PBHs}
\begin{figure}[tbp]
\begin{center}
\includegraphics[width=120mm]{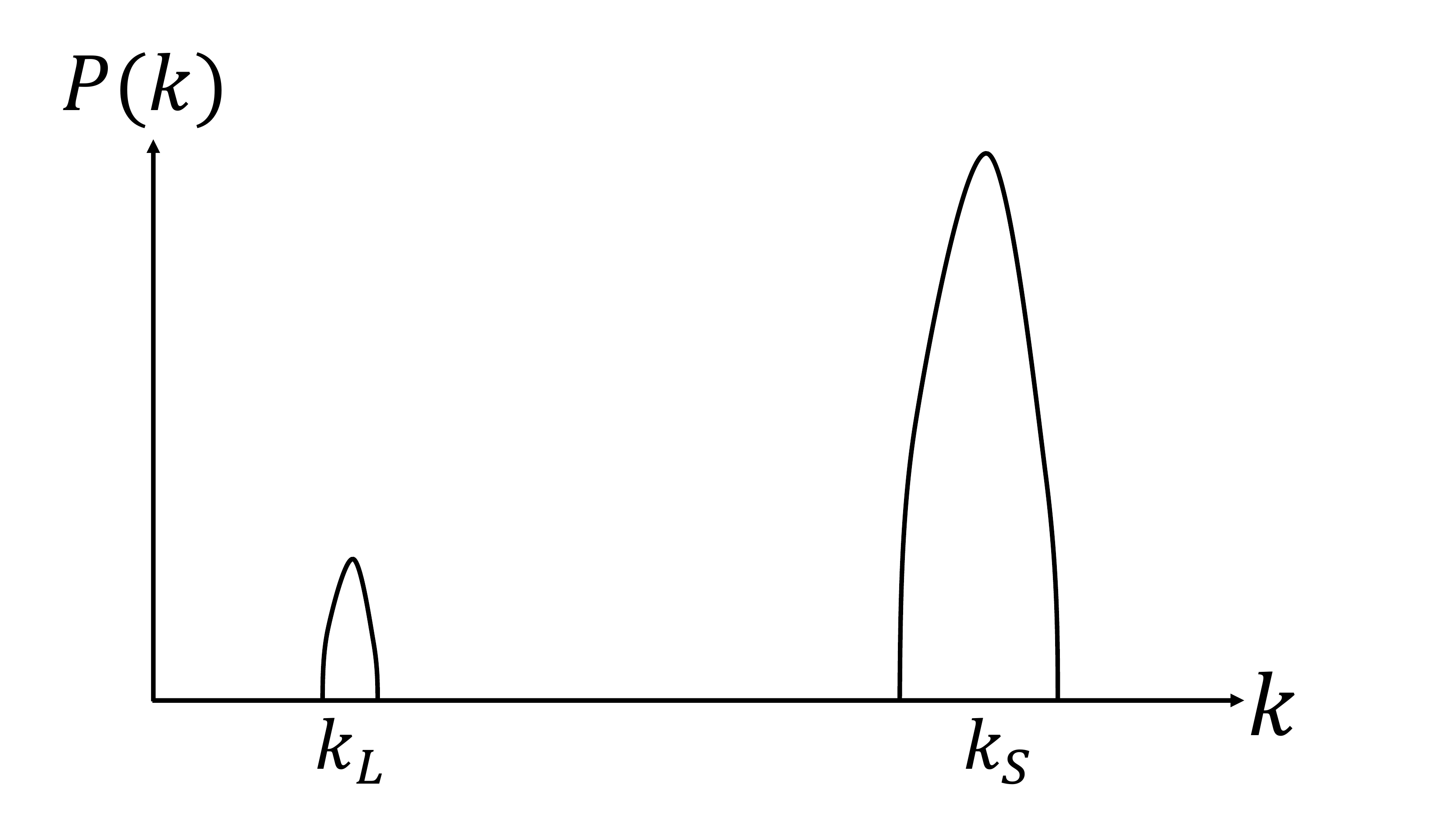}
\end{center}
\caption{Schematic figure describing the situation that primordial perturbations 
consist of short ($k_S$) and long ($k_L$) wavelength modes.
Amplitudes of the short wavelength are large and PBHs are formed out of the high-sigma peaks.}
\label{fig:clustering}
\end{figure}

Let us consider a simplified situation where primordial perturbations consist of two components, 
short and long wavelength modes,
and PBHs are formed out of the short wavelength modes (see Fig.~\ref{fig:clustering}).
We denote by $k_S$ and $k_L$ the representative comoving wavenumbers of the short and long wavelength modes, 
respectively ($k_S \gg k_L$).
At the time of the horizon crossing $k_S=aH$,
the long wavelength modes are still super-Hubble and such modes are absorbed into the homogeneous
component on the Hubble scale $a k_S^{-1}$.

If there are no correlations between the short and long wavelength modes, 
which is the case for the perturbations obeying the Gaussian statistics,
spatial distribution of PBHs at the formation time should not trace the large scale perturbations.
In fact, adopting the comoving density contrast as the perturbation variable and
the peak-background picture,
it was shown \cite{Tada:2015noa} that the effect of the large scale perturbations is significantly suppressed.
In other words, PBHs are little clustered initially on the comoving scale $k_L^{-1}$ \footnote{It has been claimed
in Ref.~\cite{Chisholm:2005vm} and later in Refs.~\cite{Clesse:2015wea, Clesse:2016vqa} that
PBHs are clustered initially even for the Gaussian density perturbations.
As far as the authors are concerned, there is not a broad consensus on this possibility.
}.
Their spatial distribution is uniform and the number density fluctuates according to the Poisson statistics.
Notice that even if the PBHs are not clustered initially, they can later cluster by the gravitational potential
created by the long modes just like the formation of dark matter halos in the $\Lambda$CDM scenario.

If, on the other hand, the short modes correlate with the long modes,
it can happen that the amplitudes of the short modes modulate by the long modes.
One example is the local-type non-Gaussianity for which the curvature variable $\psi$ on super-Hubble scales
can be written as 
\be
\psi ({\vec x})=\psi_g ({\vec x})+f_{\rm NL}\psi_g^2 ({\vec x}), \label{NG-perturbation}
\ee
where $\psi_g ({\vec x})$ is a Gaussian variable containing both short and long modes
and $f_{\rm NL}$ is the so-called non-linearity parameter \cite{Komatsu:2001rj} 
which is dimensionless and constant \footnote{The definition of $f_{\rm NL}$ in Eq.~(\ref{NG-perturbation})
is different from the standard one by a factor $3/5$. We have omitted this unessential factor to make equations simple. }.
This type of perturbation arises when the initially isocurvature modes are converted to the adiabatic modes. 
Decomposing $\psi_g$ into the short mode $\psi_{g,S}$ and the long mode $\psi_{g,L}$ as
\be
\psi_g =\psi_{g,S}+\psi_{g,S},
\ee
we find that $\psi$ can be written as
\be
\psi = \psi_S+\psi_L,
\ee
where $\psi_S$ and $\psi_L$ are defined by
\be
\psi_S=(1+2f_{\rm NL} \psi_{g,L}+f_{\rm NL} \psi_{g,S}) \psi_{g,S},~~~~~~
\psi_L=\psi_{g,L}+f_{\rm NL} \psi_{g,L}^2.
\ee
By definition, $|\partial_i \psi_S| \sim k_S \psi_S,~|\partial_i \psi_L| \sim k_L \psi_L$ hold.
Then, ignoring the terms suppressed by a factor $k_L$, 
the 3-curvature given by Eq.~(\ref{eq:3-curvature}) can be written as
\be
R^{(3)} \approx -\frac{e^{-2\psi_S}}{3a_L^2} \delta^{ij} 
(2\partial_i \partial_j \psi_S+\partial_i\psi_S \partial_j \psi_S),
\ee
where $a_L =e^{\psi_L} a$ is the local scale factor.
For $|\psi_S|={\cal O}(1)$, $R^{(3)}$ given above represents space curvature on the comoving scale $k_S^{-1}$.
Since $\psi_S$ contains the long mode $\psi_{g,L}$, the magnitude of the short mode 3-curvature 
modulates over the comoving scale $k_L^{-1}$.
As a result, PBH number density also modulates on the comoving scale $k_L^{-1}$,
and PBHs are clustered on the large scale $k_L^{-1}$.
For more details, see Refs. \cite{Young:2014oea,Tada:2015noa}.

\subsection{Abundance of PBHs}
\label{abundance-PBH}

In order to investigate the abundance of formed PBHs,
let us introduce a parameter which represents
the mass fraction (the energy density fraction)
of PBHs at the formation as $\beta$, which can be defined as
\begin{equation}
\label{def-beta}
\beta := {\rho_{\rm PBH} \over \rho_{\rm tot}} \biggr|_{{\rm at}~{\rm formation}}
= \left({H_0 \over H_{\rm form}}\right)^2 \left({a_{\rm form} \over a_0} \right)^{-3}\Omega_{\rm CDM} \, f_{\rm PBH},
\end{equation}
where $H :=\dot{a}/a$ is a Hubble parameter, $f_{\rm PBH}$ and $\Omega_{\rm CDM}$ are respectively a fraction of PBHs against the total dark matter component and a density parameter of the matter component at present, and ``form" and ``$0$", respectively, denote the values evaluated
at the formation and the present time. 
As we have mentioned in the previous subsection, the mass of PBHs formed in the radiation dominated era can approximately be evaluated to be
equal to the horizon mass, $M_{H} (:= (4 \pi / 3) \rho H^{-3}$ with $\rho$ being the total energy density of the Universe),
at the formation, and hence we have
\begin{eqnarray}
M_{\rm PBH} = \gamma \, M_{\rm H}\biggr|_{\rm at~formation} = \gamma \,{4 \pi \over 3} \rho_{\rm form} H_{\rm form}^{-3} 
&=& \gamma\,{4 \pi \over 3} {3 H_{\rm form}^2 \over 8 \pi G} H_{\rm form}^{-3} \cr\cr
&=& \gamma \, {1 \over 2 G} H_{\rm form}^{-1}\,.
\end{eqnarray}
Here, we introduce a correction factor, $\gamma$, which can be evaluated as $ \gamma \simeq 0.2$ in a simple analytic calculation~\cite{Carr:1975qj}.
By using the above relation between the mass of PBHs and the Hubble parameter
at the formation, mass fraction of PBHs, $\beta$, can be written as ({\it e.g.}, \cite{Carr:2009jm})
\begin{equation}
\beta \simeq 3.7 \times 10^{-9} \left( {\gamma \over 0.2} \right)^{-1/2} 
\left( {g_{\ast,{\rm form}} \over 10.75 } \right)^{1/4} \left( {M_{\rm PBH} \over M_\odot} \right)^{1/2} \, f_{\rm PBH} \,,
\label{eq:betaf}
\end{equation}
where $g_\ast$ is a number of relativistic degree of freedom. 
Thus, for each mass of PBHs, the observational constraint on $f_{\rm PBH}$
can be interpreted as that on $\beta$.


As we have shown in the previous subsection,
during radiation-dominated era, PBHs are basically 
considered to be formed when a sufficient overdense region,
corresponding to the density fluctuations
with a sufficiently large amplitude at a certain scale, enters the Hubble horizon.
Once the probability distribution function of the density fluctuations is given,
$\beta$ can be regarded as the probability that the density contrast is larger than the threshold for PBH formation, and we can evaluate the mass fraction $\beta$ as
\begin{equation}
\beta = \gamma \int_{\delta_{\rm th}}^{1} P(\delta) \, d\delta\,,
\end{equation}
where $\delta_{\rm th}$ is the threshold for PBH formation.
For the Gaussian distribution function, $\beta$ is approximately given by
\begin{eqnarray}
\beta (M_{\rm PBH})&=& \gamma \int^1_{\delta_{\rm th}} {d \delta \over \sqrt{2 \pi} \sigma_{M_{\rm PBH}} } \exp \left[ - {\delta^2 \over 2 \sigma^2_{M_{\rm PBH}} } \right] \cr\cr
&\approx& {\gamma \over \sqrt{2 \pi} \nu_{\rm th}} \exp \left[ - {\nu^2_{\rm th} \over 2} \right]\,,
\label{eq:beta}
\end{eqnarray}
where $\sigma_{M_{\rm PBH}}$ is the variance of the density fluctuations on the mass scale $M_{\rm PBH}$, and $\nu_{\rm th}:= \delta_{\rm th}/\sigma_{M_{\rm PBH}}$\footnote{Ref. \cite{Harada:2016mhb} found that for the PBH formation in the matter dominated era the production probability of PBHs, $\beta$, is approximately given by $\beta \approx 0.05556 \, \sigma_{\rm hor}^5$ for $\sigma_{\rm hor} \ll 1$. Here, $\sigma_{\rm hor}$ is a variance of the density fluctuations at the horizon re-entry.}.
The variance $\sigma_{M_{\rm PBH}}$ is estimated as
\begin{equation}
\sigma_{M_{\rm PBH}}^2 = \int d \ln k {\mathcal P}_\delta (k) W^2 (kR) = \int d \ln k W^2 (kR) {16 \over 81} (kR)^4{\mathcal P}_\zeta (k),
\end{equation}
where $\mathcal{P}_\delta$ and $\mathcal{P}_\zeta$ are, respectively, the power spectra of the primordial density fluctuations and the curvature perturbations on comoving slicing and $W(kR)$ is a window function smoothing over the comoving scale $R\,(\simeq 1/(a_{\rm form}H_{\rm form}) = 2 G M_{\rm PBH}/a_{\rm form} \, \gamma^{-1})$. 
Here, we assume $\sigma_{M_{\rm PBH}} \ll 1$.
Then, from the above expression, we can interpret the constraint on $\beta$ as that
on $\nu_{\rm th}$ which corresponds to the amplitude of the density fluctuations.
Furthermore, it would give a hint for constructing the successful inflationary models
with PBH formation.
As an example, if the constraint on $f_{\rm PBH}$ for $30~M_{\odot}$
would be obtained as $f_{\rm PBH} < 10^{-3}$, which is equivalent to $\beta < 3.6 \times 10^{-11}$, it could be interpreted as $\nu_{\rm th} \gtrsim 6.27$, that is,
$\sigma_{M_{\rm PBH}} \lesssim 0.08$ with $\delta_{\rm th} = 0.5$.

For the primordial power spectrum with a broad peak, it should be useful to give an expression for the mass function, which represents the fraction of PBHs with mass in $(M, M + d \ln M)$. It can be defined as
\begin{eqnarray}
{d f_{\rm PBH}(M) \over d \ln M}\, d \ln M \simeq  \nu (M)^2 \left| {d \ln \nu (M) \over d \ln M}\right| \times f_{\rm PBH} (M) \, d\ln M
\end{eqnarray}
with
\begin{equation}
f_{\rm PBH} (M) = 2.7 \times 10^8 \left( \frac{\gamma}{0.2} \right)^{1/2} 
\left( \frac{g_{*,{\rm form}}}{10.75}\right)^{-1/4}
\left( \frac{M}{M_\odot} \right)^{-1/2} \beta (M),
\end{equation}
and
\begin{equation}
\beta (M) \approx {\gamma \over \sqrt{2 \pi}\nu (M)} \exp \left[ - {\nu (M)^2 \over 2} \right]
\end{equation}
Here,
$\nu (M) : = \delta_{\rm th}/\sigma_M$, and
we have assumed the Gaussian distribution function for the primordial density fluctuations and $\nu (M) \gg 1$.

There are several works about the effect of the non-Gaussianity in
the estimation of $\beta$~\cite{Bullock:1996at,Ivanov:1997ia,Yokoyama:1998pt,Saito:2008em,Young:2013oia, Young:2015cyn,Kawasaki:2015ppx,Pattison:2017mbe}. In principle, the non-Gaussian feature of the statistics of the primordial curvature perturbations would strongly depend on the generation mechanism of the primordial fluctuations, that is, the inflation models, and we should evaluate $\beta$ for each exact form of the probability distribution function $P (\delta)$. However, it might be hard task to evaluate the exact form of the probability distribution function for each inflationary model, and hence it should be convenient to give an approximate formulation to take into account the non-Gaussian effect. One of such formulations can be considered by the cumulant expansion of the probability distribution functions \cite{Saito:2008em}.
In the standard single slow-roll inflation, the higher order cumulants are expected to be suppressed by the slow-roll parameter. However, as shown in the next subsection,
for efficient PBH formation we need to consider non-standard inflationary models with the violation of the slow-roll condition. Thus, the effect of the non-Gaussianity can be relevant for some inflationary models \cite{Young:2015cyn}.

\subsection{Generating seeds of PBH from inflation}

Here, we focus on the mechanism where the PBH formation is induced by the primordial density fluctuations, and we give a review of inflationary mechanism which could generate seeds of the PBHs,
that is, primordial curvature perturbations with large amplitudes at a certain scale.

For preparation, let us give a relation between the mass of PBHs and the comoving wavenumber of the perturbations.
Basically, as we have discussed in the previous subsection, PBH could be formed when an overdense region enters the Hubble horizon.
If such an overdense region is sourced from the primordial curvature perturbations,
the size of the overdense region should be characterized by the comoving wavenumber of the primordial perturbations, $k$.
Then, we can obtain the relation between the Hubble scale at the PBH formation and the comoving wavenumber
of the sourced primordial curvature perturbations as
\begin{equation}
a H \biggr|_{\rm at~formation} = k.
\end{equation}
During the radiation dominated era, we have $a \propto H^{-1/2}$ and hence the relation between
the comoving wavenumber, $k$, and the Hubble parameter at the formation as
$
H_{\rm form} \propto k^2  
$.
Substituting the relation between $k$ and $H_{\rm form}$ into the above expression for $M_{\rm PBH}$, we can obtain the relation between the mass of PBHs and the comoving wavenumber as~\cite{Kawasaki:2016pql}
\begin{equation}
M_{\rm PBH} (k) \simeq 30 \, M_\odot \left( {\gamma \over 0.2} \right) \left( {g_{\ast,{\rm form}} \over 10.75} \right)^{-1/6}
\left( {k \over 2.9 \times 10^5 \,{\rm Mpc}^{-1}}\right)^{-2} \,.
\end{equation}

From this equation, we find that a hierarchy between the observable scales by CMB observations and $30 \,M_\odot$-PBHs
is given by\footnote{%
Here, we use a pivot scale in Planck 2015 as a typical observable scale by CMB observations, $k_{\rm CMB}$.
}
\begin{equation}
{\mathcal N}_{{\rm CMB} - 30 M_\odot {\rm PBHs}} := \ln {k_{30\,M_\odot{\rm PBHs}} \over k_{\rm CMB} } 
= \ln {2.9 \times 10^5 \,{\rm Mpc}^{-1} \over 0.002 \,{\rm Mpc}^{-1}} \sim 20.
\end{equation}  
On the other hand, the number of $e$-folds measured from
the time when the present horizon scale exits the Hubble horizon during inflation
to the end of inflation is required to be typically $50$-$60$.
Thus, based on the above discussion,
we find that, 
if we want to form $30M_\odot$ PBHs from the primordial density perturbations,
we need to consider some mechanism which can amplify the perturbations during the inflationary phase\footnote{Refs.~\cite{Kodama:1982sf, Dolgov:1992pu, Jedamzik:1996mr,Jedamzik:1999am,Khlopov:1999ys} discussed the PBH formation during QCD phase transition era with the assumption that the phase transition is first order. However, currently the QCD phase transition has been known to be not first order but a crossover, and hence the efficient PBH formation during QCD epoch might be difficult.}.

%
\begin{figure}[htbp]
\begin{center}
\includegraphics[width=100mm]{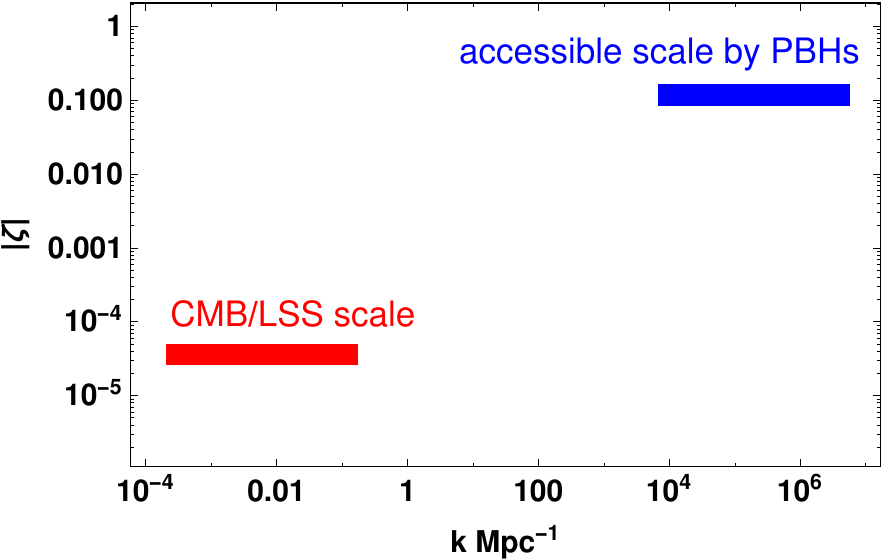}
\end{center}
\caption{Hierarchy between the CMB/LSS scale and accessible scale by PBHs. PBHs can be a powerful tool to study much smaller scales of primordial fluctuations.}
\label{fig:scale}
\end{figure}

\subsubsection{Single-field inflation}

First, let us consider the possibility of generating $30 M_\odot$ PBHs in the standard slow-roll inflationary scenario.
For the single-field standard slow-roll inflation, the power spectrum of the primordial curvature perturbations on comoving slicing is given by
\begin{equation}
\label{eq:slowroll}
{\mathcal P}_{\mathcal{R}_c} (k) \sim |\mathcal{R}_c (k)|^2 
\sim \left( {H^2 \over \dot{\phi}} \right)^2_{a H = k} ,
\end{equation}
where $\dot{\phi}$ is the time derivative of the inflaton field and the subscript ``$aH = k$" represents the
value at a time when the scale of interest exits the Hubble horizon during inflation. 
For the power-law ansatz;
\begin{equation}
\label{eq:powerlaw}
{\mathcal P}_{\mathcal{R}_c} (k) := A_{\mathcal{R}_c} \left({k \over k_\ast}\right)^{n_s - 1},
\end{equation}
Eq.~(\ref{eq:slowroll}) gives a power-law index as\footnote{
This expression is derived based on the slow-roll equations:
\begin{eqnarray}
H^2 \simeq {8 \pi G \over 3} V(\phi)~, \qquad 3 H \dot{\phi} \simeq - V_\phi \nonumber
\end{eqnarray} 
}
\begin{equation}
n_s -1 = - 6 \epsilon + 2 \eta ,
\end{equation}
where $\epsilon$ and $\eta$ are slow-roll parameters which can be characterized in terms of the potential of the inflaton field as
\begin{equation}
\epsilon := {1 \over 16 \pi G} \left({V_{\phi} \over V} \right)^2,~\eta := {1 \over 8 \pi G} {V_{\phi\phi} \over V},
\end{equation}
where $V_\phi := dV(\phi)/d\phi$ and $V_{\phi\phi} := d^2 V(\phi) / d\phi^2$.
Thus, positive large $\eta$ can realize the blue-tilted power spectrum, that is, larger amplitude for smaller scales (larger comoving wavenumber).

Current CMB observations have indicated that ${\mathcal P}_{\mathcal{R}_c}$  is about $10^{-9}$ (often called as COBE normalization) over CMB observable scales
and the PBH formation is effective enough to be observationally-interesting for ${\mathcal P}_{\mathcal{R}_c} = O(10^{-2} - 10^{-1})$. Thus, if 
we want to realize such a large amplitude at $k = k_{30\,M_\odot{\rm PBHs}} = 2.9 \times 10^5 \,{\rm Mpc}^{-1}$ by blue-tilted power spectrum, consistent with the COBE normalization,
we need
\begin{equation}
n_s - 1 = {\ln ({\mathcal P}_{\mathcal{R}_c} (k_{30\,M_\odot{\rm PBHs}}) / {\mathcal P}_{\mathcal{R}_c} (k_{\rm CMB}))
\over \ln (k_{30\,M_\odot{\rm PBHs}} /k_{\rm CMB})}  \simeq 0.85.
\end{equation}

For successful inflation, we need to require $\epsilon, \eta \ll 1$. Hence, the above value for the spectral index
seems to be large and it is not so easy to realize in the standard inflationary models.
Furthermore, we have a strong constraint on the spectral index obtained from CMB observations, e.g., from
the Planck observation, we have \cite{Ade:2015xua}
\begin{equation}
n_s = 0.968 \pm 0.006 ~~~~~~~{\rm at}~k_\ast=0.05\,{\rm Mpc}^{-1}\,.
\end{equation}
Thus, for the PBH formation, even if we can realize the blue-tilted power spectrum,
the single power-law model should be in conflict with the CMB observations.

\bigskip
\noindent
$\bullet$ {\bf Running mass inflation model}

\begin{figure}[htbp]
\begin{center}
\includegraphics[width=100mm]{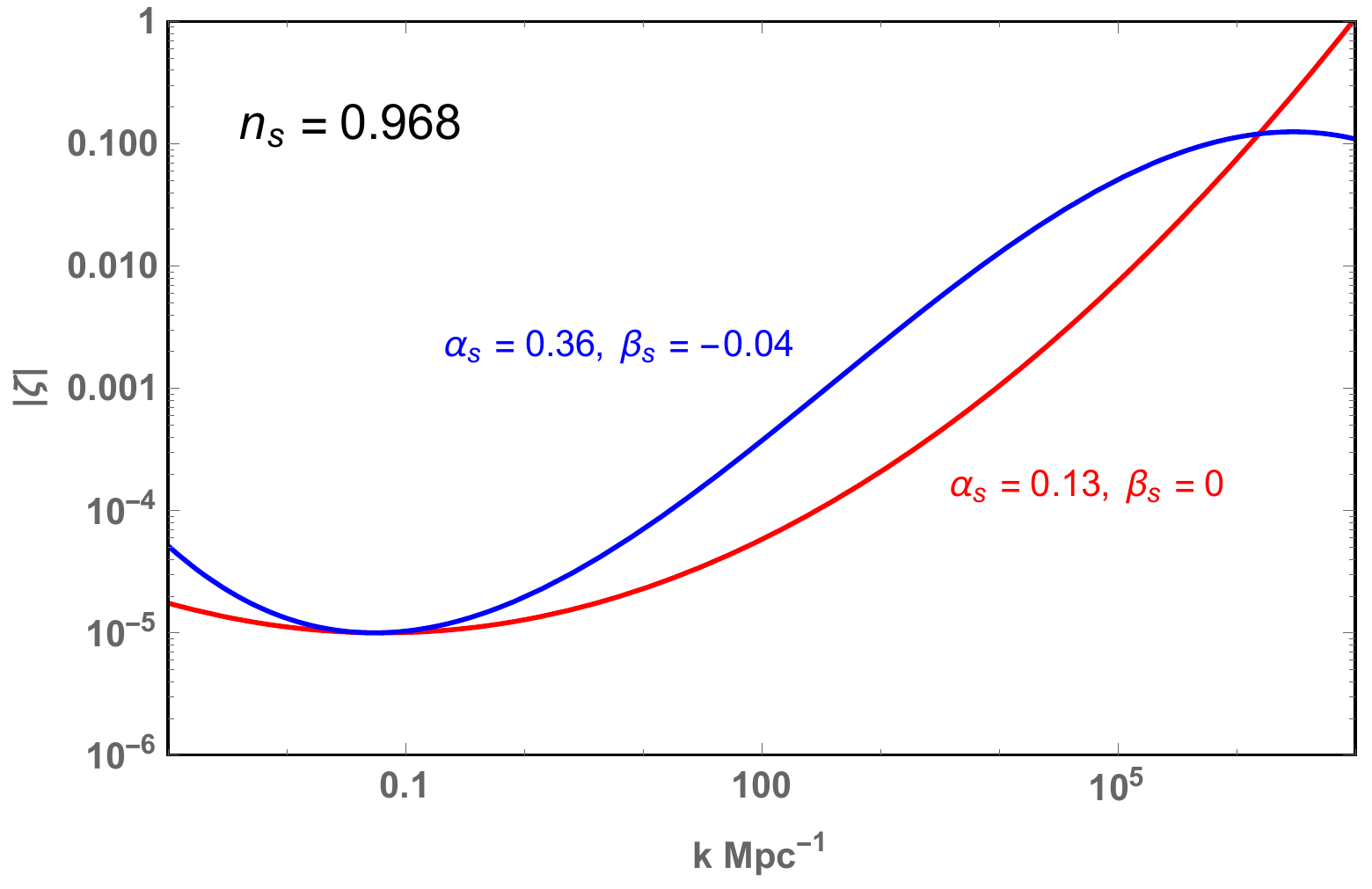}
\end{center}
\caption{Primordial power spectrum with the large running of spectral index and running of running parameters.}
\label{fig:running}
\end{figure}

One possibility of realizing relatively large blue-tilted inflationary model
consistent with the Planck result
is large running mass inflation model \cite{Alabidi:2012ex,Drees:2011hb,Drees:2011yz}\footnote{Here, we focus on inflation models which could produce PBHs with $O(1-10) {\rm M}_\odot$. Smaller PBHs formation in a running mass inflation model has been discussed also in~\cite{Leach:2000ea,Kohri:2007qn,Alabidi:2009bk,Bugaev:2008gw}.}.
In this model, the time dependence of $\eta$ during inflation could be large, that is, the scale-dependence
of $n_s$ could also become large (see, e.g., ~\cite{Stewart:1996ey,Stewart:1997wg})\footnote{Note that, in order to realize the end of inflation, this type of model needs a hybrid-inflation-type mechanism with a waterfall field other than the inflaton. In this sense, the running mass inflation model is not strictly a single field model~\cite{Stewart:1997wg}.}. Thus, when $\eta$ remains to be small 
so as to be consistent with the Planck result on CMB scales, but $\eta$ takes a  positive larger value on smaller scales,
we can realize the large amplitude of primordial curvature perturbations on an appropriate small scale
which could be seeds of PBHs.
For such kind of models, the primordial power spectrum is not a simple power-law
(\ref{eq:powerlaw}), but the one that 
includes the scale-dependence of the spectral index perturbatively:
\begin{equation}
{\mathcal P}_{{\mathcal{R}_c}} (k) := A_{\mathcal{R}_c} \left({k \over k_\ast}\right)^{n_s - 1+{1 \over 2!} \alpha_s \ln(k  / k_\ast) + {1 \over 3!} \beta_s \ln^2 (k / k_\ast) + \cdots},
\label{eq:runningpower}
\end{equation}
where $\alpha_s$ and $\beta_s$ are, respectively, called ``running of spectral index" and ``running of running" parameters. In the standard slow-roll inflation models, $\alpha_s$ and $\beta_s$ are highly suppressed by slow-roll parameters. However, in the large running mass inflation model,
these parameters would be relatively large. For the above parameterization,
in order to produce $30 {\rm M}_\odot$ PBHs with $n_s = 0.968$ on CMB scales,
we need to take $\alpha_s$ to be about 0.1  as shown by the red line in Fig.~\ref{fig:running}.
%

However, assuming the negligibly small running of running parameter $\beta_s$,
for such a large $\alpha_s$
PBHs with smaller masses (corresponding to  larger comoving wavenumbers) would be overproduced.
To avoid the overproduction of smaller PBHs,
we need to have a cutoff in the primordial power spectrum at an appropriate scale.
Based on the parameterization given as (\ref{eq:runningpower}),
such a cutoff can be realized by taking into account non-negligible running of running $\beta_s$.
As shown in Fig. \ref{fig:running}, by employing appropriately tuned positive $\alpha_s$ and negative $\beta_s$, one can realize a broader peak in the power spectrum and it predicts the broad mass spectrum of PBHs. 

Recent Planck result gives constraints not only on the spectral index $n_s$, but also on
the running of the spectral index $\alpha_s$ as~\cite{Ade:2015lrj}
\begin{equation}
\alpha_s = -0.003 \pm 0.007 ~{\rm at}~k_\ast=0.05\,{\rm Mpc}^{-1}\,.
\end{equation}
Thus, by comparing this observational constraints with the value for $\alpha_s$ shown in Fig.~\ref{fig:running},
we find that 
it is difficult to construct a viable inflationary model which can produce 
$30~{\rm M}_\odot$ PBHs based on
the perturbative scale-dependent spectral index (\ref{eq:runningpower}).

\bigskip
\noindent
$\bullet$ {\bf Inflection inflation model}
%
\begin{figure}[htbp]
\begin{center}
\includegraphics[width=100mm]{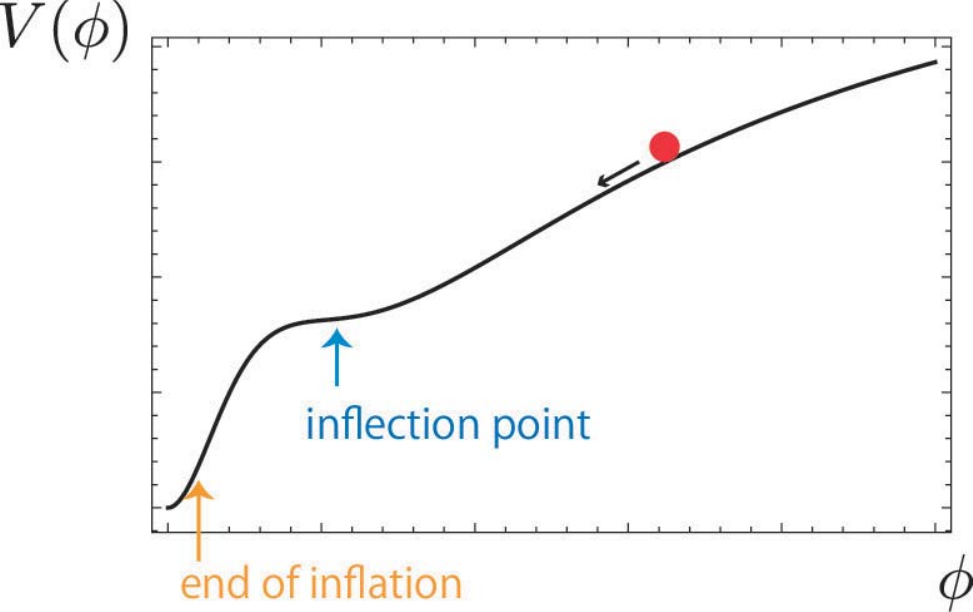}
\end{center}
\caption{A potential for the inflection model}
\label{fig:inflection}
\end{figure}

Ref.~\cite{Garcia-Bellido:2017mdw} recently proposed a single field model which can produce the primordial power spectrum with a peak. They consider the inflaton's potential with a inflection (plateau) point where
the inflaton temporarily slows down during inflationary phase.
PBH formation in such an inflationary scenario with a ``plateau" in the scalar potential was discussed in Ref.~\cite{Ivanov:1994pa}. Ref.~\cite{Ivanov:1994pa} calculated the spectrum of the adiabatic perturbations 
for the schematic representation of the scalar potential which has two breaks and a flat plateau between these breaks\footnote{Ref.~\cite{Bullock:1996at} also studied PBH formation in several toy models.}.
The reason why such a model can produce a peak in the primordial spectrum
can be easily understood as follows.
Let us recall the expression of the primordial power spectrum given by (\ref{eq:slowroll}).
This expression can be rewritten in terms of the slow-roll parameter as
\begin{equation}
{\mathcal P}_{\mathcal{R}_c} (k) = \left( {8 \pi G H^2 \over \epsilon} \right)_{a H = k}.
\end{equation}
From this expression, one can find that when the inflaton temporarily slows down,
the slow-roll parameter $\epsilon$ becomes more suppressed and the power spectrum have a peak
at the scales which exit the Hubble horizon during the slow-down phase.
However, if the plateau is completely flat there appears a problem that
the inflaton may stay too long at the plateau and the inflationary phase eternally continues, so-called eternal inflation.
To avoid this problem, in Ref.~\cite{Garcia-Bellido:2017mdw},
a ``near"-inflection point has been introduced in the inflaton potential.
Such kind of models in the context of PBH formation have also been discussed in Refs. \cite{Ezquiaga:2017fvi,Kannike:2017bxn,Germani:2017bcs,Motohashi:2017kbs}.

In fact, as pointed out in Ref.~\cite{Motohashi:2017kbs},
the standard slow-roll conditions might be generally violated near the inflection point. In the standard slow-roll approximation,
we approximate $\dot{\phi} \simeq - V_\phi / (3H)$
where $V_\phi := dV/d\phi$ and $\phi$ is an inflaton field,
and this means that in the equation of motion of the inflaton
we can neglect the acceleration of the inflaton, $\ddot{\phi}$, term. However, if $V_\phi$ becomes too small as around the inflection point, in the equation of motion $3H \dot{\phi}$ term
would become balanced with $\ddot{\phi}$ term, that is,
$|\ddot{\phi}| \approx |3 H \dot{\phi}| (\gg |V_\phi|)$.
Thus, the above discussion based on the slow-roll parameters would be violated. 
In fact, based on the detailed calculation, in such an inflection-point inflation, the amplification of the primordial density fluctuations can be realized, but ${\mathcal P}_{{\mathcal R}_c} \simeq 10^{-4}$ at most \cite{Garcia-Bellido:2017mdw,Motohashi:2017kbs}.  

%
This type of model tends to predict broader peak, $\Delta \log k \gtrsim O(10)$, in the primordial power spectrum, and hence
the mass spectrum of the formed PBHs would also be broader~\cite{Garcia-Bellido:2017mdw}.

\bigskip
\noindent
$\bullet$ {\bf Single-field chaotic new inflation model}

\begin{figure}[htbp]
\begin{center}
\includegraphics[width=100mm]{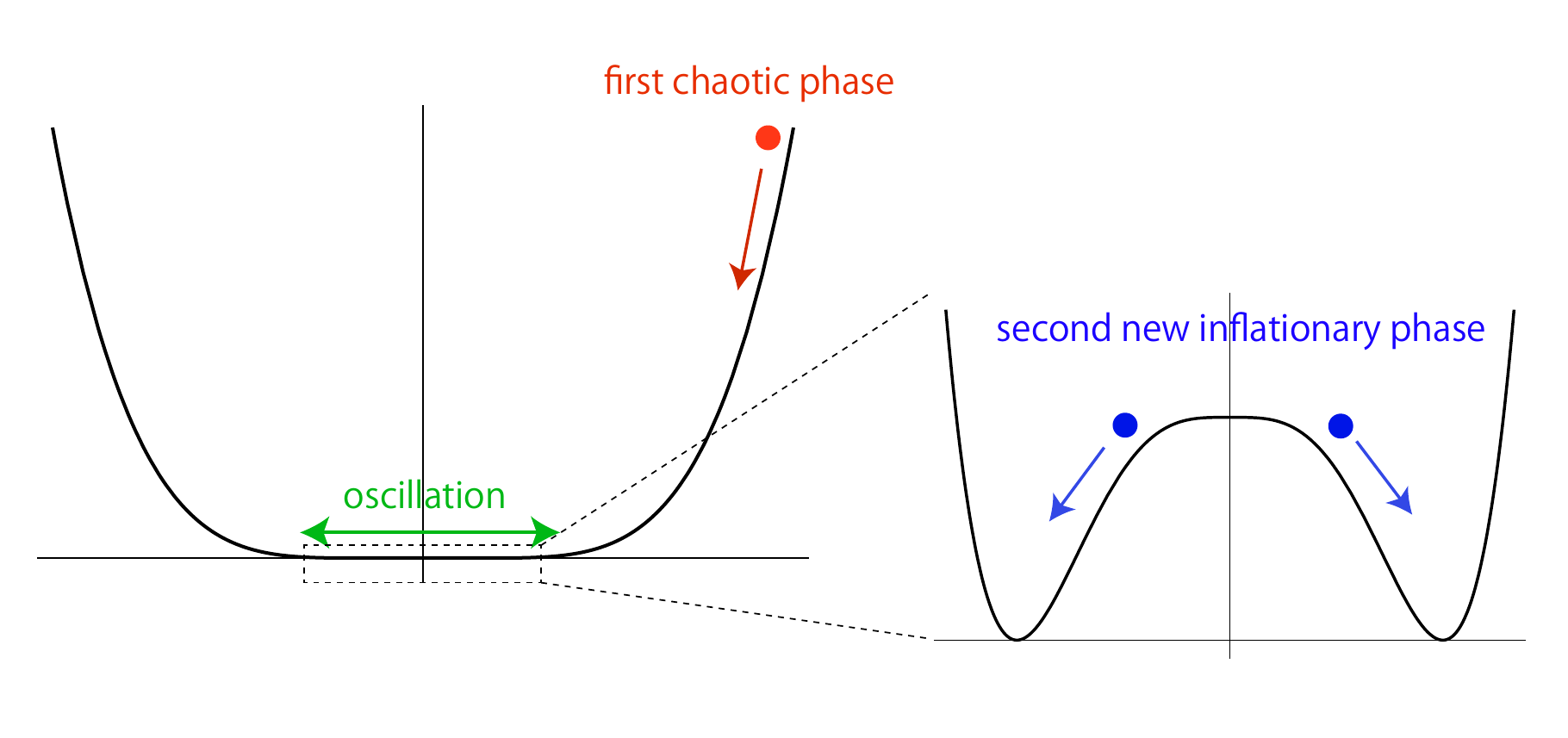}
\end{center}
\caption{A schematic representation of the scalar potential of the chaotic new inflation scenario.}
\label{fig:chaotic_new}
\end{figure}

Ref.~\cite{Yokoyama:1998pt,Saito:2008em} proposed the possibility of PBH formation
in the context of double inflation even for the single field case, called chaotic new inflation.
In this scenario, the inflaton potential is basically given by so-called Coleman-Weinberg or double-well type potential
which have been studied in the context of new inflation scenario (Fig.~\ref{fig:chaotic_new}).

In the standard new inflation scenario  inflation is caused by
an inflaton which starts slow-roll from the vicinity of 
the origin which is an unstable local maximum.
On the other hand, 
if the initial amplitude of the scalar field is Planck scale largely exceeding  the global minimum,
it causes chaotic inflation. 
For such kind of potential, if we tune the model parameters,
two inflationary phases 
could be realized.
As a first phase of  inflation,
chaotic inflation occurs with a large field value of the inflaton.
After the phase of chaotic inflation,
the inflaton oscillates around the origin as well as in the standard chaotic inflationary model. However, in this model the potential has a feature which can realize new inflation around the origin. Thus, after the oscillation, with some tuning of the model parameters, new inflation occurs as a second phase of the inflation.  


In this scenario,
the amplification of the primordial curvature perturbations
would be expected to be efficient for the modes
which exit the horizon at the transition era from the first phase to the second,
where the slow-roll conditions are temporarily violated.
As in the case of the inflection point inflation model,
around the local maximum the slow-roll conditions are temporarily violated and then the enhancement of the curvature perturbations
occurs.
This model would predict relatively-sharp peak in the power spectrum of the curvature perturbations, $\Delta \log k \lesssim O(10)$, and PBH mass spectrum  becomes even narrower\cite{Saito:2008em}.

\subsubsection{Multi-scalar inflation}

Next, let us consider multi-scalar inflation models which can produce seeds of PBHs, {\it i.e.}, the primordial curvature perturbations with large amplitudes.

\bigskip
\noindent
$\bullet$ {\bf double inflation model}

\begin{figure}[htbp]
\begin{center}
\includegraphics[width=100mm]{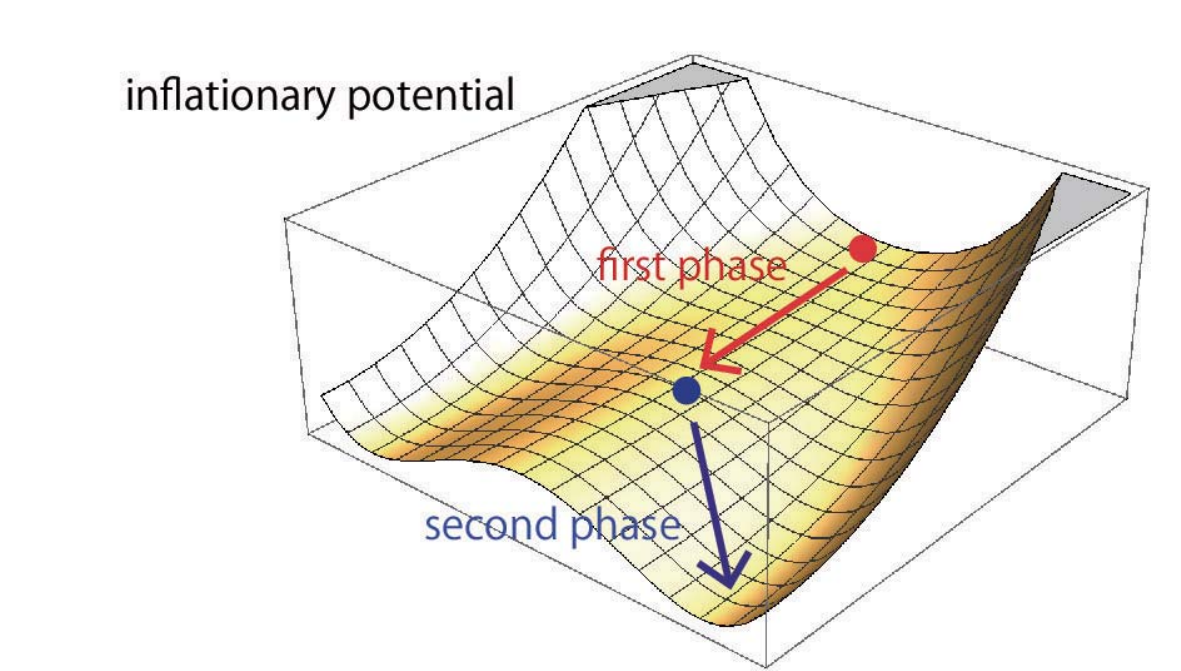}
\end{center}
\caption{A schematic picture of dynamics of the double inflation based on the hybrid model}
\label{fig:hybrid}
\end{figure}

As we have discussed in single-field inflation case,
if we have a phase where the inflaton temporarily slows down during inflation, corresponding to
the very flat region in inflationary potential,
we can realize a peak in the primordial power spectrum at an appropriate scale.
The important point in the above models to realize the enhancement of the primordial curvature perturbations
to form PBHs during inflationary era
is temporal violation of the standard slow-roll conditions.
In case of multi-scalar models, such kind of feature in the inflation dynamics
could be easily realized~\cite{Silk:1986vc,Randall:1995dj}.
Ref. \cite{GarciaBellido:1996qt} investigated the possibility of PBH production in the context of  hybrid inflation,
where  inflationary dynamics depends on two scalar fields.

In the original types of hybrid inflation, 
one scalar field which possesses a false vacuum energy is assumed to be massive,
and  inflation was driven by this false vacuum energy.
Another field slow-rolls down the flat potential,
and when it reaches a critical value, the massive field becomes tachyonic
and inflation abruptly ends. 
In general,
we can also consider the case where both fields existing in hybrid inflation are not massive
\footnote{Non-Gaussianity of the super-Hubble perturbations 
in this case was shown to be very tiny \cite{Tanaka:2010km}.}.
In such a case, we can realize two stages of inflation (double inflation) and a phase of the transition.
Basically, in the phase of the transition, fields roll down the very flat region in the potential
and hence the primordial curvature perturbations which exit the Hubble horizon during this phase
may be amplified, as we have discussed in the inflection model. 

Now lots of models in the context of double inflation have been proposed for PBH formation\cite{Clesse:2015wea}.
Refs.~\cite{Kawasaki:2016pql,Kawasaki:1997ju,Kawasaki:2006zv} discussed the construction of such double inflation models 
in supergravity, named smooth hybrid new inflation.

In this type of scenarios, the  expected power spectrum of the
curvature perturbations strongly depends on the details of the models. 
As an example, as shown in Refs. \cite{Kawasaki:2016pql,Inomata:2017bwi,Inomata:2016rbd},
the expected power spectrum has two peaks, one of which
is very sharp and the other is much broader.

\bigskip
\noindent
$\bullet$ {\bf curvaton model}

Another possibility of generating the primordial curvature perturbations with large amplitude at a certain scale
can be considered in the context of so-called curvaton scenario.
In the simple curvaton scenario, we have two scalar fields, one (called inflaton) is responsible for the accelerating expansion of the Universe, that is, inflation, and the other (called curvaton) responsible for generating the primordial curvature perturbations.
As we have mentioned, one of the difficulties of PBH formation in the single field inflation
is that we should realize the large amplitude of fluctuations at a small scale for PBH formation
and the COBE normalization on the CMB scales for the primordial curvature perturbations, simultaneously.
In some extension of the simple curvaton scenario, the primordial curvature perturbations
could be generated both from the inflaton and the curvaton fluctuations.
Thus, if we can construct a model in which the primordial curvature perturbations on the CMB scales
are generated from the inflaton fluctuations while those on small scales are generated from the curvaton fluctuations,
we can realize a successful inflationary model for PBH formation consistent 
with the CMB observations
\cite{Yokoyama:1995ex, Kawasaki:2012wr,Ando:2017veq}\footnote{In Ref.~\cite{Kohri:2012yw}, PBH formation has been discussed in the scenario where the curvaton induces a second inflationary phase.}.

\begin{figure}[htbp]
\begin{center}
\includegraphics[width=110mm]{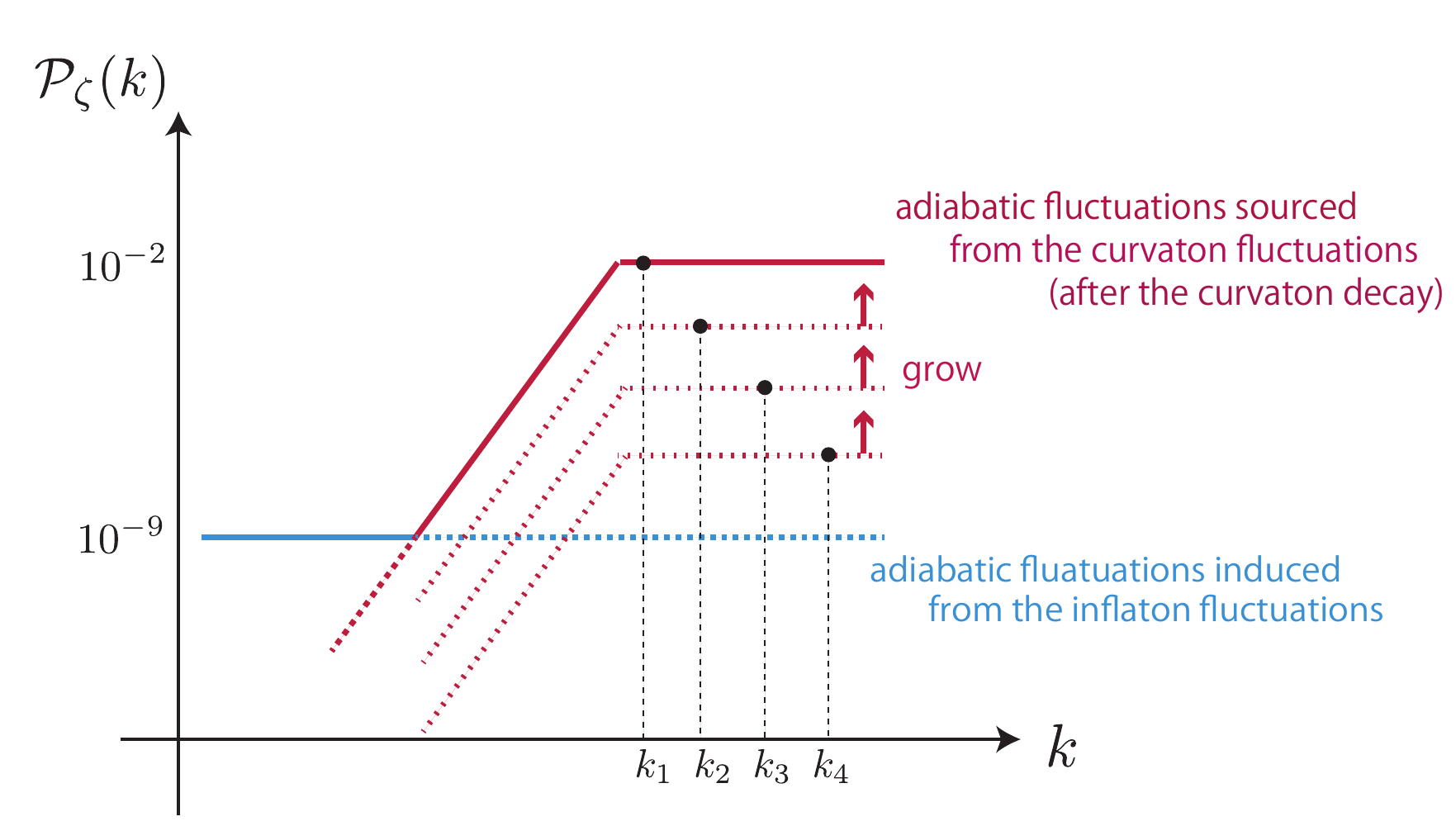}
\end{center}
\caption{A schematic feature of the primordial power spectrum in the curvaton scenario where the efficient PBH formation can be realized. In this figure, $k_1,~k_2,~k_3,$and $k_4$ correspond to the modes which re-enter the horizon at $t=t_1,~t_2,~t_3,$ and $t_4$ $(t_{\rm dec}>t_1>t_2>t_3>t_4)$, respectively. Here $t_{\rm dec}$ represents a time when the curvaton decay.}
\label{fig:curvaton}
\end{figure}

Basically, in this curvaton scenario, the power spectrum of the curvaton fluctuations (primordial isocurvature perturbations) has a cut-off on a large scale and a scale-invariant feature on smaller scales. During the radiation dominated era after the reheating, the curvaton field starts to oscillate and it behaves as a non-relativistic matter.
Thus, the energy density of the curvaton field gradually gets to contribute to the total energy density of the Universe and it means that the curvaton fluctuations,  which are initially isocurvature perturbations, are gradually transformed into the adiabatic curvature perturbations, {\it i.e.}, the adiabatic curvature perturbations grow in time. In the standard curvaton scenario, the curvaton decays into the radiation at a certain time. After the curvaton decays, the adiabatic curvature perturbations become constant in time. As shown in Fig.~\ref{fig:curvaton}, in this scenario the both components of primordial curvature perturbation spectra at the horizon re-entering are smooth. 
Thus, the mass spectrum of the formed PBHs would be broad.
In \cite{Suyama:2011pu}, it was shown that if there are multiple curvaton fields, 
they can temporarily enhance the curvature perturbations on all super-Hubble scales and 
the PBHs could be copiously produced during those modes re-enter the Hubble
horizon until the enhancement terminates \footnote{The temporal enhancement of the
curvature perturbations and the PBH formation were also studied in the case where
the non-inflaton field has a Galileon-type kinetic interaction \cite{Suyama:2014vga}.
}.
There are also several models where the primordial curvature perturbations were enhanced by non-trivial interactions
between the inflaton and some fields, e.g., a gauge field \cite{Garcia-Bellido:2016dkw}.

\newpage


\section{Observational constraints on non-evaporating PBHs $(\gtrsim 10^{15}~{\rm g})$}
\label{Observational-constraint}
In this section, we review how the upper limit on the fraction of PBHs in dark matter
$f_{\rm PBH}\equiv \Omega_{\rm PBH}/\Omega_{\rm DM}$ can be obtained for various mass
range of non-evaporating PBHs ($M_{\rm PBH} \gtrsim 10^{15}~{\rm g}$).
Broadly speaking, constraints are divided into direct and indirect ones.
The direct constraints are derived by investigating the observational effects that
PBHs directly trigger by their gravitational potential and are thus 
independent of the mechanisms of the PBH formation.
The direct constraints can be further classified into four categories by
the manners PBHs affect:
gravitational lensing, dynamical effect, the accretion, and the growth of large scale structure.
All these topics are covered in this section.
Indirect constraints are those that can be obtained by the observational effects 
that are not caused by the PBHs but something else which is deeply connected to PBHs.
Although such constraints cannot be applied to all the possible PBH scenarios,
they are powerful in excluding some of them.
Several known indirect constraints are also reviewed in the latter part of this section.

The existing constraints will be tightened (or may be replaced by the detection of PBHs) 
in the future both by the improvement of the apparatuses similar to the existing ones
and the launch of the qualitatively new observations such as $21~{\rm cm}$ lines.
Future constraints are also briefly touched at the end of this section.

All the constraints discussed in this section are based on the electromagnetic signals.
The new observable, namely gravitational waves, in the context of PBHs is the topic 
in the next section.
It is important to keep in mind that all the constraints in this section
are derived under the assumption that the PBH mass function is monochromatic,
which is valid when the width of the mass function is sufficiently narrow.
The case for a broad mass spectrum requires separate analysis, which will be briefly addressed 
at the last part of this section.

Before ending the introduction of this section,
we briefly mention the constraints on the PBHs lighter than $\sim 10^{15}{\rm g}$
that have already evaporated or are in the final state of evaporation 
by the Hawking radiation \cite{Carr:2009jm}.
Although those PBHs, depending on their mass, 
do not exist or are fading at the present epoch, 
high-energy partciles emitted from PBHs leave some signals from which we can place
the upper limit on the PBH abundance. 
They include production of the lightest supersymmetric particles (if they exist) ($10^4{\rm g} < M_{\rm PBH} < 10^9 {\rm g})$, 
entropy production in the early universe ($10^6{\rm g} < M_{\rm PBH} < 10^9 {\rm g})$,
change of the abundance of the light elements produced 
by the big bang nucleosynthesis ($10^9{\rm g} < M_{\rm PBH} < 10^{13} {\rm g})$,
extragalactic photon background ($10^{14}{\rm g} < M_{\rm PBH} < 10^{15} {\rm g})$, 
and damping of the CMB temperature anisotropies
on small scales by modifying the cosmic ionization history ($10^{13}{\rm g} < M_{\rm PBH} < 10^{14} {\rm g})$.
By comparing these effects with observations, upper limits on the PBH fraction 
$\beta$ defined by Eq.~(\ref{def-beta}) for various PBH mass (assuming monochromatic
mass function) can be obtained 
(for comprehensive study on this topic, see Refs.~\cite{Carr:2009jm,Josan:2009qn}).
Except for the constraints from the entropy production and
the primordial helium abundance,
these limits are severe in the sense that they allow only a tiny fraction 
of PBHs in dark matter at any cosmic time before the PBHs evaporate.
Assuming Gaussianity of the primordial perturbation,
these limits can be converted to the upper limit on the power spectrum 
of the curvature perturbation ${\cal R}$ as ${\cal P}_{\cal R}(k) \lesssim 10^{-2}$
for $10^9 < k/{\rm pc}^{-1} < 10^{14}$ \cite{Josan:2009qn}.
Although the upper limit on $\beta$ varies many orders of magnitude
over the PBH mass range corresponding to the above range of $k$,
the upper limit in terms of ${\cal P}_{\cal R}$ is insensitive to $k$,
which is traced back to the fact that $\beta$ depends on the ${\cal P}_{\cal R}$ exponentially.
(see Sec.~\ref{abundance-PBH}).

\subsection{Gravitational lensing}
Gravitational lensing is a very powerful method to constrain/detect PBHs.
Excellent point of the gravitational lensing is that the individual lensing signal is solely based on the
gravitational physics and does not suffer from the uncertainties that exist in the studies
of electromagnetic signals resulting from the interaction between the PBHs and
the surrounding matter.

\begin{figure}[tbp]
  \begin{center}
   \includegraphics[width=150mm]{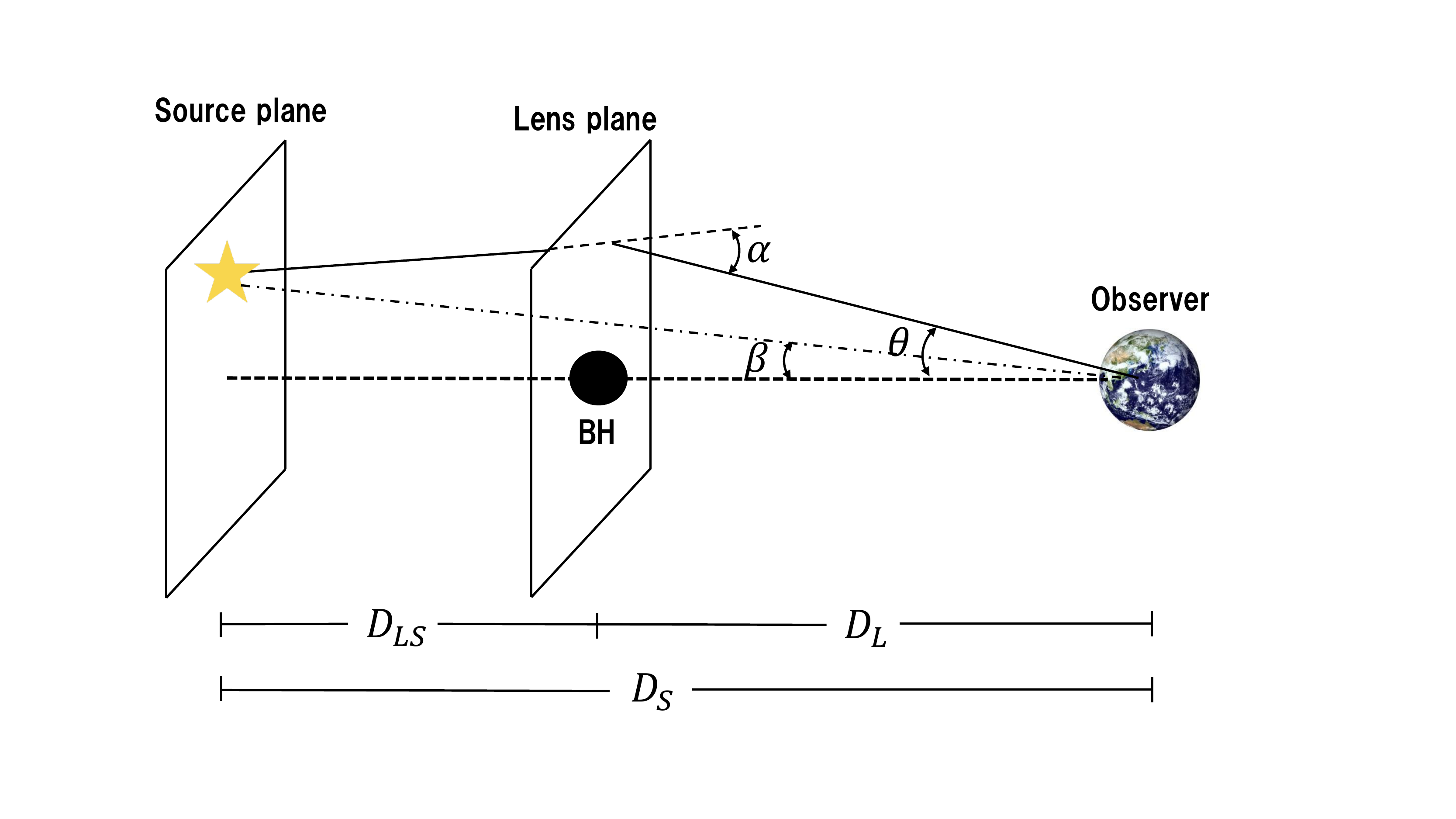}
  \end{center}
  \caption{Trajectory of light ray bent by the lens object (BH).
  $\alpha$ is the deflection angle.}
  \label{fig-GL1}
\end{figure}

If PBHs are present in the Universe, they cause the gravitational lensing on the background 
objects such as stars.
Under the thin-lens approximation in which the deflection of light occurs at a point on the lens plane
(which is a very good approximation in the astrophysical situations we are considering),
the lens equation can be written as (Fig.~\ref{fig-GL1})
\be
\theta D_S=D_S \beta+D_{LS} \alpha,~~~~~~\alpha=\frac{4GM_{\rm BH}}{D_L \theta}.
\ee
In terms of the distance on the lens plane $r=D_L \theta$, the lens equation becomes
\be
r^2-r_0 r-R_E^2=0,
\ee
where $r_0=D_L \beta$ and $R_E= \sqrt{ 4GM_{\rm BH} \frac{D_L D_{LS}}{D_S}}$ is the Einstein radius.
Thus, the positions of the lensed images are given by
\be
r_{1,2}=\frac{1}{2} \left( r_0 \pm \sqrt{r_0^2+4R_E^2} \right).
\ee
There are thus two lensed images at $r_1$ and $r_2$.
The only exception is when the source, lens object, and the observer are 
on the same line, {\it i.e}. $\beta=1$.
In this case, the image on the lens plane becomes a circle (the so-called Einstein Ring) 
with its radius given by $R_E$.
Lensing effect becomes significant when $r_0 \lesssim R_E$ and 
typical angular separation of the two images is 
\be
\Delta \sim \frac{R_E}{D_L}=\sqrt{ \frac{4G M_{\rm BH} }{D_S} \frac{1-x}{x}}
\approx 0.3~{\rm mas}~ {\left( \frac{M_{\rm BH}}{10~M_\odot} \right)}^{1/2}
{\left( \frac{D_S}{100~{\rm kpc}} \right)}^{-1/2} \sqrt{\frac{1-x}{x}}, \label{angular-separation}
\ee
where we have introduced the parametrization as $D_L=D_S x ~(0 < x <1)$.

\subsubsection{Microlensing}
\label{microlensing}
\begin{figure}[tbp]
  \begin{center}
   \includegraphics[width=120mm]{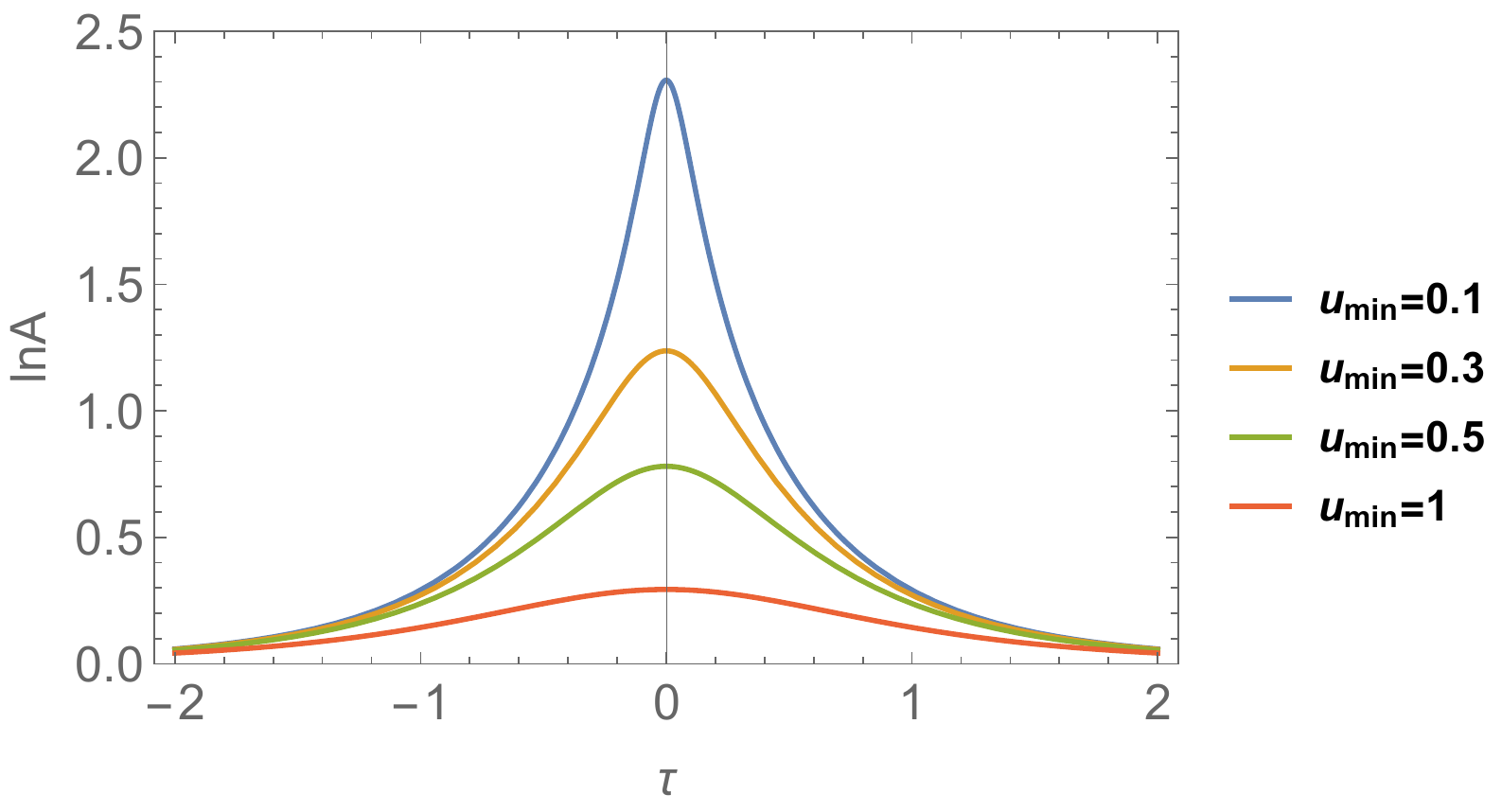}
  \end{center}
  \caption{Magnification curves for four different values of $u_{\rm min}=
  (0.1,~0.3,~0.5,~1)$ as a function of tme. The dimensionless time $\tau=vt/R_E$ is used.}
  \label{fig-lnA}
\end{figure}

Gravitational microlensing refers to the gravitational lensing event in which 
the angular separation of the images lensed by a compact object is so small that
the individual images cannot be resolved by observations \cite{Roulet:1996ur}. 
For instance, the angular separation given by Eq.~(\ref{angular-separation}) 
is much smaller than the angular resolution of the existing (or past) microlensing experiments
such as the MACHO Project and EROS collaboration which is about $0.6~{\rm as}$.
In the gravitational microlensing event, what can be observed is only the superposition
of two images which is brighter than the original source.
The magnification factor $A$, normalized to unity in the absence of the microlensing, is given by
\be
A=\frac{u^2+2}{u \sqrt{u^2+4}},~~~~~u = \frac{r_0}{R_E}. \label{magnification-A}
\ee
At $r_0=R_E$, $A=1.34$.
When the lens object is moving relative to the line of sight connecting the source
and the observer, $u$ becomes time-dependent and the magnification varies in time.
Assuming a constant tangential velocity $v_{\rm BH}$, we can write $u$ as
\be
u^2=\frac{v^2 t^2+b^2}{R_E^2}=\frac{v^2 t^2}{R_E^2}+u_{\rm min}^2,
\ee
where $b$ is the impact parameter of the source in the lens plane and
the origin of $t$ is chosen so that the source image in the lens plane becomes 
the closest to the lens object at $t=0$.
Magnification curves for various values of $u_{\rm min}$ are given in Fig.~\ref{fig-lnA}.
As we can see, the magnification is maximum at $t=0$ and symmetric about it. 
The time scale for the rise and fall-off of the magnification is given by
\be
T =\frac{R_E}{v} =\frac{\sqrt{4GM_{\rm BH}D_S x(1-x)}}{v} \approx 2~{\rm yr}~\sqrt{x(1-x)} 
{\left( \frac{M_{\rm BH}}{10~M_\odot} \right)}^{1/2} 
{\left( \frac{D_S}{100~{\rm kpc}} \right)}^{1/2}  
{\left( \frac{v}{200~{\rm km/s}} \right)}^{-1}. \label{ML-timescale}
\ee
We find that the time scale is longer for larger lens mass and larger distance to the source.

A useful quantity which is used in the context of the microlensing is the optical depth $\tau$ \cite{Vietri:1983mt, Nityananda:1984mt}.
This quantity measures how likely a background source is micro-lensed by the compact objects
with magnification factor greater than $1.34$ \footnote{The actual threshold for $A$ for the detection of
the microlensing event depends on the detection methods in individual experiments.
The value $A=1.34$ is just for convenience.}.
In other words, the optical depth, when it is much smaller than unity, 
is the probability that the light from the source passes inside 
the circle defined by the Einstein radius on the lens plane.
Assuming that all the lens objects have the same mass,
the optical depth can be written as
\be
\tau=\int_0^{D_S} dr~n_{\rm BH} (r) \pi R_E^2 (r)=4\pi G \int_0^{D_S} dD_L~\rho_{\rm BH} (D_L)
\frac{D_L (D_S-D_L)}{D_S},
\ee
where $\rho_{\rm BH}=M_{\rm BH} n_{\rm BH}$ is the energy density of PBHs.
Interestingly, the last expression of the optical depth shows that it does not depend explicitly on the mass of PBHs.
What determines the optical depth is how much PBHs constitute the entire dark matter.

In 1986, Paczy\'{n}ski pointed out that it would be possible to test the hypothesis that some fraction
of dark matter is in the form of compact object by monitoring a few million stars in the Magellanic Clouds \cite{Paczynski:1985jf}.
Light from any star in MC passes through the dark matter halo encompassing the Milky Way Galaxy before it reaches the observer on the Earth.
If the dark matter halo entirely (or partially) consists of the compact objects, they produce microlensing events
on the source stars in MC.
According to the estimate in \cite{Paczynski:1985jf}, the optical depth for the microlensing event is given by
\be
\tau \sim 10^{-6} f_{\rm PBH}.
\ee 
If $f_{\rm PBH} \sim 1$, about one star among a few million stars must produce a microlensing event.
Conversely, non-observation of microlensing events places the upper limit on the PBH abundance.
The time scale of the magnification is $\sim 2~{\rm yr}$ for $M_{\rm BH} =10~M_\odot$
and $\sim 400$ minutes for $M_{\rm BH} =10^{-6}~M_\odot$ (see Eq.~(\ref{ML-timescale})).
Thus, the few year observation of the stars in MC enables us to place constraint on the PBHs with up to
$M_{\rm BH} \sim 10~M_\odot$.

It is important to mention that there is a lower limit on the PBH mass below which  
the finite-size effect of the source changes the magnification by the microlensing from that in the point lens case.
The finite-size effect becomes relevant when the angular size of the source star projected on the lens plane
becomes comparable to or larger than the angular size of the impact parameter $r_0$.
The peak magnification becomes smaller or larger than that in the case in the point lens depending on
the magnitude relation between the projected source size and $r_0$ \cite{Witt:1994mt}.
Especially, when $r_0$ is much smaller than the source size, the maximum of $A$ becomes
\be
A_{\rm max} =\frac{\sqrt{4+r^2}}{r},~~~~~r=\frac{x R_{\rm star}}{R_E}.
\ee
Thus, $A_{\rm max}$ becomes close to unity for $r \gtrsim 1$.
In terms of the lens mass, the condition $r \gtrsim 1$ can be written as
\be
M_{\rm PBH} \lesssim \frac{D_L R_{\rm star}^2}{4GD_S^2} \sim 3 \times 10^{-9}~M_\odot
~{\left( \frac{R_{\rm star}}{R_\odot} \right)}^2 \left( \frac{D_L}{10~{\rm kpc}} \right)
{\left( \frac{D_S}{100~{\rm kpc}} \right)}^{-2}.
\ee
If the lens mass satisfies this inequality, magnification is significantly reduced
from that in the point lens case.

Following the intriguing proposal by Paczy\'{n}ski, 
several groups in the world carried out the year-scale observations of millions of stars in MCs.
Detection of the microlensing events were first reported both by the MACHO Project \cite{Alcock:1993qc}
in which one event suggesting the compact object with $\sim 0.1~M_\odot$ was detected
and the EROS (Exp\'{e}rience pour la Recherche d'Objets Sombres) collaboration \cite{Aubourg:1993wb}
in which two events suggesting the compact objects between a few $\times 10^{-2}$ and $1~M_\odot$ were detected.
Continuing monitoring the stars in the LMC, the MACHO Project finally reported the detection of $13\sim 17$ microlensing events
and the lensing optical depth as \cite{Alcock:2000ph}
\be
\tau_{\rm LMC}=1.2^{+0.4}_{-0.3} \times 10^{-7},
\ee
and suggested that about $20\%$ contribution of the compact objects in a mass range
$0.1\sim 1~M_\odot$ to the dark matter halo mass.
The EROS collaboration \cite{Tisserand:2006zx}, based on the $6.7$ yr observations of the stars in LMC and SMC, 
later reported that only one candidate microlensing event of the star in SMC was detected and the optical depth as
\be
\tau_{\rm LMC}< 0.36\times 10^{-7},~~~~~~0.085  \times 10^{-7} < \tau_{\rm SMC} < 8.0 \times 10^{-7}.
\ee
Based on the earlier estimation that the optical depth for the microlensing by objects in the SMC is 
$\tau_{\rm SMC} \sim 0.4 \times 10^{-7}$ \cite{Graff:1998ix}, it was suggested that
the detection of one event is consistent with the expectations of self-lensing by objects
in the SMC.
According to the EROS collaboration \cite{Tisserand:2006zx}, 
the severe constraint on $\tau_{\rm LMC}$ that appears to be in contradiction with the one
measured by the MACHO Project could be possibly explained as the contamination of the variable stars,
the self-lensing in the inner parts of the LMC,
blending effects (reconstructed fluxes receive contributions from more than one star),
and the possibility that the fields observed by the MACHO Project just lie behind a clumpy region of 
the compact objects that does not cover the EROS-fields.
The constraint by the EROS collaboration is shown in Fig~\ref{Constraints}.

After the EROS collaboration,
another experiment called OGLE (Optical Gravitational Lensing Experiment) also reported the detection
of two candidate microlensing events in LMC \cite{Wyrzykowski:2010mh} and three events in SMC \cite{Wyrzykowski:2011tr}
and obtained
\be
\tau_{\rm LMC}=(0.16 \pm 0.12) \times 10^{-7},~~~~~
\tau_{\rm SMC}=(1.30 \pm 1.01) \times 10^{-7}.
\ee
While the four events are consistent with contribution from the self-lensing in
the Galactic disk and SMC,
detailed analysis \cite{Dong:2007px} shows that there is a possibility that one event (OGLE-SMC-02) is a binary 
BH lens in the dark matter halo although the self-lensing scenario remains also as a possible explanation.
The upper limit on the PBH fraction by OGLE is shown in Fig~\ref{Constraints}.
However, it is important to keep in mind that this constraint is derived under
the assumption that all OGLE events are due to the self-lensing/background signals.

In \cite{Griest:2011av}, it was proposed that the data obtained by the Kepler satellite 
could be used to obtain a new upper limit on the abundance of PBHs in the mass range
$5\times 10^{-10}~M_\odot \sim 10^{-4}~M_\odot$ which is much smaller than the stellar
mass covered by the previous microlensing experiments mentioned above.
Although both the number of stars that are monitored ($\sim 0.1$ millions) and 
the distance to the stars ($\sim 1~{\rm kpc}$) are much smaller than the microlensing
experiments targeting at the Magellanic Clouds,
advantages by the high sensitivity of the photometry by the Kepler mission and 
the largeness of the microlensing cross section can compensate the above disadvantages.
It was reported in \cite{Griest:2013aaa} that after removing the background events 
no microlensing candidates have been found in the two years of Kepler data.
By the non-detection of the microlensing events, the upper limit on the PBH fraction
was obtained \cite{Griest:2013aaa}.
The resultant constraint is shown in Fig~\ref{Constraints}.
The actual constraint is much weaker than the theoretical estimate derived in \cite{Griest:2011av}.
According to \cite{Griest:2013aaa}, this is due to the optimistic assumption made in \cite{Griest:2011av}
which underestimate the flare events of short duration.

Recently, very stringent upper limit on the PBH fraction for the mass range $(10^{-13}~M_\odot, 10^{-5}~M_\odot)$
has been obtained by the observations of more than tens of million stars in M31 by 
the Subaru Hyper Suprime-Cam (HSC) \cite{Niikura:2017zjd}.
A powerful and unique feature of the HSC is its high-cadence;
observations of stars every 2 minutes (90 seconds exposure plus
about 35 seconds for readout) over about 7 hours (totally 196 exposures).
This is why HSC is sensitive to the PBH mass much smaller than the solar-mass (see Eq.~(\ref{ML-timescale})).
After eliminating the fake events, one microlensing candidate was found.
Any convincing conclusion is not obtained as for the nature of the microlensing event. 
Yet, depending on whether the microlensing candidate is attributed to PBH or not,
it is possible to derive the upper limit on the PBH abundance, which is shown in Fig~\ref{Constraints}.
The obtained constraint provides the strongest constraint for a wide mass range that fully
covers the mass range probed by the Kepler mission.

Microlensing constraint on $f_{\rm PBH}$ has been obtained in \cite{Diego:2017drh, Oguri:2017ock} 
in a manner which is very unique and different from the ones discussed above. 
Observations by Hubble Space Telescope found a fast transient 
which turned out to be a magnified star at $z=1.49$ \cite{Kelly:2017fps} 
(estimated magnification is $A >2000$, which is much bigger than $1$!). 
The star being at the vicinity of the caustic by the galaxy cluster called MACS J1149.6+2223 
located at $z=0.544$ (caustic is the place in the source plane where the magnification diverges) 
suggests that the magnification is caused by the caustic-crossing. 
However, detailed analysis showed that the observed magnification 
is not large enough to be explained solely by the caustic generated 
by the galaxy cluster for which the magnification should be as large as $A \sim 10^6$ 
(the magnification does not diverge in reality due to finite size of the source). 
A natural explanation is contribution from point mass lens in the same galaxy cluster. 
Caustic is known to be significantly distorted even by the presence of tiny point masses. 
The distortion of the caustic results in the reduction of magnification compared to 
the case where point masses are absent. 
The more the point masses are, the more the reduction of the magnification is. 
The observed magnification curve requires a certain amount of point masses. 
This does not mean that this observation yields both upper and lower limit on 
$f_{\rm PBH}$ since there are stars responsible for the intra-cluster light (ICL) 
that produce the same effect on the caustic as the PBHs do. 
Within the uncertainty of the abundance of the ICL stars, 
it was shown that the ICL stars can explain the observed fast transient. 
As a result, only the upper limit on $f_{\rm PBH}$ has been obtained. 
The constraint, which is given in \cite{Oguri:2017ock}, is shown in Fig~\ref{Constraints}. 

Quasar microlensing is also a useful method to detect/constrain the PBHs \cite{Witt:1995wn}.
Some distant quasars are observed as pairs due to the gravitational lensing by 
intervening galaxies. 
Those quasar pairs can be additionally microlensed by PBHs (or any other types of compact objects)
that reside between the quasars and the Earth.
Contrary to the microlensing surveys such as MACHO project which look for the characteristic
time variation of the source brightness, 
a salient aspect of the quasar microlensing is that we need only a single-epoch flux measurement
for each quasar pair.
The point of this method is that continuum component and the narrow emission lines 
of the photo spectrum for each quasar image of the pair show different sensitivity to the microlensing.
Emission lines are generally thought to have originated from the larger region than the region for the continuum 
component and not to be affected by the microlensing while the continuum component is microlensed by the PBHs.
This enables to extract the signal of microlensing by comparing the flux ratio of the emission lines
with that of the continuum component from the photo spectrum taken at a single epoch.
In \cite{Mediavilla:2017bok}, the above program has been applied to the microlensing data of 
24 gravitationally lensed quasars (see \cite{Mediavilla:2009um} for earlier study)
and tight upper limit on $f_{\rm PBH}$ was obtained for the stellar-mass PBHs.
However, since the upper limit on $f_{\rm PBH}$ was not explicitly given,
we do not show it in Fig~\ref{Constraints}.

\subsubsection{Millilensing}
\label{millilensing}
Supermassive PBHs ($\gtrsim 10^6~M_\odot$) are also a target of search by the gravitational lensing.
Supermassive compact objects (not necessarily PBHs) filling the Universe produce 
the two lensed images of the source at the cosmological distance
with mas scale angular separation (hence ``millilensing''). 
Press and Gunn proposed that such double images can be resolved by the VLBI maps \cite{Press:1973mt}
and estimated the probability of such events.
By using the VLBI observations of 48 compact radio sources,
the constraint on the PBH fraction $f_{\rm PBH} < 0.2$ (95 CL) for $10^7 < M_{\rm PBH}/M_\odot < 10^9$ 
was obtained under the assumption
of the Einstein-de Sitter Universe in \cite{Kassiola:1991mt}.
Later, the constraint was refined in \cite{Wilkinson:2001vv} in which VLBI
maps of 300 compact radio sources were analyzed and no evidence of the millilensing was found.
Consequently, the tighter upper limit on $f_{\rm PBH}$ was obtained for wider PBH mass range,
which is shown in Fig~\ref{Constraints}.

\subsubsection{Femtolensing}
\label{femtolensing}
PBHs with much smaller mass such as $10^{-16}~M_\odot \lesssim M_{\rm PBH} \lesssim 10^{-13}~M_\odot$,
yet in the non-evaporating regime, 
can be still probed by the gravitational lensing in a very unique manner.
For the gravitational lensing by such tiny lens mass, the wave effects of light become important. 
When the wavelength of light from the source star becomes comparable to or longer than the 
Schwarzshild radius of the lens object,
the diffraction considerably changes the amplification from that in the geometric optics approximation \cite{
Ohanian:1974ys, Bliokh:1975ys, Bontz:1981ys}.
In the quasi-geometrical optics approximation, the magnification factor is given by \cite{TTNakamura:1999ys}
\be
A=\frac{u^2+2+2 \sin (E_\gamma T_{12} )}{u \sqrt{u^2+4}}, \label{diffraction-A}
\ee 
where $E_\gamma$ is the energy of gamma-ray photons and $T_{12}$ is the difference
of arrival times between the two images,
\be
T_{12}= 4GM_{\rm PBH} \bigg[ \frac{1}{2} u\sqrt{u^2+4}+ \ln \left( \frac{\sqrt{u^2+4}+u}{\sqrt{u^2+4}-u} \right) \bigg].
\label{T12-diffraction}
\ee
This shows that the arrival time difference is the order of the
Schwarzshild radius of the lens object for the typical value of $u$ ($\sim 1$) for which
the signal of the gravitational lensing is strong.
Effect of diffraction appears as a sinusoidal oscillation part in the magnification factor.
This arises since the magnification factor is the square of the summation (not the summation of the square)
of the lensed electro-magnetic waves corresponding to the individual images
and the phase of the two waves differs by the arrival time difference multiplied by the
photon energy.
In other words, the sinusoidal part represents the interference of two images.
In the geometrical optics limit, the oscillations become extremely fine and disappear when $A$
is averaged over the energy bin $\Delta E_\gamma \sim 1/T_{12}$.
As a result, the standard formula (\ref{magnification-A}) is recovered.

Gould pointed out that if small PBHs in the mass range 
$10^{-16}~M_\odot \lesssim M_{\rm PBH} \lesssim 10^{-13}~M_\odot$ 
distribute in the Universe, they induce the oscillatory feature in the photon spectrum of
the gamma-ray bursts (GRBs) that occur at the cosmological distance \cite{Gould:1992mt}.
This PBH mass range corresponds to the interval of $E_\gamma^{-1}$ where 
$10~{\rm keV} \lesssim E_\gamma \lesssim {\rm MeV}$.
The angular separation of the source at the cosmological distance lensed by such tiny PBHs is
around the femto-as scale, thus the name ``femtolensing'' \cite{Gould:1992mt}. 
In \cite{Barnacka:2012bm}, the oscillatory feature was searched for in 
the GRB data obtained by the Fermi Gamma-ray Bust Monitor (see \cite{Marani:1998sh} for earlier
constraint derived by using the BATSE and Ulysses data).
The GRB spectra were well fitted by the standard GRB models and no
evidence of the femtolensing was found. 
This result excludes the possibility that these tiny PBHs constitute all dark matter \cite{Barnacka:2012bm}.
The constraint graph is shown in Fig~\ref{Constraints}.

\subsection{Dynamical constraints}
To a certain degree, PBHs affect any astrophysical system by the gravitational interactions.
By appropriately evaluating the impact of PBHs on the astrophysical systems and 
making comparison with observations, it is possible to put upper limit on the PBH fraction for a wide range of PBH mass.
Until present, various astrophysical systems have been considered in this context.

\subsubsection{Disruption of white dwarfs}
\label{DWD}
White dwarfs are stars with about the solar mass and of the size comparable to that of the Earth.
Due to their compactness, gravity is much stronger than that of the main sequence stars.
The strong self-gravity of white dwarfs is supported by the electron degeneracy pressure that is effective
even at the zero temperature \cite{Shapiro-Teukolsky}.
White dwarfs are believed to be the progenitors of type Ia supernovae.
When the mass of the white dwarfs grows close to the Chandrasekhar limit, {\it e.g.,} 
by the accretion,
the nuclear fusion sets in and undergoes the thermonuclear runaway.

It was pointed out in \cite{Graham:2015apa} that a passage of PBH through the white dwarf can
ignite the thermonuclear runaway that eventually makes the white dwarf well below
the Chandrasekhar limit explode.
Basic idea of \cite{Graham:2015apa} is as follows.
When the PBH passes inside the white dwarf, all the nearby particles inside the thin tube 
along the trajectory of the PBH acquire energy by the gravitational force from the PBH
for a short time. Thus, the PBH loses its kinetic energy though it is the negligible amount
compared with its initial kinetic energy.
As a result, the temperature of the thin tube increases and the nuclear fusion rate
gets significantly boosted as the fusion rate typically has strong dependence on the temperature.
If the time scale of the dissipation of the heat from the thin tube is shorter than the
time scale on which the nuclear fusion occurs, then the fusion occurs,
release energy and increase the temperature of the region outside the tube.
Since the dissipation becomes more inefficient for larger volume,
the fusion is facilitated and ends up with the explosion.

Thus, if there were too many PBHs above a certain mass, the white dwarfs with corresponding mass
cannot exist in the present Universe.
Conversely, we can constrain the abundance of the PBHs by using the observational
confirmation of white dwarfs.
We expect this argument constrain a certain PBH mass range; light PBHs do not 
give sufficient heat to the white dwarfs simply because of the weakness of gravitational effect,
and white dwarfs do not encounter much heavy PBHs simply because of their rareness.
The upper limit on the PBH fraction as a function of the PBH mass is given in Fig~\ref{Constraints}.
It is also possible to derive constraint based on the requirement that the rate of supernovae induced by the PBHs
does not exceed the observed rate, but it is found that the derived constraint is less robust compared 
to the one based on the observation of the white dwarfs \cite{Graham:2015apa}.

\subsubsection{Disruption of neutron stars}
\label{DNS}
Neutron stars are stars that are much more compact than the white dwarfs.
The density is around the nuclear density and the stars are largely supported by the
degenerate pressure of neutrons. 
Neutron stars can be also used to constrain the abundance of PBHs, although slightly
in a different manner from the case of the white dwarfs in \ref{DWD}.

As is the case in the white dwarfs, the PBH loses its kinetic energy by the dynamical friction
during its passage through the neutron star.
Effect of PBHs on neutron stars was investigated in \cite{Capela:2013yf}. 
According to \cite{Capela:2013yf}, the energy transferred from the PBH to the neutron star is given by
\be
\frac{E_{\rm loss}}{M_{\rm PBH}}\approx 6.3\times 10^{-12}~\left( \frac{M_{\rm PBH}}{10^{22}~{\rm g}} \right). \label{E-loss}
\ee
If this energy is greater than the initial kinetic energy of the PBH, then the PBH becomes gravitationally
bound to the neutron star.
After the first passage, the PBH subsequently undergoes the orbital oscillations, passes through
the neutron stars at every half oscillation period, and gradually loses the kinetic energy by transferring
it to the neutron star.
After a certain number of oscillations, the PBH finally becomes trapped inside the neutron star.
The time scale for this to happen is given by \cite{Capela:2013yf}
\be
t \sim 4\times 10^4~{\rm yr}~{\left( \frac{M_{\rm PBH}}{10^{22}~{\rm g}} \right)}^{-3/2}.
\ee
Once the PBH lies inside the neutron star, it quickly accretes the nuclear matter and destroys the star \cite{Kouvaris:2013kra}.
Thus, neutron stars must not be exposed to frequent encounters with the PBHs, from which
a certain upper limit on the PBH abundance can be derived.
 
Eq.~(\ref{E-loss}) shows that the trapping of the PBHs is effective for the low-velocity PBHs.
It was proposed in \cite{Capela:2013yf} that strong constraint on the PBH abundance can be
obtained from the observations of the neutron stars in the cores of globular clusters,
where the typical velocity dispersion of the PBHs is $\sim 10~{\rm km/s}$ (see Fig~\ref{Constraints}).
However, it is important to keep in mind that the constraint relies on the assumption
that there are PBHs as dark matter at the cores of globular clusters and it is not known observationally
whether dark matter exists in such regions.
Much stronger constraint for the same phenomena, the PBH capture by the neutron stars,
has been obtained based on different method for estimating the energy loss of the PBH in \cite{Pani:2014rca}.
The subsequent papers \cite{Capela:2014qea, Defillon:2014wla} preclude this possibility.

While references above place upper limit on $f_{\rm PBH}$, it was suggested in \cite{Fuller:2017uyd} that
PBHs in the mass range $10^{-14}<M_{\rm PBH}/M_\odot<10^{-8}$ that account a tiny fraction (a few percent)
of dark matter can explain the observed $r$-process element abundances.
According to the analysis in \cite{Fuller:2017uyd}, millisecond pulsars in dense dark matter regions 
such as the Galactic center and the dwarf spheroidal galaxies can efficiently swallow the floating PBHs.
As the eaten PBH swallows back the nuclear matter of the neutron star from the inside, 
the neutron star shrinks by its mass loss due to the PBH.
Due to the conservation of the angular momentum, the neutron star spins up as it shrinks.
For millisecond pulsars, the rotation velocity at the equator exceeds the escape velocity at some point,
and ejection of relatively cold neutron-rich material of $0.1\sim 0.5~M_\odot$ results.
Then, $r$-process nucleosynthesis occurs in the ejected material,
which may explain the observed $r$-process abundance pattern.

\subsubsection{Disruption of wide halo binaries}
\label{wide-halo-binary}
Wide halo binaries are binaries of stars with wide separation (even ${\cal O}(1)~{\rm pc}$ is possible)
residing in the Galactic halo.
Because of their weak binding energy, they are vulnerable to disruption from encounters of PBHs.
Thus, observational confirmation of wide halo binaries enables to place upper limit on the
PBH abundance.

Let us consider a situation where PBHs intermittently pass near the binary and transfer some
energy to the binary.
We denote the semi-major axis and mass of the binary by $a$ and $m_b$, respectively.
When the transfered energy by the PBH encounter becomes greater than the binding energy of the binary,
the binary is disrupted.
If the binary is disrupted by a single encounter, the disruption is in the catastrophic regime.
On the other hand, if the binary is disrupted by the multiple encounters in which the individual
encounters add energy smaller than the binding energy to the binary, the disruption is in the diffusion regime.
According to \cite{Binney-Tremaine}, the time scales for the catastrophic and diffusive disruptions are given by
\be
t_{\rm d,cat} \simeq 0.07~ \frac{m_b^{1/2}}{G^{1/2} \rho_{\rm PBH} a^{3/2}},~~~~~~~
t_{\rm d,diff} \simeq 0.002~ \frac{V m_b}{GM_{\rm PBH} \rho_{\rm PBH} a}, \label{timescale-disruption}
\ee
where $V$ is the relative velocity between the binary and the PBHs.
Both time scales coincide for the PBH mass given by
\be
M_{\rm PBH,c} \simeq 30~M_\odot ~{\left( \frac{m_b}{2~M_\odot} \right)}^{1/2} 
{\left( \frac{a}{10^4~{\rm AU}} \right)}^{1/2} \frac{V}{200~{\rm km/s}}. \label{cat-dif-equal-mass}
\ee
Above this mass, the PBH is massive enough to disrupt by the single encounter, thus in the catastrophic regime.
Below this mass, the binary is disrupted by the multiple encounters.
Roughly speaking, the constraint on the PBH abundance is obtained by requiring that the disruption time 
scale given above is longer the age of the binary.
Typical value of $t_{\rm d,cat}$ for wide binaries is given by
\be
t_{\rm d,cat} \simeq 3~{\rm Gyr}~ {\left( \frac{m_b}{1~M_\odot} \right)}^{1/2} f_{\rm PBH}^{-1}
{\left( \frac{\rho_{\rm DM}}{0.01~M_\odot/{\rm pc}^3} \right)}^{-1}
{\left( \frac{a}{0.1~{\rm pc}} \right)}^{-3/2},
\ee
which is shorter than the age of the binaries, which is the order of the age of the Universe. 
This shows that wide binaries with $a \gtrsim 0.1~{\rm pc}$ can give a meaningful upper limit on $f_{\rm PBH}$.
Interestingly, the time scale for the catastrophic regime is 
independent of the PBH mass itself.
Thus, in terms of the PBH fraction in total dark matter,
the upper limit becomes independent of the PBH mass.
On the other hand, for PBHs lighter than $M_{\rm PBH,c}$, the constraint becomes weaker for lighter PBHs.

Confirming the wide halo binaries is a challenging task.
They are rare in the first place, and hard to distinguish from the mere chance association.
In \cite{Yoo:2003fr}, analyzing the wide binary candidates given in \cite{Chaname:2003fn},
the PBH fraction was constrained to be $\lesssim 0.2$ for PBH mass larger than a few $100~M_\odot$.
Later, the authors of \cite{Quinn:2009zg} found, based on their results of the radial velocity measurements
of stars of four candidate wide binaries, that the second widest binary in \cite{Chaname:2003fn} is spurious.
Yet, the widest binary, whose separation is about $1.16~{\rm pc}$, was confirmed to be genuine.
The derived constraint is weaker than that in \cite{Yoo:2003fr} but still excludes the PBHs as all the dark matter
for the mass range similar to \cite{Yoo:2003fr} (see Fig~\ref{Constraints}.).

Finally, it is worth mentioning that the time scales given by Eq.~(\ref{timescale-disruption}) are obtained
under the impulse approximation which is valid when the time scale on which the gravitational interaction between
the PBH and the binary is effective is shorter than the dynamical time of the binary ({\it i.e.} orbital period)  \cite{Binney-Tremaine}.
The former, which is equal to the time that PBH crosses the distance of the impact parameter with the velocity $V$,
is given by
\be
t = \frac{2 b_{\rm max}}{V} \sim  \frac{2}{V} {\left( \frac{GM_{\rm PBH}^2 a^3}{V^2 m_b} \right)}^{1/4},
\ee
where $b_{\rm max}$ is the maximum impact parameter for which the encounter is catastrophic.
This time scale becomes equal to the orbital period $\sim 2\pi \sqrt{a^3/(Gm_b)}$ for
\be
M_{\rm PBH} \simeq 9\times 10^9~M_\odot.
\ee
The impulse approximation is justified for PBHs much lighter than this mass.

\subsubsection{Disruption of globular clusters}
The argument developed for the case of the wide binaries can equally apply to the 
globular clusters.
The only difference between the two is that the globular clusters involve much more stars than the binaries.
Plugging the typical values for the mass $m_{\rm gc}$ and size of the globular clusters $a_{\rm gc}$
into Eq.~(\ref{cat-dif-equal-mass}),
the PBH mass dividing the catastrophic regime and the diffusion regime becomes
\be
M_{\rm PBH,c} \simeq 10^5~M_\odot ~{\left( \frac{m_{\rm gc}}{10^5~M_\odot} \right)}^{1/2} 
{\left( \frac{a_{\rm gc}}{10~{\rm pc}} \right)}^{1/2} \frac{V}{200~{\rm km/s}}. 
\ee
The time scale of the catastrophic encounter becomes
\be
t_{\rm d,cat} \simeq 1~{\rm Gyr}~ {\left( \frac{m_{\rm gc}}{10^5~M_\odot} \right)}^{1/2} f_{\rm PBH}^{-1}
{\left( \frac{\rho_{\rm DM}}{0.01~M_\odot/{\rm pc}^3} \right)}^{-1}
{\left( \frac{a_{\rm gc}}{10~{\rm pc}} \right)}^{-3/2}.
\ee
As is the case in the wide halo binaries, for $f_{\rm PBH} \simeq 1$,
this is again shorter than the age of the globular clusters, which is the order of the age of the Universe.
Thus, the observations of the existence of the globular clusters place a meaningful constraint
on $f_{\rm PBH}$.
For more careful analysis, see \cite{Carr:1997cn}.

\subsubsection{Disruption of ultra-faint dwarf galaxies}
\label{UFDG}
Ultra-faint dwarf galaxies that are known to be present around the Galaxy and Andromeda Galaxy are also useful
to place the upper limit on the PBH abundance. 
If PBHs reside in the ultra-faint dwarf galaxies, stars inside intermittently interact with PBHs gravitationally.
Two-body interaction tends to equalize the kinetic energies of the individual objects.
This means that if PBHs are heavier than stars, stars on average acquire kinetic energy (dynamical heating).
As a result, the interaction between stars and PBHs makes the stars move faster and spread wider and wider,
which may contradict the observed distribution of stars.

Let us consider a star cluster inside an ultra-faint dwarf galaxy and
denote the cluster's half-light radius by $r_h$.
If the star cluster is immerse in the bath of PBHs, $r_h$ increases in time.
Its evolution equation is given by \cite{Brandt:2016aco}
\be
\frac{dr_h}{dt}=\frac{4 \sqrt{2}\pi G f_{\rm PBH} M_{\rm PBH} \ln \Lambda}{V} 
{\left( \alpha \frac{M_{\rm cluster}}{\rho_{\rm DM} r_h^2}+2 \beta r_h \right)}^{-1},
\ee
where $V$ is the typical PBH velocity relative to the cluster, $\rho_{\rm DM}$ is the
total dark matter density, $\ln \Lambda \approx 10$ is the Coulomb logarithm
and $\alpha \sim 0.4,~\beta \sim 10$.
$f_{\rm PBH}$ is the fraction of PBHs in the entire dark matter.
If the PBH is a subdominant component, then the rest of dark matter must be in other form,
for instance, unknown elementary particles.
In such a case, dark matter other than the PBHs drag the stars gravitationally and cool them.
However, it was shown in \cite{Brandt:2016aco} that the cooling effect is ineffective compared with
the heating concerning the upper limit on $f_{\rm PBH}$ by the considerations of the
ultra-faint dwarf galaxies.

In \cite{Brandt:2016aco}, the upper limit on $f_{\rm PBH}$ was obtained by investigating
stellar distributions in ultra-faint dwarf galaxies and
a star cluster in Eridanus II, a dwarf galaxy discovered as part of the Dark Energy Survey \cite{Bechtol:2015cbp,Koposov:2015cua}.
Eridanus II has a half-light radius of $\sim 300~{\rm pc}$ and hosts a star cluster with its
half-light radius $r_h=13~{\rm pc}$.
Because of the lack of our knowledge about the age of stellar clusters or dwarf galaxies
as well as their initial size, there are degrees of uncertainty in translating the observed stellar distributions to
the upper limit on $f_{\rm PBH}$.
The obtained constraints with the conservative assumptions are shown in Fig~\ref{Constraints}.
As is clear from the Figure, the derived constraint is dependent on the adopted requirements.
Within the range of uncertainties considered, PBHs heavier than $\sim 100~M_\odot$ as all the dark matter are excluded.
In particular, the constraints in all the cases are stronger than that obtained by
the wide binaries discussed in \ref{wide-halo-binary}.
More recently, by analyzing the Segue 1 dwarf galaxy, similar constraint was derived in \cite{Koushiappas:2017chw}.

\subsubsection{Dynamical friction on PBHs} 
\label{dynamical-friction}
If the Galactic halo is entirely or partially composed of the massive PBHs, some of them must be in the region near the Galactic center.
Such PBHs receive strong dynamical friction from the stars and the dark matter in the form of lighter PBHs
or elementary particles, lose their kinetic energy, and spiral in to the center.
If this infall time scale is shorter than the age of the Universe, accumulation of PBHs continues in the central region.
As a result, the Galactic center would be dominated by the dense cluster of the PBHs or a fewer but more massive BHs that
result from the mergers of the accumulated PBHs. 
In either case, highly concentrated region arises by the infall of the PBHs.
Since there is an upper limit on the mass in the Galactic center, this limit can be translated into the
fraction of PBHs in the Galactic halo for some PBH mass range.

A PBH moving in the surrounding matter loses its velocity at a rate given by \cite{Binney-Tremaine}.
\be
\frac{dV}{dt}=-\frac{4\pi G^2 M_{\rm PBH} \rho_s \ln \Lambda}{V^2} \left(
{\rm erf}(X)-\frac{2X}{\sqrt{\pi}} e^{-X^2} \right),~~~~~~~X=\frac{V}{\sqrt{2}\sigma},
\ee
where $\sigma$ is the velocity dispersion of particles constituting the surrounding matter
and ${\rm erf}$ is the error function.
Now, suppose that PBHs move with the speed comparable to $\sigma$
for which ${\rm erf}(X)-\frac{2X}{\sqrt{\pi}} e^{-X^2}$ is about $0.5$. Then, we have
\be
\frac{dV}{dt} \simeq -\frac{2\pi G^2 M_{\rm PBH} \rho_s \ln \Lambda}{V^2}.
\ee
This friction exerts torque on the PBHs, and the PBHs gradually lose the angular momentum.
Assuming that the orbit at each moment of time follows Keplar motion,
the orbital velocity $V$ at radius $r$ ($r$ is distance measured from the Galactic center) is given by
$V=\sqrt{GM_s(r)/ r}$,
where $M_s(r)$ is the mass of the surrounding matter contained inside the radius $r$.
Using this relation, we can convert the time evolution of $V$ into the decay rate of the orbit as
\be
\frac{1}{r} \frac{dr}{dt}=-\frac{1}{\tau_{\rm fall} (r)},~~~~~~
\tau_{\rm fall} (r) =\left( 1+\frac{4\pi r^3 \rho_s (r)}{M_s(r)} \right)
{\left( \frac{GM_s(r)}{r} \right)}^{3/2} \frac{1}{4\pi G^2 M_{\rm BH} \rho_s (r) \ln \Lambda}.
\ee
Thus, the PBH located at distance $r$ spirals to the Galactic center on the time scale $\tau_{\rm fall}$.
Near the Galactic center, the surrounding matter mainly consists of dark matter and stars in the bulge.
Let us evaluate $\tau_{\rm fall}$ for the individual component respectively.
For the stars in the bulge, we employ the Hernquist model as the density distribution \cite{Hernquist:1990be}
\be
\rho_s =\frac{m_B}{2\pi} \frac{r_B}{r} \frac{1}{ {(r+r_B)}^3},
\ee
where $r_B=0.6~{\rm kpc}$ and $M_B=2.6\times 10^{10}~M_\odot$ is the total mass.
The corresponding infall time is given by
\be
\tau_{\rm fall} \simeq 3~{\rm Gyr} ~\frac{r/r_B+3}{r/r_B+1} 
{\left( \frac{r}{1~{\rm kpc}} \right)}^{5/2} {\left( \frac{M_{\rm BH}}{10^6~M_\odot} \right)}^{-1} 
{\left( \frac{\ln \Lambda}{10} \right)}^{-1}.
\ee
For the dark matter halo, we employ the isothermal model as the density distribution,
\be
\rho_s =\frac{\sigma^2}{2\pi Gr^2},
\ee
where $\sigma =200~{\rm km/s}$.
The corresponding infall time is given by
\be
\tau_{\rm fall} \simeq 3~{\rm Gyr} ~{\left( \frac{r}{1~{\rm kpc}} \right)}^2 
{\left( \frac{M_{\rm BH}}{10^6~M_\odot} \right)}^{-1} 
{\left( \frac{\ln \Lambda}{10} \right)}^{-1}.
\ee
These estimates show that PBHs as massive as $10^6~M_\odot$ within $\sim {\rm kpc}$ from
the Galactic center fall into the center within the age of the Universe.

Let us now define the critical radius $r_{\rm df}$ below which the PBHs spiral in to the Galactic center
within the age of the Galaxy.
Since the efficiency of the dynamical friction depends on the PBH mass, 
this radius is a function of $M_{\rm PBH}$.
Then, there would be a mass concentration in the galactic center by the present time by this amount,
\be
M_{\rm con}=f_{\rm PBH} \int_0^{r_{\rm df}}~4\pi r^2 \rho_{\rm DM} (r).
\ee
In \cite{Carr:1997cn}, the bound on $f_{\rm PBH}$ was obtained by requiring that
the mass concentration $M_{\rm con}$ should not exceed the observational limit $\sim 3\times 10^6~M_\odot$
on the central dark matter.
According to their analysis, the result strongly constrains the PBH abundance for $M_{\rm PBH} > 4\times 10^4~M_\odot$,
which is shown in Fig~\ref{Constraints}.
It is important to keep in mind that there are caveats in this constraint.
Infalls of multiple PBHs into the central region may result in the ejections of PBHs 
by the slingshot mechanism, which would be efficient when the mergers of the PBHs 
in the central region by the gravitational radiation occur on the time scale longer than
the one on which another PBH falls to the center. 
If the ejections of the PBHs happen, the mass growth of the central region is impeded.
It was demonstrated by N-body simulations that this mechanism is actually effective for BHs 
with $\lesssim 10^{6.5}~M_\odot$ \cite{Xu:1994vb} when all the dark matter is assumed to be such BHs.
Another potentially important effect is the rocket effect which arise by the anisotropic emission
of the gravitational radiation at the merger stage of the PBH binaries.
According to \cite{Xu:1994vb}, a recoil velocity $\gtrsim 1500~{\rm km/s}$ is needed
to prevent the runaway for $M_{\rm PBH}=10^7~M_\odot$.
The newly formed BH will escape from the central region if the kicked velocity of the BH is significant.

\subsubsection{Disk heating}
PBHs moving randomly in the Galactic halo have chance to pass through the Galactic disk,
and the stars in the disk are pulled by the gravitational attraction and acquire velocity every time the PBH passes nearby.
Since direction of the velocity that the star gains for the individual PBH passage is random,
the time evolution of the velocity of the disk stars is described by the random walk.  
Then, the variance of the star's velocity increases in proportion to the time.
In other words, the disk stars become hotter and hotter over time.
By requiring that the velocity increased by the PBHs does not exceed the observed velocity,
we can constrain the PBH abundance for some PBH mass range. 

Under the encounters by rapidly moving PBHs, the increase of variance of the star's velocity during
the time $t$ is given by \cite{Lacey:1985mt}
\be
\sigma^2 \simeq \frac{8\pi G^2 f_{\rm PBH} \rho_{\rm DM} M_{\rm PBH} \ln \Lambda}{V} t
\simeq {(50~{\rm km/s})}^2 ~f_{\rm PBH} \left( \frac{M_{\rm PBH}}{10^6~M_\odot} \right)
{\left( \frac{V}{200~{\rm km/s}} \right)}^{-1} \left( \frac{t}{10~{\rm Gyr}} \right),
\ee
when $V^2$ is much larger than the total velocity dispersion.
Proper velocity of old stars in the solar neighborhood is around $50~{\rm km/s}$ \cite{Nordstrom:2004wd}.
Thus, the above equation shows that super-massive PBHs greater than million solar mass cannot 
be the dominant component of dark matter.
Furthermore, for PBHs heavier than $10^6~M_\odot$, we can derive a meaningful ({\it i. e. } $f_{\rm PBH} <1$)
constraint on the PBH abundance as
\be
f_{\rm PBH} < {\left( \frac{M_{\rm PBH}}{10^6~M_\odot} \right)}^{-1}.
\ee
For more detailed discussion, see \cite{Carr:1997cn}.

Formulation of how the velocity variance of the stars evolves under the influence
of the massive objects (massive gas clouds were the main focus) was developed by Lacey \cite{Lacey:1984zg}.
In \cite{Lacey:1985mt}, the idea was proposed that the dark halos are composed of the super-massive black holes ($\sim 10^6~M_\odot$) and
they could explain the origin of the observed features suggesting the disk heating.
However, recent precise measurements of the stars in the solar neighborhood disfavor this possibility.
Introducing a free parameter $\beta$ to parametrize a relation between the ages of the stars and their velocity dispersions as
\be
\sigma \propto t^\beta,
\ee
the measurements show $\beta \simeq 0.33$ with uncertainty of $\beta$ being about $\Delta \beta = 0.02\sim 0.05$
\cite{Binney:2000pe, Nordstrom:2004wd}.
The measured value does not coincide with the prediction $\beta =0.5$ in the super-massive black hole scenario
and support the spiral arms and the giant molecular clouds as the cause of the disk heating \cite{Nordstrom:2004wd}.

\subsection{Accretion constraints}
Accretion of gas onto the PBHs and its impact on the constraint of the PBH abundance 
has also been a subject of research \cite{Carr:1981zg}.
Actually, previous studies show that the accretion constraint has a potentially significant power in 
constraining PBHs for some mass range.
However, we have to keep in mind that
the physics is much more involved in the case of the accretion constraint
compared to the lensing and the dynamical constraints discussed in the previous sections.
So far, it is impossible to derive the PBH constraint from the accretion from the first-principle calculation,
and all the derived constraints that will be presented later make some assumptions and in some cases rely on
the observationally established empirical rules. 
In this section, we will review the accretion constraint on the PBH abundance based on the two different processes;
the accretion effects to the CMB that arise in the early Universe and the electromagnetic waves from
the accreted matter from the present PBHs. 

\subsubsection{Accretion effects on CMB}
\label{accretion-cmb}
Baryonic gas around the PBH is attracted by the PBH by its gravity.
As the gas falls into the central region, the gas is compressed, increases its density and temperature.
During the infall, the gas can be fully ionized either by the internal collisions of gas particles or
by the outgoing radiation. 
Near the black hole horizon, the gas temperature is enormous and intense radiation from the ionized
gas emanates outward.
This radiation ionizes or heats the gas filling the Universe and modifies the spectrum of the CMB photons
from the Planckian distribution,
the decoupling time of the CMB photons, and the ionization history.
The latter two result in changing the power spectrum of the CMB temperature and 
the polarization anisotropies.
In this way, the PBHs leave the non-standard features in the CMB observables.
Non-detection of such features is translated into the upper limit on the PBH abundance. 
Although the flow of the above logic is conceptually understandable,
it is extremely difficult to predict how much the CMB observables are modified by the PBH accretion
from the first principle and self-consistent calculations because of the complex nature of accretion.
So far, various approximations and assumptions have been adopted in obtaining the PBH constraint
from the accretion.

In early study \cite{Carr:1981zg}, the standard Bondi accretion formula was adopted for
the mass accretion rate, the efficiency of converting the accreted mass
into the luminosity of the outgoing radiation was assumed to be independent of the cosmic time,
and the spectrum of the emergent radiation was  assumed to be flat and constant in time.
In \cite{Ricotti:2007au}, the analysis was refined by taking into account the accumulation of dark matter,
relative velocity between PBHs and the baryonic gas, and the coupling between the CMB radiation and the 
baryonic gas. 
In \cite{Blum:2016cjs}, a different relative velocity between the PBHs and the baryonic gas was used.
Yet, the radiation efficiency was assumed to be a fixed value in those studies.
In \cite{Ali-Haimoud:2016mbv}, the analysis in \cite{Ricotti:2007au} was reexamined without adopting
a priori fixed value of the radiative efficiency and it was found that the derived efficiency is much
smaller than the one assumed in \cite{Ricotti:2007au}.
As a result, the strong constraint on the PBH abundance derived in \cite{Ricotti:2007au} was significantly
weakened in \cite{Ali-Haimoud:2016mbv}.
In what follows, we briefly explain the main point of the analysis in \cite{Ali-Haimoud:2016mbv}.
  
A basic picture of the accretion investigated in \cite{Ali-Haimoud:2016mbv} is as follows.
We start from considering a stationary system where a PBH is steadily swallowing the surrounding baryonic gas
on the cosmological background.
In other words, we assume that the accretion time scale is shorter than the Hubble time and the accretion 
follows adiabatically the cosmic expansion.
It was shown in \cite{Ricotti:2007au} that this picture is valid for $M_{\rm PBH} \lesssim 3\times 10^4~M_\odot$,
and we focus on this mass range.
Far from the PBH, the baryonic gas is little attracted by the PBH and its density becomes
equal to the cosmological density.
Down to a certain radius (below the Bondi radius), the cooling of the baryonic gas by the background CMB
is efficient and the temperature of the gas remains the same as that of CMB.
Below this radius, the cooling by the CMB becomes negligible and the gas temperature increases as $r^{-1}$
by the adiabatic compression as the gas falls.
When the temperature exceeds $\sim 10^4~{\rm K}$, the collisional ionization starts to be important.
During this phase, the increase of the internal energy gained by the infall is consumed as the ionization energy
and the gas temperature remains constant.
After the gas is completely ionized, the temperature again increases as $r^{-1}$ until the electrons become
relativistic.
The rate of temperature increase changes to $r^{-2/3}$ after the electrons become relativistic.
Finally, the gas is swallowed by the PBH at the event horizon.
Near the event horizon, the gas temperature is enormous and the bremsstrahlung radiation
produces the intense outgoing radiation which eventually heats or ionizes the background gas.
It is possible that the gas ioniziation is caused not by the collisions of gas particles but by the
radiation emanating from the vicinity of the BH horizon.
In \cite{Ali-Haimoud:2016mbv}, two extreme cases in which the complete ionization is achieved
solely by either the collisional ionization or the photoionization.

\begin{figure}[t]
  \begin{center}
   \includegraphics[width=110mm]{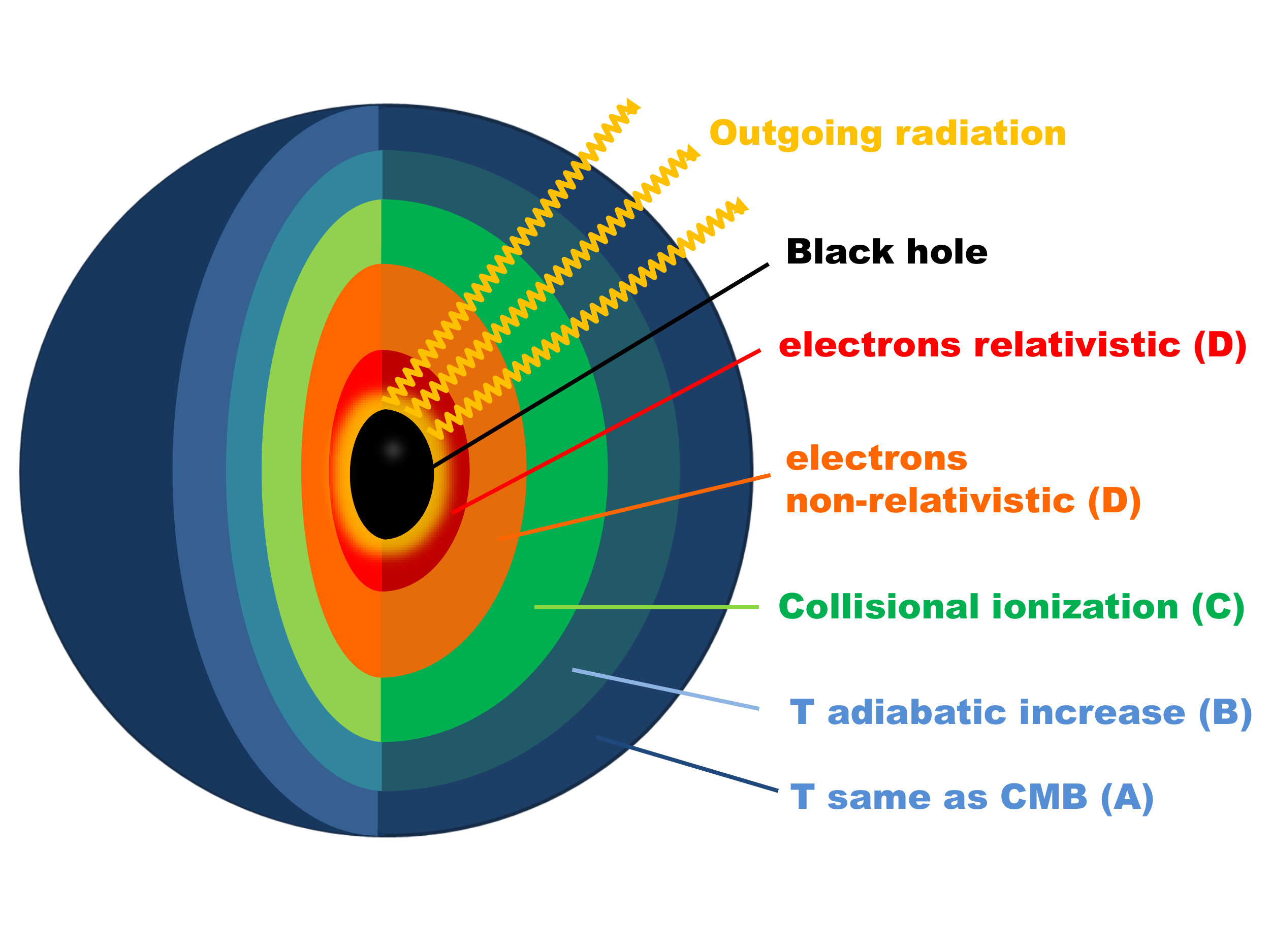}
  \end{center}
  \caption{Schematic picture of the state of the gas falling onto the central PBH. In the outermost region A,
  the gas is efficiently coupled by CMB and the gas temperature remains the same as that of CMB.
  In the region B, the gas is compressed adiabatically and the temperature increases as the gas falls.
  In the region C, the gravitational energy gained by the infall is used for the ionization and the temperature does not increase.
  In the regions D and E, the gas is again compressed adiabatically. At the vicinity of the BH, the gas temperature is high and
  intense radiation emanates by the bremsstrahlung radiation.}
  \label{PBH-accretion-picture}
\end{figure}

Basic equations describing the accretion in the outer region where the ionization fraction
coincides with that on the cosmological background (${\bar x_e}$) are given by
\begin{eqnarray}
&&4\pi r^2 \rho |v|={\dot M}={\rm const.}, \\
&&v\frac{dv}{dr}=-\frac{GM}{r^2}-\frac{1}{\rho} \frac{dP}{dr}-\frac{4}{3} \frac{{\bar x_e} \sigma_T
\rho_{\rm CMB} }{m_p} v, \\
&&v\rho^{2/3} \frac{d}{dr}\left( \frac{T}{\rho^{2/3}} \right)=
\frac{8 {\bar x_e} \sigma_T \rho_{\rm CMB}}{3m_e (1+{\bar x_e})} (T_{\rm CMB}-T),
\end{eqnarray}
where $v<0$ is the radial gas velocity.
The last terms in the second and the third equations represent the drag force and the cooling by the CMB, respectively.
From these equations, the mass accretion rate ${\dot M}$ is determined.
Then, the continuity equation gives the electron number density at the vicinity of the event horizon as
\be
n_e =\frac{\dot M}{4\pi m_p r^2 |v|}=\frac{\dot M}{4\pi m_p r_g^2} {\left( \frac{r}{r_g} \right)}^{3/2},
\ee
where we have used the free fall velocity $v \approx \sqrt{r_g/r}$.
The gas temperature at the vicinity of the event horizon can be also determined by adiabatically extending the profile $T(r)$
derived by solving the above basic equations down to the horizon.
From the knowledge of the electron number density and the temperature, the luminosity $L$ of the bremsstrahlung 
radiation is determined as
\be
L=\int 4\pi r^2 j dr,~~~~~~j=n_e^2 \alpha \sigma_T T {\cal J}(T/m_e),
\ee
where ${\cal J}$ is a certain function fixed by the elementary process of the bremsstrahlung radiation.
Then, the energy injection rate per volume of the radiation produced by the PBHs is given by
\be
{\dot \rho_{\rm inj}}=f_{\rm PBH} \frac{\rho_{\rm DM}}{M_{\rm PBH}} \langle L \rangle,
\ee
where $\langle L \rangle$ is the average of the luminosity over the PBH peculiar velocity
relative to the baryonic gas.
In \cite{Ali-Haimoud:2016mbv}, it was assumed that the luminosity in the case where the PBH moves relative to the
baryonic gas was obtained by simply adding the relative velocity to the sound speed at infinity. 
The injected radiation interacts with the background gas by the Compton scattering.
Time evolution of the energy actually deposited to the gas is given by \cite{Ali-Haimoud:2016mbv}
\be
a^{-7} \frac{d}{dt} (a^7 {\dot \rho_{\rm dep}}) \approx 0.1~{\bar n_H} \sigma_T
( {\dot \rho_{\rm inj}}-{\dot \rho_{\rm dep}} ),
\ee
which is approximate, and derived under the assumption that most of the energy is injected 
near $~0.1-10~{\rm MeV}$, corresponding to the characteristic temperature of the gas near the BH horizon
It was assumed that the deposited energy in this manner is distributed to the increase of the gas temperature,
ionization fraction, and the excitation of the hydrogen atoms (to the first excited states) as \cite{Chen:2003gz}
\be
\Delta {\dot T_{\rm gas}}=\frac{2}{3 n_{\rm tot}} \frac{1+2{\bar x_e}}{3} {\dot \rho_{\rm dep}},~~~
\Delta {\dot {\bar x}_e}=\frac{1-{\bar x_e}}{3} \frac{{\dot \rho_{\rm dep}}}{E_I n_H},~~~
\Delta {\dot x_2}=\frac{1-x_e}{3} \frac{{\dot \rho_{\rm dep}}}{E_2 n_H},
\ee
where $E_I=13.6~{\rm eV},~E_2=10.2~{\rm eV}$.
These equations describe how the temperature and the ionization history are modified by the accreting PBHs.
For instance, increase of the ionization fraction enhances the CMB optical depth, which results in the damping of the
small scale CMB fluctuations and the enhancement of the polarization power on large angular scales compared
to the standard case. 
Also, the shift of the redshift of the last scattering changes the phase of the acoustic oscillations in the
CMB spectrum.
In \cite{Ali-Haimoud:2016mbv}, the modified CMB fluctuations were computed by using the recombination
and the Boltzmann codes \cite{AliHaimoud:2010dx, Blas:2011rf} and they were compared with
the CMB TT, TE and EE power spectra provided by the Planck collaboration \cite{Aghanim:2016cps}.
The resultant constraint on the PBH abundance is shown in Fig~\ref{Constraints}.

Accretion of gas onto the PBHs in the early Universe (even before the CMB decoupling) also induces the CMB
spectral distortions since the produced photons by the accreting PBHs are not completely thermalized.
Photons generated in the redshift $5\times 10^4 < z<2\times 10^6$ achieve the kinetic equilibrium and
yield the non-vanishing chemical potential ($\mu$-distortion).
Here the upper and the lower limit on the redshift correspond to the temperature above which the photon-number
changing process efficiently occurs to make the distribution Planckian
and the temperature below which the kinetic equilibrium is no longer maintained, respectively. 
Photons generated in the redshift $200 <z<5\times 10^4$ yield the Compton-$y$ distortion.
Here the lower limit corresponds to the temperature at which the baryonic gas decouples from the CMB.
Contrary to the meaningful constraint on $f_{\rm PBH}$ by the temperature
and the polarization anisotropies of CMB, 
it was also found in \cite{Ali-Haimoud:2016mbv} that the induced amplitudes of both the $\mu$
and $y$-distortions are too small to be relevant to present and future observations.

So far, the discussion has been based on the assumption that the accretion is spherical.
The accretion is approximately spherical when the angular momentum of the gas
at the Bondi radius is smaller than the one determined by the Keplerian orbit at the 
innermost stable circular orbit.
If this is not satisfied, the gas motion becomes Keplerian before it is absorbed by the PBH and
the accretion disk forms around the PBH.
While the radiation in the case of the spherical accretion originates from the bremsstrahlung radiation,
the radiation is more efficiently produced by the viscous heating in the presence of the accretion disk \cite{Shapiro-Teukolsky}.
As a result, the radiative efficiency is typically higher for the accretion disk than the spherical accretion. 
In \cite{Poulin:2017bwe}, it was claimed that the spherical assumption may be violated after the recombination.
In this reference, based on the analytical estimation,
it was suggested that the accretion disk forms around the PBH.
Performing the similar analysis for the CMB temperature and the polarization anisotropies described above
for the case of the accretion disk, 
it was found that the corresponding constraint on $f_{\rm PBH}$ becomes much stronger than the one obtained 
in \cite{Ali-Haimoud:2016mbv}.
The derived constraints are shown in Fig~\ref{Constraints}.

\subsubsection{X-rays and radio from the present-day PBHs}
\label{accretion-p}
In the previous subsubsection, we have considered the gas accretion onto the PBHs in the early Universe
and its effects on the CMB.
PBHs in the present Universe also accrete the surrounding gas if they are in the dense environment,
and the constraint on the PBH abundance can be obtained by the comparison between the observational data
and the theoretical predictions of the electromagnetic waves from the accreting PBHs.

In \cite{Fujita:1997fh}, radiation spectra originating from the gas accretion onto the PBHs of $0.5~M_\odot$ 
in the solar neighborhood was computed based on the assumption that the accretion is modeled by the
advection-dominated accretion flow (ADAF).
This research was motivated by the microlensing observations that hint the PBHs of mass $\sim 0.5~M_\odot$
as dominant component of dark matter.
It was found that the radiation spectra from the PBHs could be at the detectable
level at the IR-optical band and is negligibly small at the X-ray band.
This implies that a meaningful upper limit on the PBH abundance could be obtained from the IR-optical band only.
However, since the main interest in \cite{Fujita:1997fh} was the detectability of the PBH signal
by the accretion, explicit number for the upper limit on $f_{\rm PBH}$ was not given.
Since there are contaminations from other astrophysical sources,
for instance young stellar objects at the IR-optical band,
it is important to keep in mind that it is crucial to do the consistency check over the multi-wavelength bands
when one claims the positive detection of the PBHs from the accretion.

\begin{figure}[htp]
  \begin{center}
   \includegraphics[width=160mm]{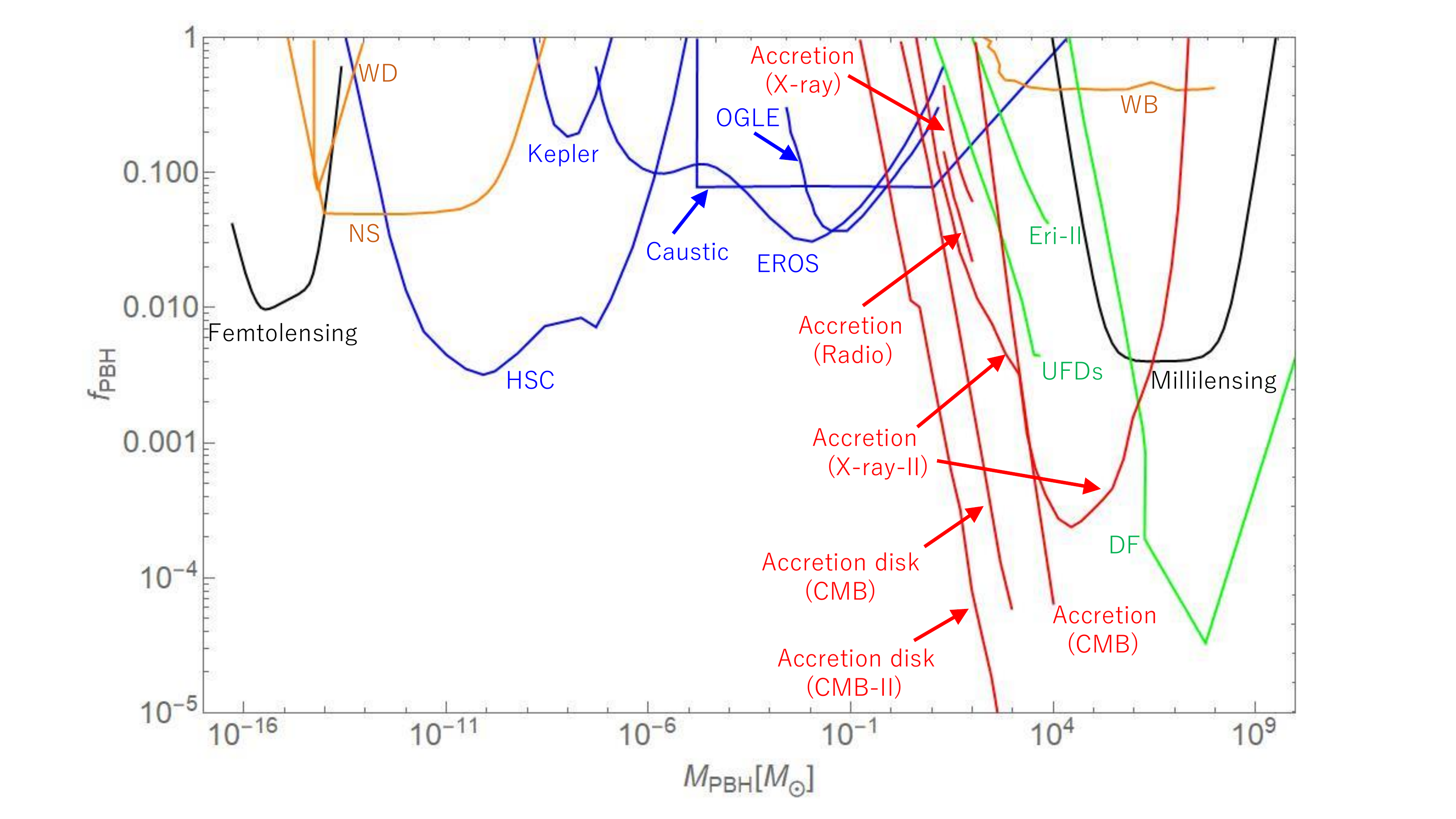}
  \end{center}
  \caption{Upper limit on $f_{\rm PBH}=\Omega_{\rm PBH}/\Omega_{\rm DM}$ for various PBH mass (assuming monochromatic
  mass function). Blue curves represent lensing constraints by EROS \cite{Tisserand:2006zx}, 
  OGLE \cite{Wyrzykowski:2011tr}, 
  Kepler \cite{Griest:2013aaa}, HSC \cite{Niikura:2017zjd} and 
  Caustic \cite{Oguri:2017ock} (see \ref{microlensing}). 
  Black curves represent constraints by the millilensing \cite{Wilkinson:2001vv} (\ref{millilensing}) 
  and the femtolensing \cite{Barnacka:2012bm} (\ref{femtolensing}).
  Orange curves represent dynamical constraints obtained by requiring that existent compact objects such as 
  white dwarfs (WDs) \cite{Graham:2015apa} (\ref{DWD})
  and neutron stars (NSs) \cite{Capela:2013yf} (\ref{DNS}) as well as the wide binaries (WBs) \cite{Quinn:2009zg} (\ref{wide-halo-binary}) 
  are not disrupted by PBHs.
  Green curves represent constraints by the dynamical friction (DF) 
  on PBHs \cite{Carr:1997cn} (\ref{dynamical-friction}), 
  the ultra-faint dwarfs (UFDs) \cite{Brandt:2016aco}, and Eridanus II \cite{Brandt:2016aco} (\ref{UFDG}). 
  Red curves represent constraints by the accretion onto the PBHs such as 
  CMB for the case of the spherical accretion \cite{Ali-Haimoud:2016mbv} and the
  case of the accretion disk \cite{Poulin:2017bwe} with two opposite situations where the sound speed of the
  baryonic matter is greater (labeled by CMB) or smaller (labeld by CMB-II) than the relative baryon-dark matter velocity
  (\ref{accretion-cmb}), 
  radio, and X-rays \cite{Gaggero:2016dpq, Inoue:2017csr} (\ref{accretion-p}).
  }
  \label{Constraints}
\end{figure}

In \cite{Gaggero:2016dpq}, the upper limit on the PBH abundance for $10 <M_{\rm PBH}/M_\odot <100$
was obtained by using the Very Large Array (VLA) radio and the Chandra X-ray observational data.
The flow of the argument is as follows.
First, the PBH was assumed to accrete the surrounding gas with the mass accretion rate given by
\be
{\dot M}=4\pi \lambda {(GM_{\rm PBH})}^2 \frac{\rho_{\rm gas}}{{(v_{\rm PBH}^2+c_s^2)}^{3/2}},
\label{accretion-dotm}
\ee
where the value of $\lambda$, which measures the amount of mass accretion rate normalized by the
Bondi accretion rate, was chosen to be $\lambda=0.02$, 
which is consistent with X-ray observations of dwarf nova and
the nuclear accretion in galaxies (see also \cite{Fender:2013ei}).
The radiative efficiency, that converts the mass accretion rate to the bolometric luminosity from
the accreting gas, was then assumed to be given by
\be
L_B=\eta {\dot M},~~~~~~~\eta=0.1~\left( \frac{\dot M}{0.01~{\dot M}_{\rm Edd}}\right),
\ee
where ${\dot M}_{\rm Edd}$ is the mass accretion rate corresponding to the Eddington limit. 
That the radiative efficiency is proportional to the mass accretion rate for ${\dot M}<0.01~{\dot M}_{\rm Edd}$,
which is the region of our current interest,
is suggested by the observational data \cite{Kording:2006sa}.
In order to obtain the X-ray luminosity out of the bolometric luminosity, 
a relation
\be
L_X=0.3 L_B,
\ee
was adopted. 
It is known that many accreting BHs accompany jets from which ${\rm GHz}$ radio waves are emitted.
Then, by making use of the so-called fundamental plane \cite{Merloni:2003aq}, 
which is an empirical relation among the mass of BH accompanying jets, 
X-ray luminosity $L_X$, and radio luminosity $L_R$,
the radio luminosity was computed under the assumption that PBHs accreting the gas produce jets.
With this formalism to compute $L_X$ and $L_R$ from the given PBH mass (delta function mass spectrum), 
the Monte Carlo simulation has been performed by distributing the positions and velocities of PBHs
according to the Navarro-Frenk-White (NFW) profile \cite{Navarro:1995iw} and the Maxwell-Boltzmann distribution, respectively,
and the expected detectable number of X-ray and radio sources from the PBHs have been evaluated
for the Chandra \cite{Muno:2008qy} and the VLA observations \cite{Lazio:2008ax}.
It was found that both X-ray and the radio observations result in meaningful constraint $f_{\rm PBH}<1$.
These constraints are shown in Fig~\ref{Constraints}.

The PBH constraint by the present-day accretion has also been obtained in \cite{Inoue:2017csr}
by using the luminosity function of X-ray binaries derived in \cite{Mineo:2011id} in which
29 nearby star-forming galaxies
in Chandra, Spizer, GALEX and 2MASS archives were analyzed.
In \cite{Inoue:2017csr}, assuming the value $\lambda=1$ in Eq.~(\ref{accretion-dotm})
and $L_X=L_B$ for the X-ray luminosity, and introducing the dark matter disk in galaxies
which has been suggested in simulations \cite{Read:2008fh, Read:2009iv} to the PBH distribution,
luminosity function of the X-ray sources powered by PBHs
was computed and was compared with the observationally determined one given in \cite{Mineo:2011id}.
It was found that the PBHs in the mass range a few $\sim 10^7~M_\odot$
are tightly constrained. This result is shown in Fig~\ref{Constraints}.

As just described, there are theoretical uncertainties in modeling the accretion and the emitted radiation.
Yet, these studies demonstrate that the accretion to the present-day PBHs has a potential to provide a 
meaningful constraint on the PBH abundance.
Finally, before closing this section, we mention that searching electromagnetic signals from the isolated astrophysical
black holes ({\it i.~e.}, non-PBH) has also been a target of active research (e.~g.~, \cite{Fujita:1997fh, Agol:2001hb, Fender:2013ei, Matsumoto:2017adh}).

\subsection{Large scale structure constraint}
\label{Lyalpha}
PBHs randomly distributing in space in the early Universe generate primordial density perturbations 
by their Poisson fluctuations on scales larger than the PBH mean distance,
as first noticed in \cite{Meszaros:1975ef}. 
The mean comoving distance of PBHs is given by
\be
\ell_{\rm mean}={\left( \frac{M_{\rm PBH}}{f_{\rm PBH} \rho_{\rm DM}} \right)}^{1/3}=
0.3~{\rm Mpc}~{\left( \frac{M_{\rm PBH}}{10^6~M_\odot} \right)}^{1/3}
{\left( \frac{f_{\rm PBH}}{10^{-3}} \right)}^{-1/3}. \label{PBH-mean-distance}
\ee
Denoting by $n_{\rm PBH}$ the comoving PBH number density, the number of PBHs in the comoving volume $\lambda^3$ 
fluctuates typically as $\sim N_\lambda^{1/2}$, 
where $N_\lambda= n_{\rm PBH} \lambda^3$ is the average PBH number in that volume. 
The fluctuations are isocurvature perturbations since they are present on the hypersurface 
where the radiation energy density looks uniform.
Thus, in the presence of the PBHs, on top of the standard nearly scale-invariant adiabatic perturbations, 
the dark matter density contrast has the isocurvature perturbations whose variance is given by
\be
\langle \delta_{\rm DM}^2 \rangle=\bigg\langle {\left( \frac{\delta \rho_{\rm PBH}}{\rho_{\rm DM}} \right)}^2 \bigg\rangle
=f_{\rm PBH}^2 N_\lambda^{-1}=\frac{f_{\rm PBH} M_{\rm PBH}}{\rho_{\rm DM} \lambda^3},
\ee
on the comoving scale $\lambda$.
The scaling of the variance as $\propto \lambda^{-3}$ shows that the produced dark matter perturbations are more enhanced 
on smaller scales.
Equivalently, in terms of the dimensionless power spectrum, the PBHs yield the contribution
\be
{\cal P}_{\delta_{\rm DM}} (k)=\frac{k^3}{2\pi^2} \frac{f_{\rm PBH} M_{\rm PBH}}{\rho_{\rm DM}}, \label{Ly-PBH-isocurvature}
\ee
for which the spectral index is 3, namely, extremely blue spectrum. 

Based on this observation that the Poisson fluctuations in the PBH distribution enhance the 
dark matter perturbations on small scales, their impact on the Ly$\alpha$ forest observations
was investigated in \cite{Afshordi:2003zb}. 
The basic idea is as follows.
Spectra of distant quasars and galaxies show many absorption lines known as the Ly$\alpha$ forest.
This arises due to the intervening neutral hydrogen in the intergalactic medium between 
the quasars and the Earth (typically $z=2\sim 5$) that absorbs photons from quasars by the Ly$\alpha$ transition  
($n=1$ to $n=2$, where $n$ is the principal quantum number of the hydrogen atom).
Although this epoch is after the reionization, 
there is still a tiny fraction of the neutral hydrogen in the intergalactic medium because of the balance
between the photoionization by the surrounding UV radiation and the recombination
and tiny amount of the neutral hydrogen is sufficient to produce the Ly$\alpha$ forests. 
The wavenumber corresponding to this transition is $\lambda_{\rm Ly\alpha}=1216~$\AA.
When the photons are partially absorbed by the neutral hydrogen at redshift $z_c$, then
the absorption line appears at $\lambda=\lambda_{\rm Ly\alpha}(1+z_c)$ in the observer frame. 
The optical depth of the Ly$\alpha$ transition to the direction of a quasar is given by (e.~g.~, \cite{McQuinn:2015icp})
\be
\tau_{\rm Ly\alpha} (\lambda) \simeq1.3~\Delta_b (z_c) \left( \frac{x_{\rm HI} (z_c)}{10^{-5}} \right)
{\left( \frac{1+z_c}{4} \right)}^{3/2},
\ee
where $\Delta_b=n_b/{\bar n_b}$ is the baryon density normalized by the average value,
and $x_{\rm HI}$ is the fraction of the neutral hydrogen. 
This equation shows that the Ly$\alpha$ absorption is more efficient in the site where baryon is denser.
Because of this, many absorption lines in the spectra, namely, fine fluctuations of $\tau_{\rm Ly\alpha}$ for
the relevant range of $\lambda$, reflect the inhomogeneous distribution of baryonic matter.
Given that the baryon perturbations are affected by the dark matter perturbations gravitationally, 
statistical properties of the optical depth encode those of the dark matter perturbations.
Conversely, observational analysis of the optical depth allows to probe the matter perturbations
on small scales down to $\sim {\rm Mpc}$, where the minimum scale is determined by the thermal 
broading of the spectra of the Ly$\alpha$ forest.
These scales are mildly non-linear in the Ly$\alpha$ epoch.

In \cite{Afshordi:2003zb}, superposing the dark matter isocurvature perturbation given by Eq.~(\ref{Ly-PBH-isocurvature})
(assuming $f_{\rm PBH}=1$) on the standard adiabatic perturbations, 
the resultant Ly$\alpha$ clouds were computed for various PBH masses by the use of hydro-dynamical cosmological simulations.
Then, the power spectrum $P_F(k)$, which is a two-point function of the Fourier transform of the transmitted flux $F(\lambda)$
defined by $F=e^{-\tau_{\rm Ly\alpha}}$ was evaluated and compared with the observational data presented in \cite{Croft:2000hs}.
It was found that the PBHs heavier than $\sim 10^4~M_\odot$ comprising all the dark matter produce too much power
of the transmitted flux and are inconsistent with the Ly$\alpha$ observations. 
Although the explicit upper limit on $f_{\rm PBH}$ is not provided in \cite{Afshordi:2003zb},
simple analytical expression for the upper limit consistent with the results of \cite{Afshordi:2003zb}
was given in \cite{Carr:2009jm}.
The main point of \cite{Carr:2009jm} is that the matter perturbations sourced by Eq.~(\ref{Ly-PBH-isocurvature})
should not exceed ${\cal O}(1)$, typical amplitude of the matter perturbation, at the Ly$\alpha$ epoch, which yields
\be
f_{\rm PBH} < {\left( \frac{M_{\rm PBH}}{10^4~M_\odot} \right)}^{-1}. \label{Lyalpha-con}
\ee
This constraint becomes meaningless when it implies that 
the PBH mean distance (\ref{PBH-mean-distance}) is larger than the size of the 
Ly$\alpha$ clouds $\ell_{\rm Ly\alpha}$.
Thus, Eq.~(\ref{Lyalpha-con}) makes sense when
\be
f_{\rm PBH}> 3\times 10^{-4} \left( \frac{M_{\rm PBH}}{10^7~M_\odot} \right) {\left( \frac{\ell_{\rm Ly\alpha}}{1~{\rm Mpc}} \right)}^{-3},
\ee
is satisfied.
Equivalently, Eq.~(\ref{Lyalpha-con}) can be also obtained by requiring that the power spectrum of Eq.~(\ref{Ly-PBH-isocurvature})
at the Ly$\alpha$ scale ($k\sim 1~{\rm Mpc}^{-1}$) is less than the value of the standard one $\sim 10^{-10}$ based on 
the consistency between the prediction of the standard adiabatic perturbations and the observed Ly$\alpha$ forests.

In \cite{Kashlinsky:2016sdv}, inspired by the first detection of GWs by LIGO,
interesting suggestion was made that PBHs with mass ${\cal O}(10)M_\odot$ comprising all the dark matter
can solve the potential tension of the observed near-IR cosmic infrared background (CIB) anisotropies.
CIB is accumulated emission of luminous objects throughout the history of the Universe (e.~g.~\cite{Kashlinsky:2004jt}) 
and was detected for the first time by COBE DIRBE instrument \cite{Hauser:2001xs}.
Intriguingly, measured anisotropies of CIB obtained by removing the foreground stars and galaxies
are larger than expected and cannot be explained only by the contribution from remaining faint galaxies \cite{Kashlinsky:2005di}.
This potential tension basically arises due to rareness of halos where luminous stars form in high redshift.
Denoting by $f_*$ efficiency that baryons inside each halo convert to luminous sources,
the required value of $f_*$ to explain the observed amplitude of the CIB fluctuations is estimated as \cite{Kashlinsky:2016sdv} 
\be
f_*\simeq 0.1 {\left( \frac{f_{\rm Halo}}{0.01} \right)}^{-1} \left( \frac{z_{\rm eff}}{10} \right),
\ee
where we assumed that a mass-fraction $f_{\rm Halo}$ of dark matter is contained in the halos that have luminous sources,
and $z_{\rm eff}$ is the effective average redshift.
How to achieve such high value $f_* \simeq 0.1$ is a challenging task if $f_{\rm Halo} \simeq 0.01$.
Based on the idea explained above that the additional dark matter fluctuations sourced by the PBHs  
increase the minihalos in high redshift, 
$f_{\rm Halo}$ was computed in the presence of PBHs and was found to become
much larger than that in the conventional $\Lambda$CDM case.
Hence, in this scenario, the required value of $f_*$ is reduced to very modest one,
which is the main point in \cite{Kashlinsky:2016sdv} that PBHs can explain the CIB fluctuations.

\subsection{Indirect constraints}
So far, we have considered the cosmological and astrophysical effects caused by the PBHs and 
discussed how the PBH abundance can be constrained by them. 
Those constraints are direct in the sense that all the effects are directly triggered by the PBHs. 
A nice point of the direct constraints is that they do not resort to the formation mechanism of PBHs. 
Any model predicting the PBH formation must satisfy the direct constraints in order for it to be considered as a consistent model.
 
In addition to the direct constraints, there are indirect ones, which are the subjects of this section. 
As we have discussed in great detail in Sec.~2, the most popular mechanism of the PBH formation is the direct
gravitational collapse of the primordial density perturbations,
which can be naturally embedded in the framework of inflation in which the quantum fluctuations of
the scalar fields are the ultimate origin of the primordial density perturbations.
In this scenario, PBHs form only at high-$\sigma$ (typically $\sim 10\sigma$) regions that are extremely rare. 
Although the other regions are not inhomogeneous enough to produce PBHs, they are still inhomogeneous enough 
to induce effects that are already excluded by or marginally consistent with observations, depending
on the statistical properties of the primordial density perturbations.
Those effects, which we are going to discuss in detail, 
are not sourced by the PBHs but by the density perturbations that seed the PBHs, and hence 
the resultant constraints on the PBHs are indirect.
We have to keep in mind the underlying assumptions when one tries to constrain a particular
inflation model predicting the PBHs by using the indirect constraints.

\subsubsection{Stochastic gravitational waves from the primordial density perturbations}
\label{indirect-SGW}
The first effect that the primordial density perturbations seeding the PBHs cause is the stochastic
gravitational waves produced by the mode-mode coupling of the density perturbations.
Density perturbations, which are classified as the scalar-type, evolve independently of the
gravitational waves (GWs), which are classified as the tensor-type, at the linear order in the perturbation (e.g., \cite{Liddle-Lyth}).
This independence no longer holds beyond the linear order, and the GWs are sourced by the 
density perturbations at the second order by their mode-mode couplings \footnote{
Conversely, stochastic GWs also source the density perturbations by the mode-mode couplings. 
From the requirement that the resultant density perturbations do not lead to the overproduction of PBHs, 
upper limit on the amplitude of the stochastic GWs can be obtained \cite{Nakama:2015nea, Nakama:2016enz}.}.

Assuming that primordial density perturbations existed on super-Hubble scales,
the production of the GWs by the second-order effect happens dominantly at the time when the density perturbations
re-enter the Hubble horizon \cite{Ananda:2006af, Baumann:2007zm}.
In other words, GWs are mainly generated at the same epoch as the PBH formation.
Once produced, those GWs freely propagate in the subsequent epochs and are still permeating the present Universe. 
Combining that typical frequency of such GWs at the formation time is comparable to the Hubble horizon
and that the horizon radius at that time is comparable to the size of the PBHs allows to
relate the PBH mass to the frequency of the GWs at present time as \cite{Saito:2008jc}
\be
f_{\rm GW} \simeq 1\times 10^{-9}~{\rm Hz}~ {\left( \frac{M_{\rm PBH}}{30~M_\odot} \right)}^{-1/2}.
\ee
Thus, primordial density perturbations producing the stellar-mass PBHs generate ultra-low frequency GWs in
${\rm nHz}$ band.
Quite interestingly, those low-frequency GWs are severely constrained by the pulsar timing experiments.

Pulsars, which are rapidly rotating neutron stars emitting beam of radio waves, 
are useful to probe low-frequency GWs (e.g. \cite{Hobbs:2017oam}).
In particular, millisecond pulsars, which rotate with period of ${\cal O}({\rm ms})$, are observed to be significantly 
more stable than the normal pulsars.
The arrival times of each pulse from the millisecond pulsars have been measured accurately and
are compared with the predictions of the pulsar timing model. 
Just as the expansion of space,
{\it i.e.} time-varying space-space components of the metric, causes the cosmological redshift,
the pulse frequency $\nu=1/T$,
where $T$ is the time interval between the arrival times of the successive pulses,
is modulated if GWs are present.
The modulation of the pulse frequency is given by \cite{Detweiler:1979wn}
\be
\frac{\delta \nu}{\nu}=-H^{ij} \bigg[ h_{ij} (t,{\vec x}_e)-h_{ij} (t-D,{\vec x}_p ) \bigg], \label{modulation-pulse-nu}
\ee
where $H^{ij}$ is a geometrical factor depending on the propagation direction of the GWs relative to
the direction of the pulsar, $h_{ij}$ represent GWs, $D$ is the distance to the pulsar, and
${\vec x}_e$ and ${\vec x}_p$ is position of the Earth and the pulsar, respectively.
This induces the timing residuals of the pulse arrival time at time $t$ (with reference to time $0$) as
\be
R(t)=-\int_0^t ~\frac{\delta \nu}{\nu}dt. \label{pulsar-residual}
\ee

A few hundred of millisecond pulsars have been observed.
It is reasonable to think that the timing residuals caused by the GWs (\ref{pulsar-residual}) for the pulsars
are strongly correlated since the frequency modulations due to the first term
in Eq.~(\ref{modulation-pulse-nu}) become almost identical.
It was shown by Hellings and Downs \cite{Hellings:1983fr} that the correlation function (normalized by the GW amplitude) 
of the timing residuals by the stochastic GWs between the pulsars
separated by the sky angle $\theta$ is given by
\be
c(\theta)=x \ln x-\frac{x}{6}+\frac{1}{3},~~~~~~~x=\frac{1-\cos \theta}{2}.
\ee
By searching this type of correlation in the measured data for many pulsar pairs,
which is known as the pulsar timing array (PTA) experiment,
GWs can be detected if they exist by amount
more than the sensitivity the experiments can reach.
Otherwise, the upper limit on the abundance of GWs is obtained.

Currently, there are three major PTA projects,
the Parkes Pulsar Timing Array (PPTA) \cite{Manchester:2012za},
the North American Nanohertz Observatory for Gravitational Waves (NANOGrav) \cite{McLaughlin:2013ira},
and the European Pulsar Timing Array (EPTA) \cite{Kramer:2013kea},
that aim to detect the ultra-low frequency GWs in the ${\rm nHz}$ range.
So far, no detection of the stochastic GWs has been reported, and the upper limit on the GWs has been derived.

Now, let us return to the theoretical computations of the secondary GWs produced by the primordial density perturbations.
First, let us write the metric representing the scalar-type perturbations
and the sourced tensor-type perturbations on the flat FLRW spacetime \cite{Ananda:2006af, Baumann:2007zm};
\be
ds^2=a^2 (\eta) \bigg[ -(1+2\Phi)d\eta^2+(1-2\Psi) \left( \delta_{ij}+\frac{1}{2}h_{ij} \right) dx^i dx^j \bigg],
\ee 
where $\Phi$ and $\Psi$ are scalar-type perturbations (gravitational potential and curvature perturbation, respectively)
and $h_{ij}$ are the tensor-type perturbations satisfying the transverse and traceless conditions $\partial_i h^i_{~j}=h^i_{~i}=0$, 
where $h^i_{~j}=\delta^{ik} h_{kj}$.
In the absence of the anisotropic stress, which is a good approximation \cite{Baumann:2007zm}, we have $\Phi=\Psi$.
Here, we do not include $h_{ij}$ generated during or after inflation by some other mechanisms 
since they do not affect the GWs we are interested in in this section.
The evolution of $\Psi$ at the linear order in the radiation dominated era is given by
\be
\Psi''+\frac{4}{\eta} \Psi'-\frac{1}{3} \triangle \Psi=0.
\ee
The solution of this equation in the Fourier space with the requirement of vanishing decaying mode is given by
\be
\Psi_k (\eta)=D_k (\eta) \Psi_k (0),~~~~~D_k (\eta)=\frac{3}{{(k\eta)}^2} \bigg[ \frac{\sqrt{3}}{k\eta} \sin \left( \frac{k\eta}{\sqrt{3}} \right)-
\cos \left( \frac{k\eta}{\sqrt{3}} \right) \bigg], \label{2ndgw-sol-Psi}
\ee
where the initial condition $\Psi_k (0)$ is fixed either by the assumed inflation model or
by the assumed power spectrum of $\Psi$.
The evolution of $h_{ij}$ is also obtained by expanding the Einstein equations. 
At the second-order in the scalar-type perturbations, we have \cite{Ananda:2006af, Baumann:2007zm}
\be
h_{ij}''+2{\cal H}h_{ij}'-\triangle h_{ij}=-4 {\hat {\cal T}}_{ij}^{rs} S_{rs}, \label{2ndgw-ev-h}
\ee
where ${\cal H}=a'/a$, ${\hat {\cal T}}_{ij}^{rs}$ is a projection operator that produces the transverse
and traceless quantity, and the source term consisting of the products of scalar-type perturbations is given by
\be
S_{ij}=2\Psi \Psi_{,ij}- \left( \Psi_{,i}+{\cal H}^{-1} \Psi_{,i}' \right) \left( \Psi_{,j}+{\cal H}^{-1} \Psi_{,j}' \right). 
\ee
GWs can be Fourier-transformed as
\be
h_{ij}(\eta,{\vec x})= \int \frac{d^3k}{ {(2\pi)}^{3/2}} e^{i {\vec k}\cdot {\vec x}} 
\sum_{I=+,\times} e_{ij}^I ({\vec k}) h_I (\eta, {\vec k}),
\ee
where $e_{ij}^{+,\times}$ are polarization tensors normalized by $e_{ij}^+ e^{+ ij}=e_{ij}^\times e^{\times ij}=1$.
Knowing the time evolution of $S_{ij}$ by Eq.~(\ref{2ndgw-sol-Psi}), the time evolution of $h_I$ is
determined as 
\be
h_I (\eta,{\vec k})=\frac{1}{a(\eta)} \int_0^\eta ~G_k (\eta,\eta') a(\eta') S_I(\eta',\vec{k}) d\eta',
\ee
where $G_k(\eta,\eta')=\sin (k(\eta-\eta'))/k$ is the Green's function of Eq.~(\ref{2ndgw-ev-h}) and $S_I(\eta,{\vec k})$
is a Fourier transform of the right-hand side of Eq.~(\ref{2ndgw-ev-h}) \cite{Inomata:2016rbd},
\be
S_I (\eta,{\vec k})=\int \frac{d^3q}{ {(2\pi)}^{3/2}} ~4e_I^{ij} ({\vec k}) q_i q_j f({\vec q},{\vec k}-{\vec q},\eta)
\Psi_{\vec q}(0) \Psi_{{\vec k}-{\vec q}}(0),
\ee
where 
\be
f({\vec k}, {\vec q},\eta)=2D_k (\eta) D_q (\eta)+\left( D_k (\eta)+\frac{D_k' (\eta)}{\cal H} \right)
\left( D_q (\eta)+\frac{D_q' (\eta)}{\cal H} \right).
\ee

The perturbations $\Psi$ being stochastic, the induced GWs are also stochastic.
A useful quantity parameterizing the amount of GWs is $\Omega_{\rm GW}$, the energy density
of GWs per logarithmic frequency normalized by the critical density.
Using the formula of the GW energy density \cite{MTW}
\be
\rho_{\rm GW}=\frac{1}{128\pi G} \langle {\dot h_{ij}} {\dot h_{ij}} \rangle,
\ee
and the definition of the GW power spectrum 
\be
\langle h_I ({\vec k}) h_J ({\vec q}) \rangle=\frac{2\pi^2}{k^3} {\cal P}_h (k) \delta ({\vec k}+{\vec q}) \delta_{IJ},
\ee
we have
\be
\frac{\rho_{\rm GW}}{\rho_{c,0}}=\int d\ln f~\Omega_{\rm GW}(f),~~~~~~~
\Omega_{\rm GW}(f)=\frac{1}{24 H_0^2} k^2 {\cal P}_h (k),
\ee
with $k=2\pi f$. 
Then, assuming the Gaussianity of the scalar-type perturbations, for which
we can decompose the power spectrum of $S_I$ into the product of the power
spectrum of $\Psi$, $\Omega_{\rm GW}$ can be written as
\begin{align}
\Omega_{\rm GW}(f)=\frac{k^5}{3\pi H_0^2} &\int_0^{\eta_0} d\eta' \int_0^{\eta_0} d\eta''~
G_k (\eta_0,\eta')G_k (\eta_0,\eta'') \frac{a(\eta') a(\eta'')}{a^2 (\eta)} \nonumber \\
&\times \int d^3q ~\frac{ {(e^{ij}({\vec k}) q_i q_j)}^2}{q^3 {|{\vec k}-{\vec q}|}^3} 
f({\vec q},\vec{k}-{\vec q},\eta') f({\vec q},\vec{k}-{\vec q},\eta'') {\cal P}_\Psi (q) {\cal P}_\Psi (|{\vec k}-{\vec q}|). \label{2ndgw-Omega}
\end{align}
Except for ${\cal P}_\Psi$, all the functions such as $G_k,~a(\eta)$ and $f$ are known.
By using this equation, we can compute $\Omega_{\rm GW}$ from the power spectrum of the
scalar-type perturbations ${\cal P}_\Psi$.

Then, assuming the delta-function type power spectrum for ${\cal P}_\Psi$ and its amplitude to the one 
predicting sizable amount of PBHs in the present Universe,
it was claimed that the resultant $\Omega_{\rm GW}$ conflicts with the upper limit set by the
pulsar timing experiments \cite{Saito:2008jc}.
The analysis was then generalized in \cite{Saito:2009jt} to the case where ${\cal P}_\Psi$ 
has a silk hat type spectrum with a finite width.
It was found that the predicted $\Omega_{\rm GW}$ becomes smaller for larger width (see also \cite{Bugaev:2009zh}).
In \cite{Inomata:2016rbd, Orlofsky:2016vbd}, computations of $\Omega_{\rm GW}$ for some concrete inflation models 
predicting the PBHs were performed and comparison with the data of the three PTA experiments 
(EPTA \cite{Lentati:2015qwp}, PPTA \cite{Shannon:2015ect}, and NANOGrav \cite{Arzoumanian:2015liz}) 
were made (see also \cite{Garcia-Bellido:2016dkw}).
Although not all the regions in the parameter space of inflation models are excluded, it was shown that
pulsar timing places stringent constraints on inflation models.

So far, the primordial density perturbations sourcing the stochastic GWs have been assumed to be Gaussian.
In \cite{Nakama:2016gzw}, it was suggested that the constraints mentioned above can weaken for the non-Gaussian 
density perturbations, which may be typical when the amplitude of the density perturbation is significantly large.
A basic observation behind this conclusion is that the variance of the density perturbation,
which determines the magnitude of $\Omega_{\rm GW}$,
can be smaller in the non-Gaussian perturbation than in the Gaussian case for a fixed PBH abundance.
In \cite{Nakama:2016gzw}, two different types of non-Gaussian density perturbations were considered.
The first one is specified by the probability density function of the curvature perturbation 
smoothed over the Hubble horizon at the time of the PBH formation, which is given by
\be
P(\zeta)=\frac{1}{2^{3/2} {\tilde \sigma} \Gamma (1+1/p)} \exp \bigg[
-{\left( \frac{|\zeta|}{\sqrt{2} {\tilde \sigma}} \right)}^p \bigg],
\ee
where $p$ and ${\tilde \sigma}$ are free parameters.
When $p=2$, the perturbations are Gaussian.
As explained in Sec.~\ref{abundance-PBH}, the PBH fraction at the formation time is given by
the probability that the perturbation exceeds the formation threshold $\zeta_{\rm th}$,
\be
\beta=\int_{\zeta_{\rm th}}^\infty ~P(\zeta)d\zeta.
\ee
For fixed $\beta$, the variance of $\zeta$ becomes smaller than the Gaussian case for $p<2$ \cite{Nakama:2016kfq}.
The second example is the local-type non-Gaussian perturbation for which the 
smoothed curvature perturbation in the position space is written as
\be
\zeta=\zeta_G+\frac{3}{5}f_{\rm NL}\zeta_G^2,
\ee
where $\zeta_G$ is Gaussian, and 
$f_{\rm NL}$ is a free parameter that parametrizes the significance of the non-Gaussian contribution.
For positive $f_{\rm NL}$, the variance of $\zeta$ is suppressed than the Gaussian case \cite{Byrnes:2012yx}.
According to the analysis in \cite{Nakama:2016gzw}, the PTA constraint can be safely evaded 
if $p \lesssim {\cal O}(1)$ or $f_{\rm NL} \gtrsim {\cal O}(1)$ (precise value depends on the theoretical uncertainties about the PBH formation).
Similar conclusion has been obtained in \cite{Garcia-Bellido:2017aan}, where
the curvature perturbation obeying the $\chi^2$ statistics,
motivated by the rolling axion scenario \cite{Namba:2015gja},
was considered.

\subsubsection{CMB spectral distortions from the primordial density perturbations}
\label{cmb-distortion}
The second indirect constraint can be obtained from the generation of the
CMB spectral distortion of the primordial density perturbations \cite{Carr:1993aq, Carr:1994ar}.
The basic argument is as follows.

The perturbations of photons and baryonic gas that re-entered the Hubble horizon prior to
the CMB decoupling $(z \simeq 1100)$ undergo the acoustic oscillations due to the tight coupling
between photons and baryons.
These oscillations are eventually erased by the photon diffusion, {\it i.e.} imperfect coupling between photons and free electrons, 
known as the Silk damping.
Using the photon mean free path $\ell_{\rm mp}=1/(\sigma_T n_e)$,
where $n_e$ is the number density of free electrons, 
and that the diffusion is described by random walks, 
the comoving wavenumber of a perturbation is related to the damping time of the perturbation as \cite{Dodelson}
\be
k=\frac{1}{1+z} \sqrt{n_e \sigma_T H(z)} \simeq 4 \times 10^{-6} {(1+z)}^{3/2}~{\rm Mpc}^{-1}. \label{silk-damping-k}
\ee  
This shows that the Silk damping occurs earlier for smaller scale perturbations. 
Through the damping, the oscillation energy of the perturbations is transfered to
the background homogeneous plasma gas.
As briefly discussed in \ref{accretion-cmb}, if the Silk damping occurs before $z \approx 2\times 10^6$,
the photon-number changing interactions occur frequently and the injected energy is completely consumed 
for the thermalization. As a result, the net effect is just the slight increase of the entropy
per unit comoving volume.
If the perturbations undergo the Silk damping after that redshift but before $z \approx 5\times 10^4$,
only the kinetic equilibrium is achieved.
As a result, the photons acquire chemical potential and the distribution of photons becomes
the Bose-Einstein distribution ($\mu$-type).
Using the above equation (\ref{silk-damping-k}), the interval of wavenumber of perturbations that yield
the $\mu$-type distortion is $50 \lesssim k/{\rm Mpc}^{-1} \lesssim 10^4$.
If the perturbations dissipate after $z\approx 5\times 10^4$, even the kinetic equilibrium is no longer
reached, and the distribution of photons is characterized by the Compton-$y$ parameter ($y$-type).
Thus, the perturbations of smaller comoving wavenumber than $50~{\rm Mpc}^{-1}$ generate the
$y$-type distortion in the CMB spectrum.

Now, let us consider for simplicity the delta-function power spectrum of the density perturbation
at $k=k_*$ with total power ${\cal A}$.
To produce PBHs, ${\cal A}={\cal O}(0.01-0.1)$ is required. 
Then, the produced $\mu$-type distortion is given by \cite{Chluba:2012we, Kohri:2014lza}
\be
\mu \simeq 2\times {\cal A}
\bigg[ \exp \left( -\frac{k_*}{5400~{\rm Mpc}^{-1}} \right)-
\exp \left( -{\bigg[ \frac{k_*}{31.6~{\rm Mpc}^{-1}} \bigg]}^2 \right) \bigg]. \label{mu-delta}
\ee
So far, spectral distortion of the CMB has not been detected, 
and the strongest upper limit is placed by the COBE/FIRAS experiment as
$\mu \lesssim 9\times 10^{-5}$ \cite{Fixsen:1996nj}.
Using this limit and the fiducial value $A=0.02$, 
the level of the produced $\mu$-distortion (\ref{mu-delta}) is inconsistent with the COBE/FIRAS bound
for $2 \lesssim k/{\rm Mpc}^{-1} \lesssim 3\times 10^4$.
This interval is only logarithmically sensitive to the choice of ${\cal A}$.
In terms of the PBH mass, non-detection of the $\mu$-type distortion excludes PBHs in the mass range
\be
2\times 10^4~M_\odot \lesssim M_{\rm PBH} \lesssim 2\times 10^{13}~M_\odot, \label{pbh-mass-mu}
\ee
if PBHs are produced from the direct collapse of the nearly Gaussian primordial density perturbations \cite{Kohri:2014lza}
(see also \cite{Kawasaki:2012kn}).

In \cite{Nakama:2016kfq}, it was pointed out that the above conclusion that
the PBHs in the mass range (\ref{pbh-mass-mu}) are excluded can in principle be circumvented 
if the primordial density perturbations are strongly non-Gaussian such that amplitude of the density perturbations
in all regions other than the sites of PBH formation is too tiny to induce sizable $\mu$-distortion.
Such a situation can be realized if only patches that later convert into PBHs experience different expansion
history during inflation (see Fig.~3 of \cite{Nakama:2016kfq}).

\subsubsection{Big-bang nucleosynthesis}
From the last discussion in \ref{cmb-distortion}, it may appear that
it is impossible to constrain the amplitude of the primordial perturbation
for $k \gtrsim 10^4~{\rm Mpc}^{-1}$, translating to $M_{\rm PBH} \lesssim 2\times 10^4~M_\odot$, 
since the Silk damping of such small
scale perturbations does not leave any spectral feature in the CMB.
Yet, big-bang nucleosynthesis (BBN) has a potential to exclude smaller PBHs than the above mass.

Primordial density perturbations in the range $10^4 \lesssim k/{\rm Mpc}^{-1} \lesssim 10^5$ dissipate
by the Silk damping in the redshift range $2\times 10^6 < z< 10^7$, which is after the BBN era but 
much before the decoupling time of the CMB (perturbations with $k\gtrsim 10^5~{\rm Mpc}^{-1}$ dissipate
prior to the neutrino decoupling due to the neutrino diffusion).
Injected energy by the Silk damping of those perturbations results in the increase of the temperature
of the plasma gas.
Because of the conservation of the baryon number, only the photon number density is increased by this process.
In other words, the baryon-to-photon ratio $\eta$ defined by the ratio between the two number density as $\eta \equiv n_b/n_\gamma$
decreases through the Silk damping.
This means that $\eta$ during BBN era is bigger than that in the CMB era, namely $\eta_{\rm BBN} > \eta_{\rm CMB}$.
Abundance of light elements produced by BBN is controlled by $\eta_{\rm BBN}$,
and the measurement of the light elements in the present Universe can determine $\eta_{\rm BBN}$.
Acoustic peaks in the CMB temperature power spectrum is controlled by $\eta_{\rm CMB}$,
and the measurement of the CMB anisotropies can determine $\eta_{\rm CMB}$.
The difference between $\eta_{\rm BBN}$ and $\eta_{\rm CMB}$ depends on how much energy was injected
by the Silk damping, {\it i.e.} the amplitude of the primordial density perturbations.
Therefore, the determination of $\eta_{\rm CMB}$ and $\eta_{\rm BBN}$ by independent observations
constrains the amplitude of the primordial density perturbations in the relevant length scale \cite{Nakama:2014vla}.
From observations, we have $\eta_{\rm BBN}=(6.19\pm 0.21)\times 10^{-10}$
and $\eta_{\rm CMB}=(6.11\pm 0.08)\times 10^{-10}$ \cite{Steigman:2014pfa}.
Using these values, the upper limit on the total power ${\cal A}$ for the delta-function type perturbation was
obtained as \cite{Nakama:2014vla}
\be
{\cal A} \lesssim 0.06.
\ee

In \cite{Inomata:2016uip}, more stringent upper limit on ${\cal A}$ was obtained by investigating the freezeout
neutron fraction during BBN in the presence of the (adiabatic) primordial density perturbations (see also \cite{Jeong:2014gna}).
Those density perturbations, which dissipate after BBN, produce local temperature fluctuations during BBN
since $\rho_\gamma \propto T^4$.
In local regions where the temperature is higher than the average, interactions persist longer
and the freezeout is delayed, resulting in smaller amount of neutrons.
On the other hand, more neutrons remain in colder regions.
Because of two enhancement effects in hotter regions that more baryons are available ($\delta n_b \propto \delta T$) 
and that physical volume is also enlarged by the metric perturbation
compared to colder regions for fixed comoving volume,
the averaged neutron fraction is biased toward the higher temperature region. 
As a result, the predicted helium abundance becomes smaller than the standard case.  
The derived upper limit on ${\cal A}$ is \cite{Inomata:2016uip}
\be
{\cal A} \lesssim 0.02.
\ee
Although this value does not immediately exclude the production of PBHs,
this falls into the typical power required for PBHs.
This suggests that reduction of errors in future measurements may find a PBH signal
or eventually rule out the PBHs in the relevant mass range.

\subsection{Future constraints}
So far, all the constraints we have discussed are established constraints in the sense that
they are obtained by the existing observational data.
Thus, as long as the data and theory translating to the constraint are correct, 
the derived constraint is real and must be taken into account when one tests
any early Universe model predicting the PBH formation.

In the future, those constraints will be improved by the new experimental apparatus similar
to what is existing/existed but with better sensitivity.
Furthermore, appearance of new technology will enable us to probe PBHs by using completely new methods.
In this section, we briefly review several proposals to probe PBHs by future experiments.

\subsubsection{Fast radio bursts}
Fast radio bursts (FRBs) are radio transients that last only for ${\cal O}({\rm ms})$ \cite{Lorimer:2007qn}. 
Since its discovery in 2007 \cite{Lorimer:2007qn}, many FRBs have been detected.
The measured values of dispersion measures suggest that FRBs occur at cosmological distances \cite{Cordes:2002wz}.
The origin of FRBs is not known yet.

In \cite{Munoz:2016tmg}, it was pointed out that lensing of the FRB signals by the intervening PBHs
can be used to probe PBHs with $\gtrsim 20~M_\odot$.
The idea is as follows.
Gravitational lensing causes appearance of two source images. 
Although the angular separation of the two images is too small to be resolved experimentally, 
the difference of arrival times of the images becomes the order of 
\be
T_{12} \simeq 1~{\rm ms}~\left( \frac{M_{\rm PBH}}{30~M_\odot} \right),
\ee
for $u=1$ (see Eq.~(\ref{T12-diffraction})).
Thus, PBHs heavier than $\sim 10~M_\odot$ produce double bursts separated longer than the burst width,
hence can in principle be resolved.
Considering the two redshift distributions of FRBs,
constant comoving number density and the one that follows the star formation rate,
the lensing optical depth has been evaluated in \cite{Munoz:2016tmg}.  
It was found that planned experiment the Canadian Hydrogen Intensity Mapping Experiment (CHIME),
which is expected to measure $700\sim 15000$ FRBs per year \cite{Connor:2016rhf},
may detect tens of double bursts caused by PBHs if PBHs comprise all the dark matter and
provide constraint $f_{\rm PBH} \lesssim {\cal O}(0.01)$ if no such events are detected.

\subsubsection{Pulsar timing array (PTA) experiments}
As explained in \ref{indirect-SGW}, 
primordial density perturbations that generated PBHs from high-$\sigma$ peaks 
also produce the stochastic GWs by the mode-mode coupling.
The present frequency of the GWs corresponding to the PBHs in the stellar mass range
is in the ${\rm nHz}$ range which PTA experiments are sensitive to.
Square Kilometer Array (SKA) is a planned giant radio telescope consisting of thousands of
receptors with total collecting area being about one square kilometer .
With better sensitivity, SKA is expected to detect more pulsars and
improve the constraint on $\Omega_{\rm GW}$ by $3\sim 4$ orders of magnitude
stronger than the current limit if no GWs are detected \cite{Kramer:2010tm}.
Such constraint, if achieved, will further tighten the constraint in \ref{indirect-SGW}
much more severely.

While the above constraint is indirect in the sense that the constraint is derived by looking at 
the GWs from the primordial density perturbations that {\it failed} to turn into PBHs, 
pulsar timing also has a potential to give a direct constraint on PBH abundance by measuring
the effect of the Shapiro time delay caused by the PBHs intervening between pulsars and the Earth \cite{Schutz:2016khr}.
A PBH moving relative to the line of sight produces non-stationary change of the arrival times of pulses as
\be
t(n) \approx nP_{\rm obs}+\frac{n^2}{2} P_{\rm obs} {\dot P}_{\rm obs}+n^3 P_{\rm obs}^3
\frac{4GM_{\rm PBH} v_r^3 }{3r_L^3},
\ee
for the $n$-th pulse \cite{Schutz:2016khr}.
Here $P_{\rm obs}, {\dot P}_{\rm obs}, v_r, r_L$ are the observed pulse period, its time derivative, 
(constant) velocity of the PBH toward the line of sight, and the distance between the PBH and the line of sight, respectively.
The observable effect of the PBH appears at ${\cal O}(n^3)$ term.
According to \cite{Schutz:2016khr}, non-detection of such effect in the long term observations (${\cal O}(10)$ years)
of known pulsars or pulsars that would be newly detected by SKA can constrain PBHs as $f_{\rm PBH}\lesssim 0.01 \sim 0.1$.

\subsubsection{21cm}
SKA is also expected to detect cosmological 21{\rm cm} line from neutral
hydrogen in the dark age before reionization.
Observations of such 21{\rm cm} line have a potential to constrain PBHs.

21{\rm cm} line is radiation emitted/absorbed by the transition between
the two levels in the hydrogen 1s ground states.
Energy of the spin singlet state  
is lower than that of the spin triplet states
by $T_* = 5.9\times 10^{-6}{\rm eV}=0.068{\rm mK}$.
Relative number density between singlet ($n_0$) and triplet ($n_1$) hydrogen atoms 
is parametrized by the spin temperature $T_s$ as
\be
\frac{n_1}{n_0}=3 \exp \left( -\frac{T_*}{T_S} \right).
\ee
There are three ingredients to determine the spin temperature as
\be
T_S=\frac{T_{\rm CMB}+y_\alpha T_\alpha+y_c T_K}{1+y_\alpha+y_c},
\ee
where the second and the third term in the numerator represents the
Lyman-$\alpha$ pumping, namely 21{\rm cm} transition through ambient
Lyman-$\alpha$ photons as singlet/triplet $\to$ 2p $\to$ singlet/triplet,
and the collisions among hydrogen atoms, respectively \cite{Madau:1996cs}. 
The 21{\rm cm} intensity is normally expressed in terms of the brightness temperature $T_b$.
With reference to the CMB brightness temperature,
the differential temperature measured at the Earth is given by
\be
\delta T_b= {(1+z)}^{-1} (T_S-T_{\rm CMB}) (1-e^{-\tau}),
\ee
where $\tau$ is the optical depth for the 21{\rm cm} photons \cite{Madau:1996cs}. 

In \cite{Tashiro:2012qe}, the effect of X-rays emitted by the accreting PBHs on
the 21{\rm cm} fluctuations has been investigated.
Radiation emitted from the accreting PBHs in the dark age ionizes and
heats the surrounding neutral hydrogen gas.
Ionization reduces the amount of neutral hydrogen, Lyman-$\alpha$ photons
from PBHs affects the spin temperature through the $T_\alpha$ term,
and the heating of neutral hydrogen by emanating radiation also affects
the spin temperature by changing the kinetic temperature $T_K$. 
Assuming radiation intensity corresponding to $10 \%$ of the Eddington luminosity
with a power-law X-ray spectrum, brightness temperature fluctuations was computed.
It was found that with SKA-like experiments upper limit on PBH will be 
$\Omega_{\rm PBH}=10^{-5} {(M_{\rm PBH}/10^3M_\odot)}^{-0.2}$ at $z=30$
and $\Omega_{\rm PBH}=10^{-7} {(M_{\rm PBH}/10^3M_\odot)}^{-0.2}$ at $z=20$
for $10^2 < M_{\rm PBH}/M_\odot < 10^8$.

In \cite{Gong:2017sie}, based on the observation discussed in \ref{Lyalpha}
that the random distribution of PBHs adds isocurvature component on top
of the adiabatic perturbations,
formation of minihalos sourced by the PBHs and their effect on 21{\rm cm}
fluctuations has been studied.
Minihalos can change the 21{\rm cm} signals since the gas temperature and
the density of neutral hydrogen inside the minihalos differ from the background values.
According to the analysis in \cite{Gong:2017sie}, the PBHs with $\gtrsim 10M_\odot$
leave enhancement of the brightness temperature that is detectable for SKA 
for $f_{\rm PBH} > 10^{-3}-10^{-4}$. 

\subsection{Constraints for the extended PBH mass function}
So far, we have assumed that the PBH mass function is monochromatic,
{\it e.g.,} all the PBHs have the same mass.
Thus, the derived constraints are valid only when the PBH mass function is
sufficiently narrow, and can no longer be trusted when the mass function is broad.
In this subsection, we briefly mention how to generalize the constraint on
the PBH abundance to the case of the extended PBH mass function (see \cite{Carr:2016drx, Green:2016xgy, Carr:2017jsz}
for more details).

A simple formalism for deriving the PBH constraint for the extended mass
function was given in \cite{Carr:2017jsz}.
Let $\psi (M)$ be the PBH mass function and consider an observable $A$ which
PBHs produce or contribute.
Only in this subsection, we define the normalization of the PBH mass function as
\be
f_{\rm PBH}=\int \psi (M_{\rm PBH}) dM_{\rm PBH}.
\ee
The quantity $A$ depends on the details of experiments as well as the 
astrophysical phenomena one is interested in.  
For instance, in the microlensing experiment,
$A$ is the expected number of the microlensing events detected by
a particular experiment and depends on the sensitivity of the detectors.
Generally, $A$ is a functional of the mass function and can be expressed as
\be
A[\psi (M)]=A_0+\int dM~\psi (M) K_1 (M)+\int dM_1 dM_2~\psi (M_1) \psi (M_2) K_2 (M_1,M_2)+\cdots, \label{expansion-of-A}
\ee
where $\cdots$ are higher order terms in $\psi (M)$.
Here $A_0$ represents any contribution other than from PBHs.
For the microlensing experiments, $A[\psi (M)]$, the total expected number of the 
microlensing events during the observation period, contains only the $K_1 (M)$-term
as the PBH contribution, which is given by \cite{Griest:1990vu,Green:2016xgy}
\be
K_1(M)=E \int_0^\infty ~\frac{32D_S u_T \epsilon(t_e)}{t_e^4 v_c^2 M}
\int_0^{x_h} \rho_{dm} (x) R_E^4 (x) e^{-Q(x)}dx,~~~~~~Q(x) \equiv \frac{4R_E^2 (x) u_T^2}{t_e^2 v_c^2},
\ee
where $E$ is the number of stars multiplied with the observation period, 
$u_T$ is the maximum $u$ (see Eq.~(\ref{magnification-A})) below which the
microlensing magnification becomes greater than the threshold,
$v_c \approx 220~{\rm km/s}$ is the circular velocity of the Sun,
and $\epsilon (t_e)$ is the probability
that the detector detects the microlensing events that last for the period $t_e$.
As we will see in the next section, if $A$ is the merger event rate of the PBH binaries
that are formed in the early Universe,
the expansion (\ref{expansion-of-A}) starts at the $K_3$ term if only the third BH is taken into
account as the dominant source of the tidal force and starts at even higher-order terms 
if more distant PBHs are also included.
On the other hand, for PBH binaries that are formed in the low-redshift Universe by the close encounters,
$A$ is given by the $K_2$-term.
These examples show that the order in $\psi$ at which the expansion of (\ref{expansion-of-A}) starts
varies for different observables.

Let us consider the simplest case where $A$ is dominantly given by the $K_1$ term.
Suppose the observations place an upper bound on $A$ as
\be
A[\psi (M)] \leq A_{\rm exp}.
\ee
Then, for the monochromatic mass function $\psi (M)=f_{\rm PBH} (M_{\rm PBH}) \delta (M-M_{\rm PBH})$, 
the above constraint becomes
\be
f_{\rm PBH} (M_{\rm PBH}) \leq \frac{A_{\rm exp}-A_0}{K_1(M_{\rm PBH})} \equiv f_{\rm max} (M_{\rm PBH}),
\ee
where $f_{\rm max} (M_{\rm PBH})$ is the maximally allowed value of $f_{\rm PBH}$ by the observation under
consideration when all the PBHs have mass $M_{\rm PBH}$.
Replacing $K_1$ by $f_{\rm max}$, we finally obtain
\be
\int \frac{\psi (M_{\rm PBH})}{f_{\rm max} (M_{\rm PBH})} \leq 1. \label{extended-formula}
\ee
This formula enables us to derive the constraint on the extended mass function once we know
the upper limit $f_{\rm PBH}$ for the monochromatic mass function.
This simple conversion does not hold in general when higher order terms contribute to $A$.

In \cite{Carr:2017jsz}, based on the above formalism, the constraints on the PBH abundance
for several types of the extended mass function were obtained.
The results show that the constraints generally become stringent for the extended mass function compared to the case
of the monochromatic mass function due to the combination of the multiple observational limits for different PBH masses.

\newpage

\section{PBHs as sources of gravitational waves (GWs)}
\label{GW-PBH}
As we have discussed in detail in Sec.~\ref{Observational-constraint},
before the direct detection of GWs by LIGO, searching for the PBHs in the Universe
to a varying degree relied on the electromagnetic waves.
For instance, gravitational lensing uses the background electromagnetic sources such as
stars and quasars, and the dynamical constraints are derived by the observations of stars
which PBHs affect.
So far, although there are a couple of studies which attributed some unexplained observational signals to 
the PBHs, no observational searches for PBHs by the electromagnetic waves have detected 
inarguable evidence for the existence of PBHs.

Direct detection of GWs by the laser-interferometers is a completely novel method to
search for PBHs that does not rely on the electromagnetic waves.
Soon after LIGO announced the first detection of GWs in February 2016, 
which are caused by the merger of two BHs in a binary, 
several groups pointed out the possibility of the scenario that 
the observed BHs are PBHs \cite{Bird:2016dcv, Clesse:2016vqa, Sasaki:2016jop, Kashlinsky:2016sdv}.
Thus, exciting possibility has arisen that we might have discovered PBHs 
for the first time by the direct observation of GWs (not just constraint!).

Explaining the LIGO event by the PBHs is not trivial in two aspects.
First, since the GWs are emitted from the BH binaries, formation mechanism of the PBH binary must be considered
in order to test the PBH scenario with GW observations.
Secondly, as discussed in Sec.~\ref{Observational-constraint}, there are existing constraints on the PBH abundance
for the mass around the observed BH mass $\sim 30~M_\odot$. 
It needs to be checked if the PBH scenario does not conflict with those constraints.
These issues, which we will address in detail later, are the first main topic in this section.
It will turn out that those issues are cleared and the PBH scenario can be considered 
as a reasonable candidate scenario for the observed BH merger event.

There are also astrophysical explanations for the observed heavy stellar-mass BHs (see \cite{TheLIGOScientific:2016htt}
and references therein): 
isolated field binary scenario in which two stars in an isolated binary collapse to BHs 
in decreasing order of star mass, and dynamical formation scenario in which isolated BHs
in dense stellar environment form BH binaries at the core and eventually are ejected by the three-body interactions.
In both cases, studies \cite{Belczynski:2009xy, Mapelli:2012vf, Spera:2015vkd} suggest that heavy 
BHs about $30~M_\odot$ are born out of stars with low metallicity (at most half of solar-metallicity) . 
Although there are many astrophysically uncertain factors in estimating the merger event rate,
these scenarios are roughly consistent with the GW observations.
Thus, our next task is to clarify how to discriminate those scenarios of the binary BHs
and to pin down the correct one.

The GW astronomy has just began, and its future is bright.
In the coming decade, the sensitivity of the existing GW experiments will greatly improve and 
many merger events of the BH binaries will be detected.
Looking at further future, new detectors with a larger armlength 
will be build both on ground and in space,
which can probe GWs in different frequency bands. 
By those experiments, we will obtain much information of BH binaries such as mass distribution,
space distribution, redshift distribution and spin distribution.
These information will definitely help us discriminate different scenarios of the binary BHs.
How such information can be used for this purpose is the second topic in this section.
Concerning this point, discussions about the astrophysical scenarios are beyond the scope of this monograph
and we will concentrate on the PBH scenario in what follows.

\subsection{Formation of PBH binaries}
Here, we will briefly review two distinct formation mechanisms of the PBH binaries.
It is important to remark here that the two mechanisms are not incompatible,
{\it i.e.,} not like the relationship between oil and water, but operate at different epoch in the cosmic history.
Thus, what matters is to figure out which mechanism is more efficient to make PBH binaries that merge by present time.

\subsubsection{PBH binary formation in the early Universe}
\label{PBH-binary-EU}
The first mechanism we are going to discuss operates in the epoch when
the Universe was dominated by radiation.
This mechanism was proposed in \cite{Nakamura:1997sm} to investigate the detectability
of binary mergers of MACHO PBHs ($\sim 0.5~M_\odot$).
In \cite{Nakamura:1997sm}, it was assumed that $f_{\rm PBH}=1$, all the PBHs have the same mass,
and PBHs are initially distributed randomly in space (Poisson distribution). 
In the following, we consider the monochromatic PBH mass function and uniform distribution of PBHs
and treat $f_{\rm PBH}$ as a free parameter.

Just after the PBHs were formed in the very early Universe, they were distributed sparsely in space,
that is, mean distance at that time is much longer than the Hubble horizon.
Because of the rapid cosmic expansion, they are on the expansion flow and the mean distance
grows in proportion to the scale factor $a(t) \propto t^{1/2}$.
Since the Hubble horizon grows as $H^{-1} (t) \propto t$, the mean distance relative to the Hubble 
horizon decreases as the Universe expands.
Denoting by $\ell_{\rm PBH} (z)$ the mean PBH distance at redshift $z$,
its length normalized by the Hubble horizon is given by
\be
H(z) \ell_{\rm PBH} (z)=H(z) n_{\rm PBH}^{-1/3} {(1+z)}^{-1} \simeq 6\times 10^{-6}~
f_{\rm PBH}^{-1/3} \left( \frac{1+z}{1+z_{\rm eq}} \right)
{\left( \frac{M_{\rm PBH}}{30~M_\odot} \right)}^{1/3},
\ee
for $z>z_{\rm eq}$, where $z_{\rm eq}$ is the redshift of the matter-radiation equality
and $n_{\rm PBH}$ is the comoving (initial) PBH number density.
Thus, for the stellar-mass PBHs, unless $f_{\rm PBH}$ is extremely tiny as
$f_{\rm PBH} \lesssim 10^{-15}$, there is period in the radiation dominated epoch
in which there are typically more than one PBHs in the Hubble horizon.
Range of $f_{\rm PBH}$ of our interest is much larger than this value.

Let us focus on a PBH and the PBH closest to it, and let their comoving distance be $x$.
The physical separation at redshift $z$ is then $x/(1+z)$.
The cosmic expansion acts as a force that pulls two PBHs away from each other.
Two PBHs are also pulled by the gravitational force acting between them.
The corresponding free-fall time becomes shorter than the Hubble time at a certain
time during the radiation dominated epoch if the comoving distance is shorter than
\be
x <x_{\rm max} \equiv f_{\rm PBH}^{1/3} \ell_{\rm PBH}(z=0). \label{condition-bound}
\ee
PBH pair satisfying this condition decouples from the cosmic expansion and
becomes gravitationally bound \footnote{This picture has been confirmed to be correct by numerically 
solving the Newtonian equations of motion \cite{Ioka:1998nz,Ali-Haimoud:2017rtz}.}. 
Conversely, PBH pair with comoving distance longer than $x_{\rm max}$ never
becomes gravitationally bound since a ratio of the free-fall time to the Hubble time
remains constant in the subsequent matter dominated epoch.
The decoupling redshift $z_{\rm dec}$ is given by
\be
1+z_{\rm dec}=(1+z_{\rm eq}) {\left( \frac{x_{\rm max}}{x} \right)}^3.
\ee

\begin{figure}[t]
\begin{center}
\includegraphics[width=120mm]{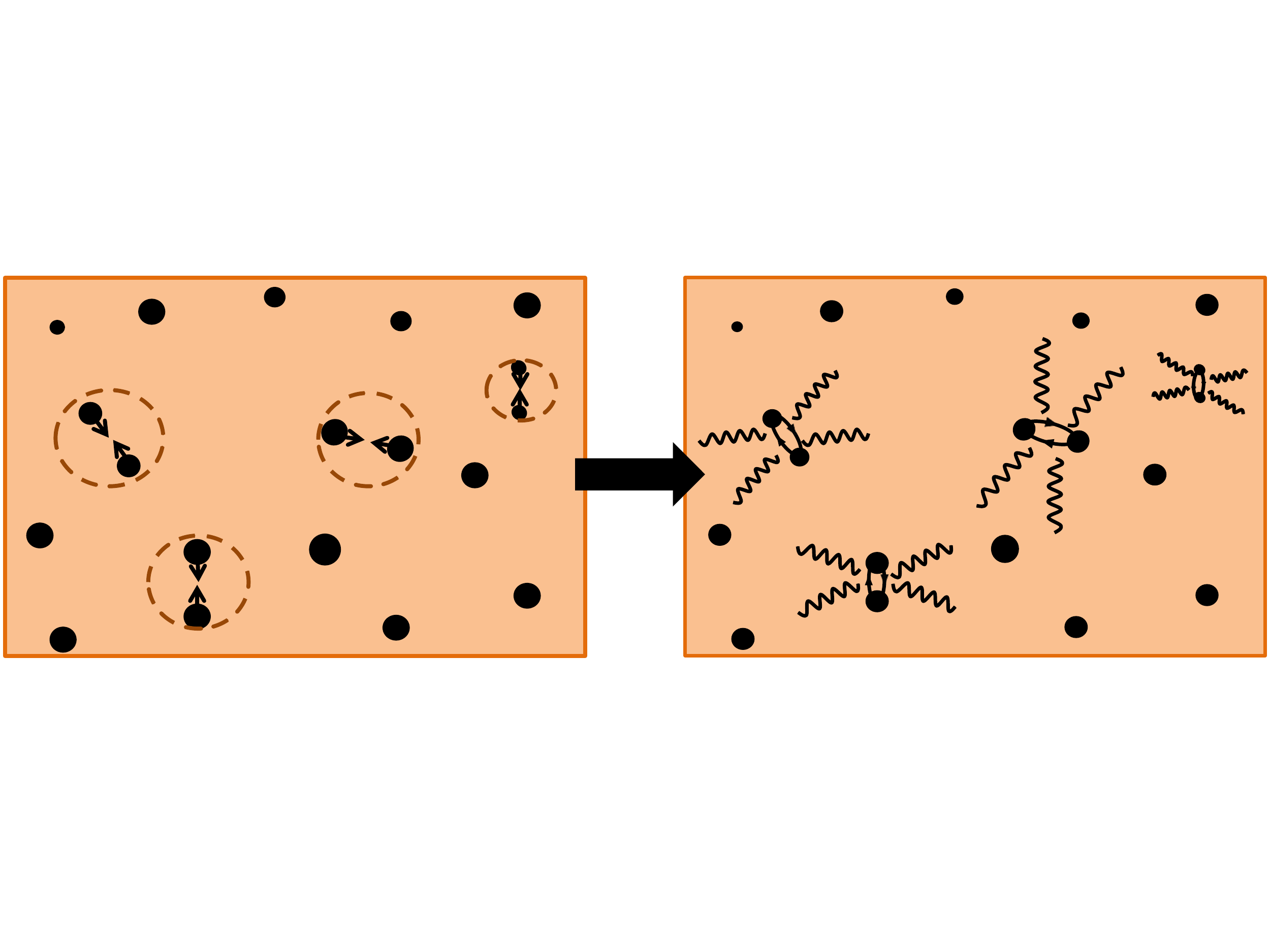}
\end{center}
\caption{A schematic picture of the formation of PBH binaries in the radiation dominated epoch.}
\label{Binary-RD}
\end{figure}

\begin{figure}[t]
\begin{center}
\includegraphics[width=70mm]{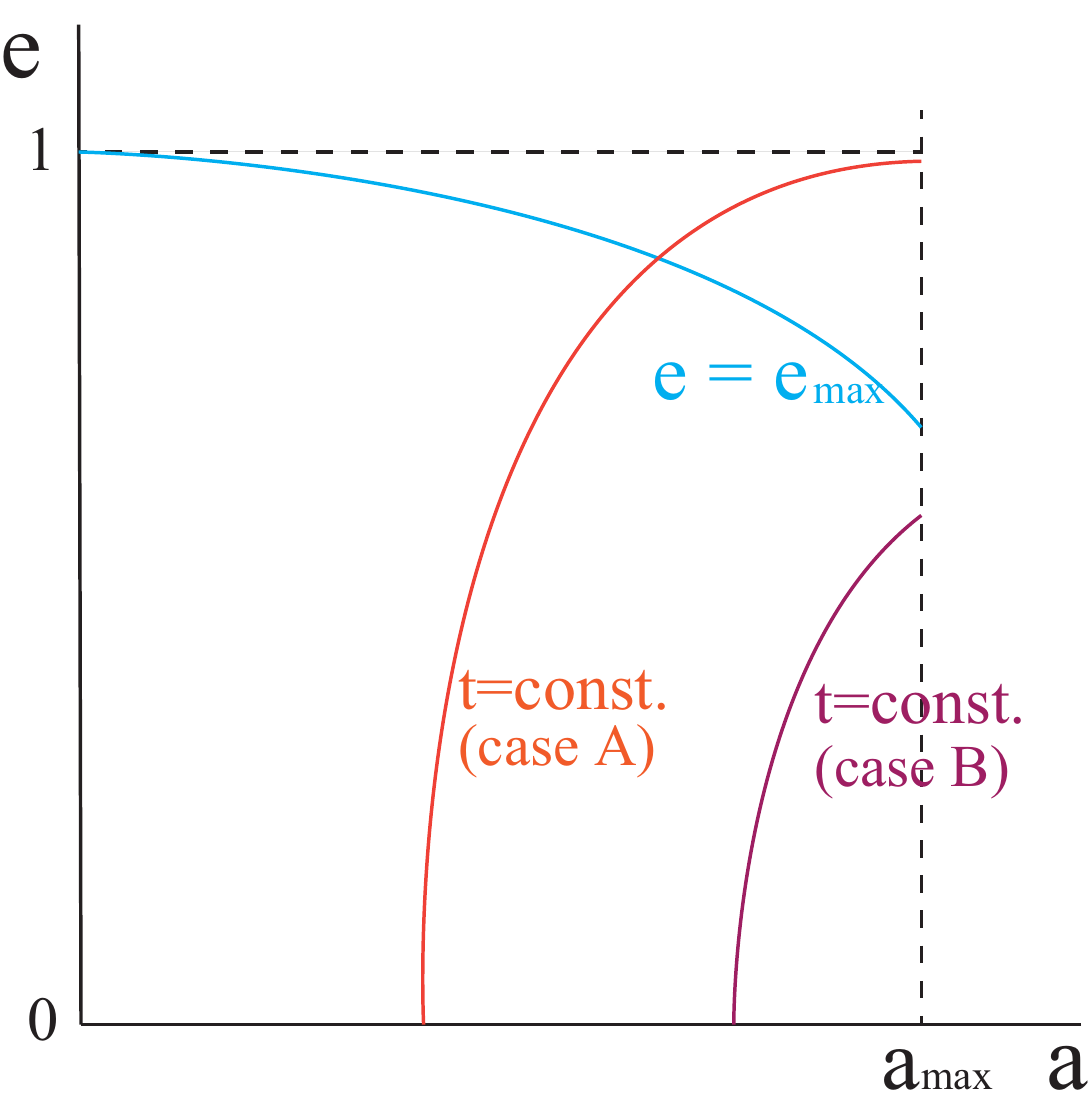}
\end{center}
\caption{A schematic picture of $e=e_{\rm max}$ as a function of $a$.}
\label{fig-emax}
\end{figure}

During the two PBHs come closer, the surrounding PBHs, especially the nearest one,
exert torques on the bound system.
As a result, the two PBHs avoid a head-on collision and form typically a highly eccentric binary (see Fig.~\ref{Binary-RD}).
The major axis $a$ of the binary orbit is equal to $x/(1+z_{\rm dec})$.  
The angular momentum $J$ of the binary is estimated by multiplying the exerted torque from the
nearest PBH by the free-fall time and is given by
\be
J \simeq t_{ff} G M_{\rm PBH}^2 (1+z_{\rm dec}) \frac{x^2}{y^3},
\ee
where $y$ is the comoving distance to the nearest PBH and a factor of ${\cal O}(1)$
has been ignored.
Assuming the Keplerian motion after forming the binary, the angular momentum is related
to the eccentricity $e$ of the orbit as $J^2=G\mu^2 Ma (1-e^2)$,
where $\mu$ and $M$ is the reduced mass and total mass.
We can convert these equations in terms of $x$ and $y$ as
\be
a=\frac{\rho_{c,0} \Omega_{\rm DM}}{(1+z_{\rm eq}) M_{\rm PBH}} x^4,~~~~~~
e=\sqrt{1- {\left( \frac{x}{y} \right)}^6}. \label{formula-a-e}
\ee
Because of the random distribution of PBHs, the probability that comoving distances are
in the intervals $(x,x+dx)$ and $(y,y+dy)$ is given by
\be
dP=\frac{4\pi x^2dx}{n_{\rm PBH}^{-1}} \frac{4\pi y^2dy}{n_{\rm PBH}^{-1}}
\exp \left( -\frac{4\pi y^3}{3n_{\rm PBH}^{-1}} \right) \Theta (y-x).
\ee
Instead of dealing with this probability distribution, the simplified one as
\be
dP=\frac{4\pi x^2dx}{n_{\rm PBH}^{-1}} \frac{4\pi y^2dy}{n_{\rm PBH}^{-1}} \Theta (y-x)
\Theta (y_{\rm max}-y),~~~~~y_{\rm max}={\left( \frac{4\pi}{3}n_{\rm PBH} \right)}^{-1/3},
\ee
was used in \cite{Nakamura:1997sm}.
Notice that because of $y_{\rm max}$, there is an upper limit on the eccentricity for fixed $x$ (and hence for fixed $a$) as
\be
e^2_{\rm max}=1-{\left( \frac{4\pi}{3} n_{\rm PBH} \right)}^2 
{\left( \frac{(1+z_{\rm eq}) M_{\rm PBH}}{\rho_{c,0} \Omega_{\rm DM}}a \right)}^{\frac{3}{2}}.
\ee
Schematic graph of $e=e_{\rm max}$ as a function of $a$ is shown in Fig.~\ref{fig-emax}
as a blue curve.
Notice that there is also a maximum for $a$ as $a_{\rm max}=x_{\rm max}/(1+z_{\rm eq})$.

Knowing the probability distribution of $(x, y)$, we can translate it in terms of $(a, e)$
by using the formula (\ref{formula-a-e}) as
\be
dP=\frac{4\pi^2}{3}n_{\rm PBH}^{1/2} {(1+z_{\rm eq})}^{3/2}
f_{\rm PBH}^{3/2} a^{1/2} e {(1-e^2)}^{-3/2} da de. \label{rd-dp1}
\ee
This equation tells us how much PBH binaries with orbital parameters in $(a,a+da)$,
$(e,e+de)$ exist in the Universe at formation time \footnote{It is worth mentioning that exact distribution
function of the initial angular momentum taking into account all the distant PBHs was obtained in \cite{Ali-Haimoud:2017rtz}.}.
After PBH binaries were formed in the radiation dominated epoch,
each PBH binary continuously emit gravitational waves and finally merge much later.
The estimate of the merger rate will be discussed in the subsequent subsection.

\subsubsection{PBH binary formation in the present Universe}
\label{bphbfpu}
In addition to the PBH binary formation in the radiation dominated epoch,
PBHs can form binaries in the present Universe, which we review in this subsection.

\begin{figure}[t]
\begin{center}
\includegraphics[width=120mm]{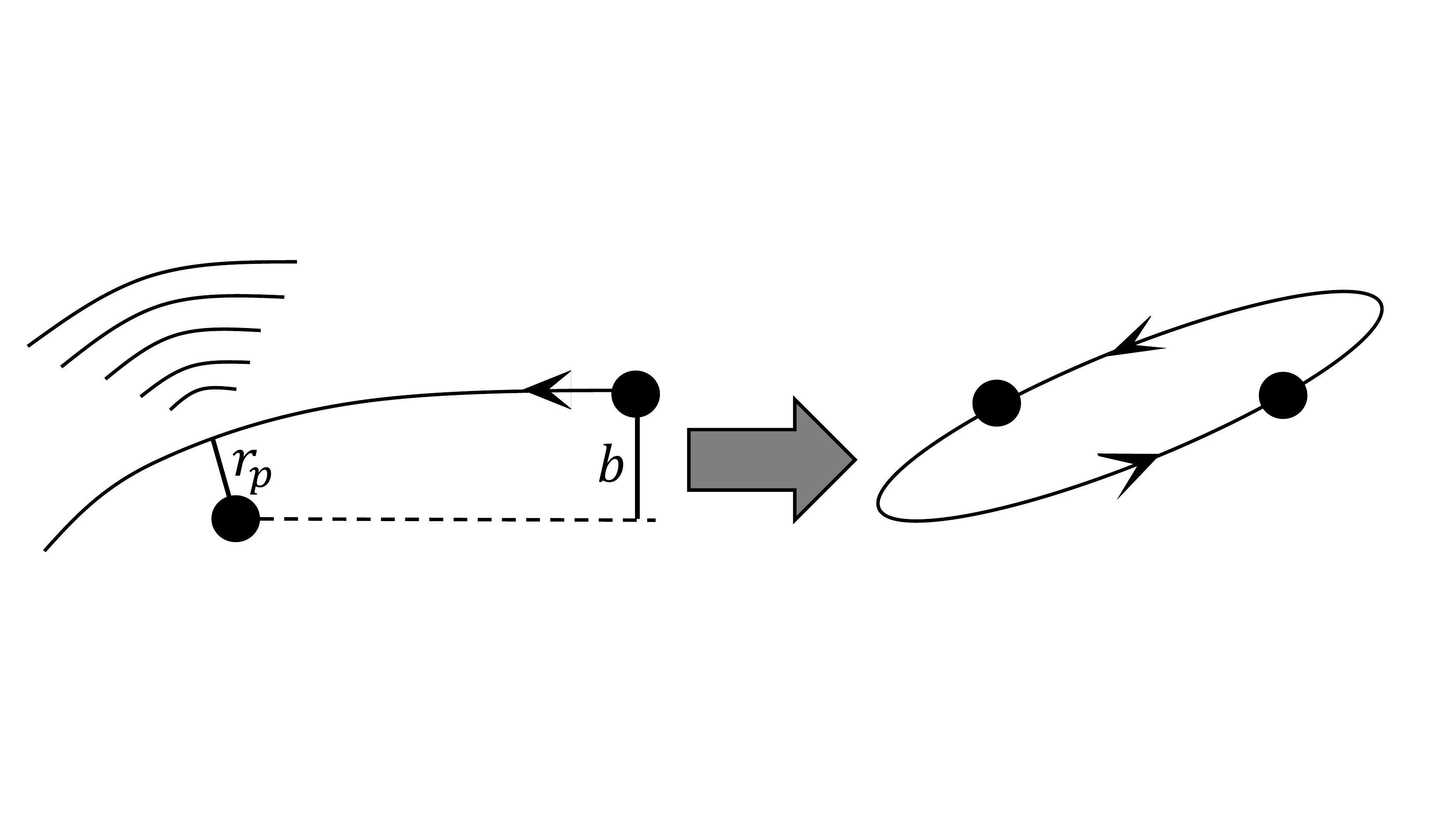}
\end{center}
\caption{A schematic picture of the close encounter of PBHs}
\label{PBH-encounter}
\end{figure}

Let us consider a situation where a PBH traveling in space accidentally 
has a near-miss with another PBH.
These PBHs may be concentrated in local region like inside larger dark matter halo 
or simply moving freely in space.
For the moment, we do not make a particular assumption on how PBHs are distributed in the region
of our interest.
Fig.~\ref{PBH-encounter} shows a schematic picture of the close encounter with impact parameter $b$
and periastron $r_p$.
Near the periastron, relative acceleration of the PBHs becomes the largest and dominant emission of 
gravitational radiation occurs.
If the amount of energy of the emitted GWs is greater than the kinetic energy of PBHs, then
the PBHs cannot escape to infinity any more and form the bound system.
Since the direct head-on collision is probabilistically unlikely, the binary typically results.

Let us investigate this problem more quantitatively \cite{Quinlan:1989xy}.
We assume that the power of GWs is estimated by the unperturbed trajectory (without backreaction due to GW emission) 
in the Newtonian approximation.
According to Peters \cite{Peters:1964zz}, the time-averaged energy loss rate of the binary in the Keplerian orbit
due to gravitational radiation is given by
\be
\bigg\langle \frac{dE}{dt} \bigg\rangle =-\frac{32}{5} \frac{G^4 m_1^2 m_2^2 (m_1+m_2)}{a^5 {(1-e^2)}^{7/2}} 
\left( 1+\frac{73}{24}e^2+\frac{37}{96} e^4 \right).
\ee
Thus, the energy loss during one orbital period $T$ becomes
\be
\Delta E=-T \bigg\langle \frac{dE}{dt} \bigg\rangle =
\frac{64 \pi \sqrt{G(m_1+m_2)} G^3 m_1^2 m_2^2}{5 r_p^{7/2} {(1+e)}^{7/2}}
\left( 1+\frac{73}{24}e^2+\frac{37}{96} e^4 \right).
\ee
where we have used the Kepler's third law and $r_p=a(1-e)$.
We can approximate the trajectory of the close encounter by the ellipse with $e=1$
since the true trajectory is physically indistinguishable from the ellipse near the periastron
where the dominant emission of GWs occurs.
Then, the energy loss by the close-encounter is obtained by plugging $e=1$
into the above equation,
\be
\Delta E=\frac{85 \pi \sqrt{G(m_1+m_2)} G^3 m_1^2 m_2^2}{12 \sqrt{2} r_p^{7/2}}.
\ee
If this energy is greater than the kinetic energy $\mu v^2/2$,
where $\mu$ is the reduced mass and $v$ is the relative velocity at large separation, 
then the PBHs form a binary. This imposes a condition on $r_p$ as
\be
r_p < r_{p,{\rm max}}=\bigg[ \frac{85 \pi}{6\sqrt{2}}
\frac{G^{7/2} {(m_1+m_2)}^{3/2} m_1 m_2}{v^2} \bigg]
\ee
In the Newtonian approximation, relation between $b$ and $r_p$ is given by
\be
b^2(r_p)=r_p^2+\frac{2GMr_p}{v^2}.
\ee
The encounter with the impact parameter less than $b(r_{p,{\rm max}})$ yields a binary.
In the limit of the strong gravitational focusing ($r_p \ll b$), which we are interested in,
the cross section for forming a binary becomes
\be
\sigma=\pi b^2(r_{p,{\rm max}}) \simeq {\left( \frac{85\pi}{3} \right)}^{2/7} \frac{\pi {(2GM_{\rm PBH})}^2}{v^{18/7}}. \label{encounter-cs}
\ee
Contrary to the PBH binaries that are formed in the radiation dominated epoch,
the PBH binaries produced by the present mechanism merge in less than the age of the Universe \cite{Cholis:2016kqi}.

\subsection{Merger event rate of PBH binaries}
Having explained two mechanisms of the formation of PBH binaries,
we review the expected merger event rate of PBH binaries formed
in each mechanism, separately.

\subsubsection{PBH binaries formed in the early Universe}
The following discussion is based on \cite{Sasaki:2016jop}.
PBH binaries that are formed in the radiation dominated epoch continuously emit gravitational waves,
gradually shrink, and finally merge.
Since the initial orbital parameters of the binaries are stochastic, some binaries merge in the past,
some other at present epoch, and the others in the future.
According to Peters \cite{Peters:1964zz}, a binary consisting of point masses $m_1$ and $m_2$
with orbital parameter $(a,e)$ merges due to gravitational radiation after time $t$ given by
\be
t=\frac{15}{304} \frac{a^4}{G^3 m_1 m_2 (m_1+m_2)} {\Bigg[ \frac{(1-e^2)}{e^{\frac{12}{19}}}
{\left( 1+\frac{121}{304} e^2 \right)}^{\frac{870}{2299}} \Bigg]}^4
\int_0^e de'~\frac{e'^{\frac{29}{19}}}{{(1-e'^2)}^{-\frac{3}{2}}}
{\left( 1+\frac{121}{304} e'^2 \right)}^{\frac{870}{2299}}.
\ee
The lower limit of integral is set to $0$ by the assumption that the binary is almost circular ($e'=0$)
just before the binary merges.
Because of the smallness of the tidal force from the outer PBHs compared to the gravitational force
between the PBHs that form the binary, 
the orbital eccentricity at the binary formation time is typically close to unity.
When $e \approx 1$, the above formula can be simplified as
\be
t=\frac{3}{85} \frac{1}{G^3 m_1 m_2 (m_1+m_2)} {(1-e^2)}^{7/2} a^4. \label{binary-life-time}
\ee
This shows that highly eccentric binary merges in shorter time by a factor $\frac{768}{425}{(1-e^2)}^{7/2}$
than the circular binary with the same semi-major axis,
which simply reflects that the binary radiates GWs dominantly around the periastron.
It is this factor that makes PBHs binaries formed in the radiation dominated epoch
efficiently merge in the age of the Universe and yields the merger event rate that can
even exceed the one estimated by LIGO when PBHs comprise all the dark matter.

\begin{figure}[t]
\begin{center}
\includegraphics[width=120mm]{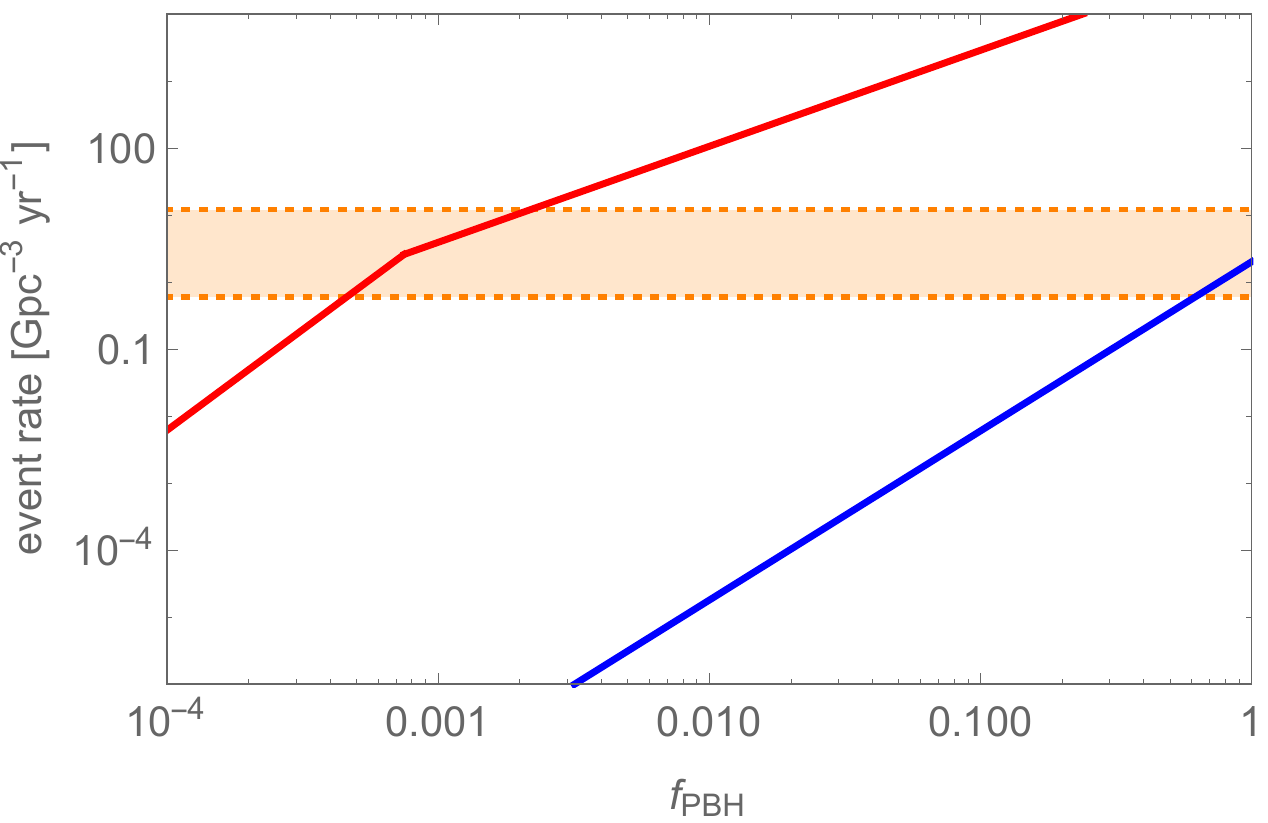}
\end{center}
\caption{Expected merger rate of PBH binaries at present time in two different binary formation mechanisms.
Red curve represents the merger rate for PBH binaries formed in the radiation dominated epoch Eq.~(\ref{eu-merger-rate}) \cite{Sasaki:2016jop},
and the blue curve for the ones formed in the present Universe Eq.~(\ref{present-er}) with $\alpha=1$ \cite{Bird:2016dcv}.
The orange band is the estimated merger rate $0.6 - 12~{\rm Gpc}^{-3} {\rm yr}^{-1}$ by LIGO \cite{TheLIGOScientific:2016pea}.}
\label{merger-rate}
\end{figure}

For simplicity, let us consider the case where all the PBHs have the same mass.
In this case, Eq.~(\ref{binary-life-time}) becomes
\be
t(a,e)=Q{(1-e^2)}^{7/2} a^4,~~~~~~~Q=\frac{3}{170} \frac{1}{G^3 M_{\rm PBH}^3}. \label{binary-life-time2}
\ee
From this equation, we can express $a$ as a function of $t$ and $e$ as $a=a(t,e)$.
Then, we can rewrite the probability (\ref{rd-dp1}) as
\be
dP=\frac{4\pi^2}{3}n_{\rm PBH}^{1/2} {(1+z_{\rm eq})}^{3/2} f_{\rm PBH}^{3/2}
a^{1/2} e {(1-e^2)}^{-3/2} \frac{\partial a}{\partial t} dt de. \label{rd-dp2}
\ee
Since the observations do not measure the initial eccentricity of the binaries, 
we integrate this probability over $e$ along the $t={\rm const}$ curve
up to either a point where the line intersects the curve $e=e_{\rm max}$ 
or a point where the line intersects the line $a=a_{\rm max}$, 
whichever comes first (see Fig.~\ref{fig-emax}).
Plugging $\partial a/\partial t=a(t,e)/(4t)$ and integrating over $e$, we find
\be
dP=\frac{3}{58} {\left( \frac{t}{T} \right)}^{\frac{3}{8}} 
\Bigg[ \frac{1}{ {(1-e_{\rm upper}^2)}^{\frac{29}{16}}}-1 \Bigg] \frac{dt}{t},~~~~~~~~~~
T \equiv Q {\left( \frac{3 y_{\rm max}}{4\pi f_{\rm PBH} (1+z_{\rm eq})}\right)}^4, \label{rd-dp3}
\ee
where $e_{\rm upper}$ is defined by
\be
e_{\rm upper}=\begin{cases}
    \sqrt{1-{\left( \frac{t}{T} \right)}^{\frac{6}{37}}}, ~~~~~{\rm for}~t<t_c\\
    \sqrt{1-{\left( \frac{4\pi f_{\rm PBH}}{3} \right)}^2 {\left( \frac{t}{t_c} \right)}^{\frac{2}{7}}},~~~~~{\rm for}~t\ge t_c,
  \end{cases}
\ee
and $t_c$ is defined by $t_c=T {\left( \frac{4\pi f_{\rm PBH}}{3} \right)}^{\frac{37}{3}}$.
This gives the probability that a given PBH forms a binary and merges at a time in $(t,t+dt)$.
Thus, the merger event rate ${\cal R}$ per unit volume per unit time (at time $t$) is given by
\be
{\cal R}=n_{\rm PBH} \frac{dP}{dt}=\frac{3n_{\rm PBH}}{58} {\left( \frac{t}{T} \right)}^{\frac{3}{8}} 
\Bigg[ \frac{1}{ {(1-e_{\rm upper}^2)}^{\frac{25}{16}}}-1 \Bigg] \frac{1}{t}. \label{eu-merger-rate}
\ee
The red curve in Fig.~\ref{merger-rate} shows ${\cal R}$ for $M_{\rm PBH}=30~M_\odot$,
BH mass close to the first event (GW150914) detected by LIGO, and for $t=14~{\rm Gyr}$
as a function of $f_{\rm PBH}$.
We find that the merger rate largely exceeds the LIGO's observation when $f_{\rm PBH} \simeq 1$
and lies in the band estimated by LIGO for $f=5\times 10^{-4} \sim 2 \times 10^{-3}$.
Thus, in the present mechanism of the PBH binary formation, the case that PBHs constitute
only a fraction of dark matter is observationally relevant.

The knee at $f\simeq 7\times 10^{-4}$ corresponds to the bifurcation point $t=t_c$.
Physically, this bifurcation can be understood as follows.
Since there is an upper limit on the distance between neighboring PBHs that form binary,
Eq.~(\ref{binary-life-time2}) tells that there is also an upper limit on the eccentricity for fixed $t$.
Since the eccentricity is caused by the tidal force from the outer BH,
the eccentricity becomes closer to unity as the outer BH is more separated.
Probability of the location of the outer BH peaks at the mean separation of PBHs.
If $f_{\rm PBH}$ falls below about $7\times 10^{-4}$, the eccentricity caused by the
outer BH at the mean separation exceeds the mentioned upper limit on the eccentricity
and the outer BH must be located closer than the mean distance to cause 
the merger at time $t$. 
As a result, the merger rate is suppressed by the volume factor than the case where
the outer BH is at the mean distance, which produces the knee.

\begin{figure}[t]
\begin{center}
\includegraphics[width=120mm]{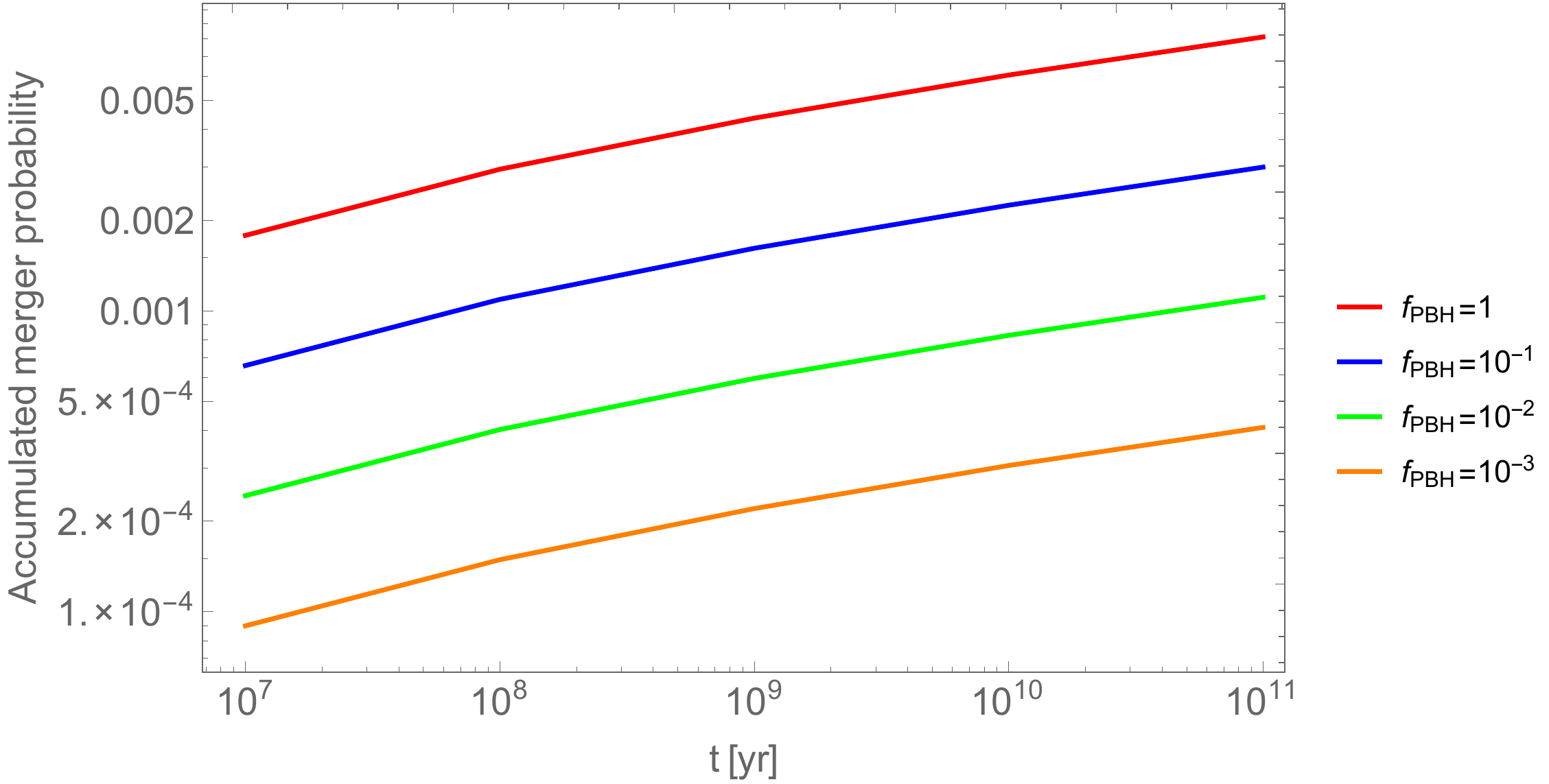}
\end{center}
\caption{The figure shows the probability that a PBH of $M_{\rm PBH}=30~M_\odot$
forms a binary and merges by the cosmic time $t$ for four different values of $f_{\rm PBH}$.}
\label{accumulated-merger-prob}
\end{figure}

One may wonder if the mergers of PBH binaries that have accumulated over the age of the Universe
significantly modify the PBH mass function in the present Universe.
Fig.~\ref{accumulated-merger-prob} shows the probability, which is obtained by integrating Eq.~(\ref{eu-merger-rate}) over the
cosmic time up to $t$, that any PBH undergoes a merger by the time $t$.
From this figure, we find that even in the case with $f_{\rm PBH}=1$ the probability that
a PBH has merged by the present time is about $0.01$.
Thus, in this case the PBH mass function in the present universe is reduced by $\sim 0.01$ at $M=M_{\rm PBH}$
and has a little spike at $2M_{\rm PBH}$.
For smaller $f_{\rm PBH}$, the time evolution of the mass function is more moderate.

From the accumulated merger probability $P$ given in
Fig.~\ref{accumulated-merger-prob}, we can also estimate the accumulated merger rate per unit time
as $\sim H_0^{-2} n_{\rm PBH} P \sim 10^{-4}~{\rm s}^{-1}$ for $M_{\rm PBH}=30M_{\odot}, f_{\rm PBH}=10^{-3}$,
and $t \simeq 10^{10}$ yr.
Thus, the typical time interval between the successive merger events that occurred in the Hubble volume
is much longer than the duration of a single merger event,
which is classified as the "{\it shot noise}" type according to Ref.~\cite{Regimbau:2011rp}.

We have assumed that PBHs are not clustered initially. 
The effect of clustering on the merger rate (\ref{eu-merger-rate}) was addressed in \cite{Raidal:2017mfl}.
According to their result, the merger rate is enhanced by the clustering,
which is a natural consequence since the clustering effectively increases the PBH comoving number density.

There are several effects that have been ignored in deriving the merger rate (\ref{eu-merger-rate}).
These include tidal force from the outer PBHs other than the nearest one,
subsequent capture of the outer BH by the already formed binaries, 
initial peculiar velocities of PBHs, 
gravitational perturbation by the surrounding non-PBH dark matter inhomogeneities,
subsequent accretion of dark matter and baryonic gas onto the PBH binaries.

The first three effects have been investigated in \cite{Ioka:1998nz} in which $f_{\rm PBH}=1$ was assumed.
Because the tidal force is inversely proportional to the distance cubed and
each outer PBH exerts torque to random direction, only a few outer PBHs mostly
contribute to the tidal force.
As a result, the merger event is reduced by at most $60\%$.
The present mechanism of the formation of the bound system by the decoupling from the cosmic expansion 
works as long as the distance between PBHs is smaller than $x_{\rm max}$ (see Eq.~(\ref{condition-bound})).
For $f_{\rm PBH}=1$, $x_{\rm max}$ becomes the mean PBH distance. 
This means that the outer PBH, which generated angular momentum of the inner PBH binary,
is commonly trapped by the binary later, resulting in the hierarchical triple system.
As the three-body problem is difficult to analyze, fate of such triple system is hard to predict.
In \cite{Ioka:1998nz}, such a case was left aside and another case that the distance
to the nearest outer PBH is greater than $x_{\rm max}$ was investigated.
Restricting the PBH mergers to the latter case, it was found that the merger rate is reduced by about $40\%$,
which is not a significant impact.
If $f_{\rm PBH} <1$, which is the case relevant to LIGO observation,
we expect that the probability of forming the triple system is more suppressed than the
case of $f_{\rm PBH}=1$ since the fraction of PBHs satisfying the condition Eq.~(\ref{condition-bound})
is reduced by a factor $f_{\rm PBH}$. 
Initial peculiar velocities, that are randomly directed for each PBH, naturally yields angular momentum of the binary.
If the angular momentum from the peculiar velocity is larger than the one from the tidal force we discussed,
the prediction of the merger rate will be modified.
According to \cite{Ioka:1998nz}, the effect of the peculiar velocity is to reduce the merger rate by
at most $30\%$ even if the initial peculiar velocity is equal to the speed of light.

In \cite{Eroshenko:2016hmn}, the additional tidal force from the adiabatic perturbation
of non-PBH dark matter (valid for $f_{\rm PBH}<1$) was taken into account and 
its effect on the merger rate was investigated.
Since we do not know the power of dark matter perturbations on relevant scales, smaller 
than the ones probed directly by the CMB observations,
straightforward extrapolation of the Planck results was adopted to define the
dark matter perturbations.
It was found that the inclusion of this effect reduces the merger rate by at most
a factor of $2$.
Effects of the tidal force from the non-linear structures of dark matter on the PBH binaries have
been estimated in \cite{Ali-Haimoud:2017rtz} and were found to be insignificant as well.

The accretion of dark matter and baryonic gas onto the PBH binaries was investigated in \cite{Hayasaki:2009ug}.
According to this study, the continuous accumulation of matter onto the binaries can rapidly
decrease the binary radius by the dynamical friction and the merger may happen in the early Universe.
More recent study in \cite{Ali-Haimoud:2017rtz}, based on the simple analytic calculation, 
suggests that the baryonic mass accumulated around the PBHs is likely overestimated in \cite{Hayasaki:2009ug} 
and the baryonic effect is not significant enough to change the evolution of the PBH binaries. 
Yet, more detailed studies are needed to clarify the effectiveness of this scenario.

\subsubsection{PBH binaries formed in the present Universe}
The merger event rate of PBH binaries formed in the present Universe was estimated in \cite{Bird:2016dcv}.
The cross section of forming PBH binaries by the close encounter given by Eq.~(\ref{encounter-cs}) shows
that the binary formation is effective for low relative velocity.
The encounters, which are accidental, are more frequent in high density region than the low density region.
These facts suggest that the PBH binary formation occurs efficiently inside the low-mass dark halos,
which are dense and have small virial velocity. 
The merger rate inside a halo with mass $M_h$ is given by
\be
{\cal R}_h(M_h)=\int_0^{R_{\rm vir}} dr~4\pi r^2 
\frac{1}{2} {\left( \frac{\rho_{\rm PBH} (r)}{M_{\rm PBH}} \right)}^2 
\langle \sigma v_{\rm PBH} \rangle,
\ee
where $\rho_{\rm PBH}(r)$ is density profile of PBHs inside the halo,
and $\langle \sigma v_{\rm PBH} \rangle$ denotes the average over relative velocity distribution
with $\sigma$ given by Eq.~(\ref{encounter-cs}).
In \cite{Bird:2016dcv}, it was assumed that $\rho_{\rm PBH}(r)$ and the velocity distribution 
is given by the Navarro-Frenk-White profile and the Maxwell-Boltzmann distribution, respectively. 
Then, the total merger rate per unit volume and unit time is given by
\be
{\cal R}=\int_{M_{\rm min}} dM_h~\frac{dn}{dM_h} {\cal R}_h (M_h),
\ee
where $\frac{dn}{dM_h}$ is the halo mass function and $M_{\rm min} \sim 400~M_\odot f_{\rm PBH}^{-1}$ 
is the minimum mass of halos that have not yet evaporated by the present time.
Three different mass functions, one obtained by the Press-Schechter formalism, one
based on the simulations by Tinker {\it et al.} \cite{Tinker:2008ff}, and the other that has cutoff at small mass 
from Jenkins {\it et al.} \cite{Jenkins:2000bv}, 
were employed in computing ${\cal R}$.
The result for $M_{\rm PBH}=30~M_\odot$ is given by
\be
{\cal R} \approx 2\alpha f_{\rm PBH}^{\frac{53}{21}}~{\rm Gpc}^{-3} {\rm yr}^{-1}, \label{present-er}
\ee  
where $\alpha \approx 1,~0.6, 10^{-2}$ for the Press-Schechter, Tinker {\it et al.},
and Jenkins {\it et al.}, respectively.

The blue curve in Fig.~\ref{merger-rate} shows ${\cal R}$ given by Eq.~(\ref{present-er}) with $\alpha=1$.
We find that the expected merger rate is consistent with the LIGO observation 
when $f_{\rm PBH} \approx 1$.
The fact that there is a physically interesting region $f_{\rm PBH} \le 1$ consistent with the observations by LIGO 
makes the binary formation path by the present mechanism interesting.
This formation path becomes important if the mergers of PBH binaries formed in the radiation dominated epoch
are significantly reduced by some mechanism.

If PBHs cluster in the vicinity of the supermassive BHs at galactic centers,
they can efficiently form binaries by the close encounter mechanism \cite{Nishikawa:2017chy}.
According to the study in \cite{Nishikawa:2017chy},
the expected merger rate exceeds the one given by Eq.~(\ref{present-er}) 
for some possible range of the slope of the PBH density profile around the supermassive BHs.
In such a case, PBHs in the galactic center may become the dominant channel
to form binaries which emit GWs.

\subsection{Constraint on the PBH abundance from the GW observations}
\begin{figure}[t]
\begin{center}
\includegraphics[width=100mm]{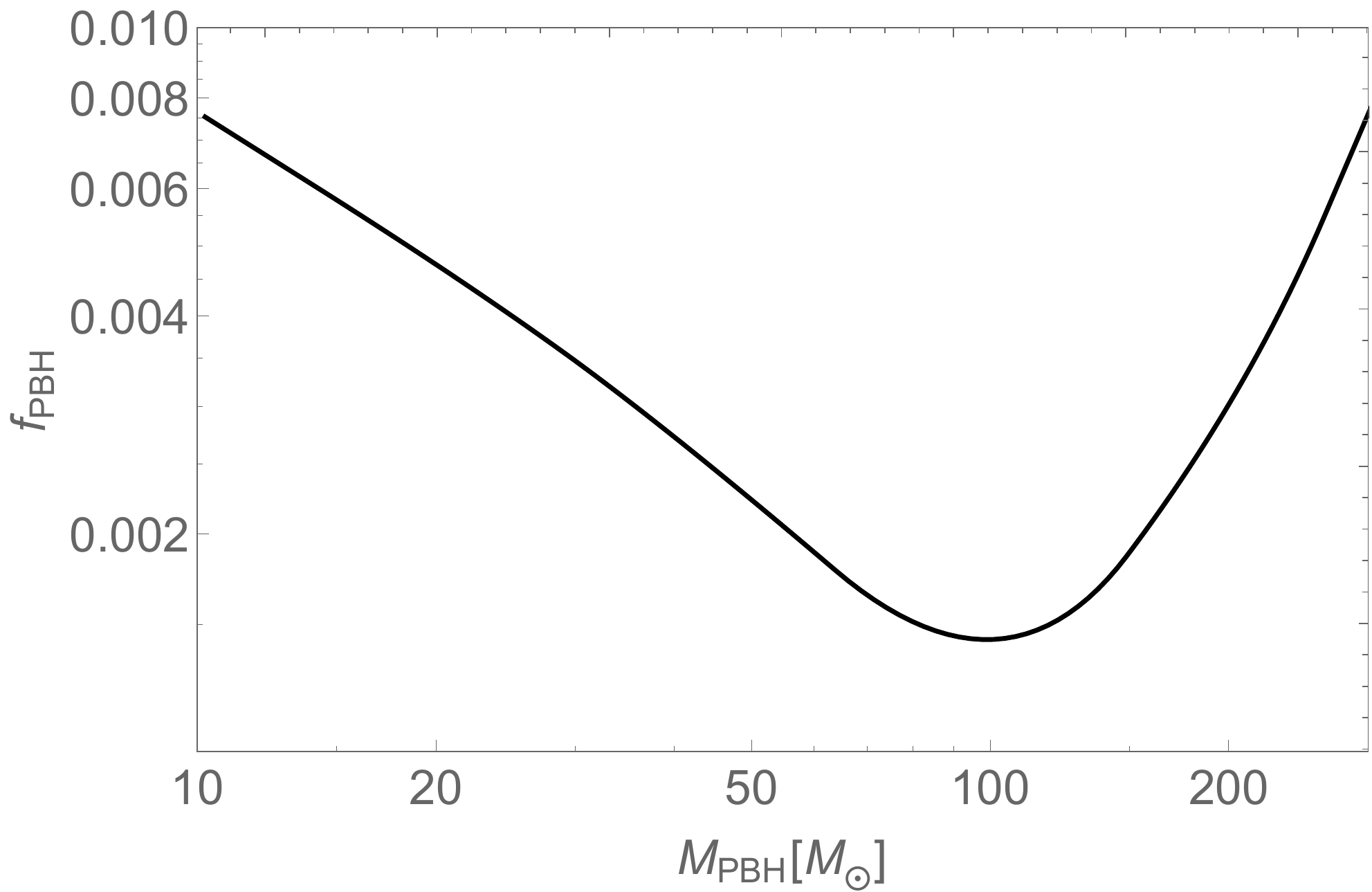}
\end{center}
\caption{Upper limit on $f_{\rm PBH}$ obtained by requiring that the merger event rate
of the PBH binary formed in the radiation dominated era does not exceed the one estimated
by the LIGO O1 \cite{Ali-Haimoud:2017rtz}. Monochromatic mass function is assumed.}
\label{merger-constraint}
\end{figure}

In the previous subsection, we have seen that $30M_\odot-30M_\odot$ PBH binaries 
that are formed in the radiation dominated era merge with frequency consistent with the one
estimated by the LIGO observations if $f_{\rm PBH}\sim 10^{-3}$ and merge
much more frequently if $f_{\rm PBH} \sim 1$.
This means we can use the GW observations to place an upper limit on $f_{\rm PBH}$ irrespective
of whether the observed mergers of BH binaries are attributed to PBHs or not.
This was done in \cite{Ali-Haimoud:2017rtz} for various PBH masses with the monochromatic mass function.
The derived constraint is shown in Fig~\ref{merger-constraint}.
We see that the derived constraint $f_{\rm PBH} \lesssim 10^{-3}-10^{-2}$ excludes stellar mass PBHs
as the dominant component of dark matter.
This result demonstrates that GWs observations provide a novel tool to probe/constrain PBHs
independently of the electromagnetic observations.

\subsection{Distinguishing from other scenarios}
In the previous subsections, we have reviewed the two mechanisms of the PBH binary formation
that work at different cosmic epochs and discussed that the LIGO events can
be explained by the mergers of the PBH binaries.
Of course, the PBH scenario is not the only explanation of the LIGO events and
there are several astrophysical scenarios that have been proposed as the origin of the LIGO events.
So far, due to the small number of the detected events, those scenarios are
allowed observationally.
The next obvious task is to clarify how we can test the PBH scenario and
discriminate it from the others by using the future observations that will bring much more information.
This is the main topic of this subsection.

\subsubsection{Stochastic GW background from PBH binaries}
\label{SGWB-PBH}
At the time of writing this article (autumn 2017), LIGO has detected five GW-events from BH-BH mergers.
These GW-events were close enough so that they were {\it heard} as single events by the detectors.
In addition to these loud events, there are other merger events that occur at more distant places.
Although GWs from such mergers pass through the Earth more frequently than the louder ones,
they are tiny, buried in noise, and may not be identified as single events.
Yet, those events may be detected as a whole by taking correlations of the GW signals among different
detectors and integrating it over some time.  
If such tiny GWs exist, after the time integration, the GW signal will emerge.  
Such GWs are referred to as stochastic GW background.

\begin{figure}[t]
\begin{center}
\includegraphics[width=120mm]{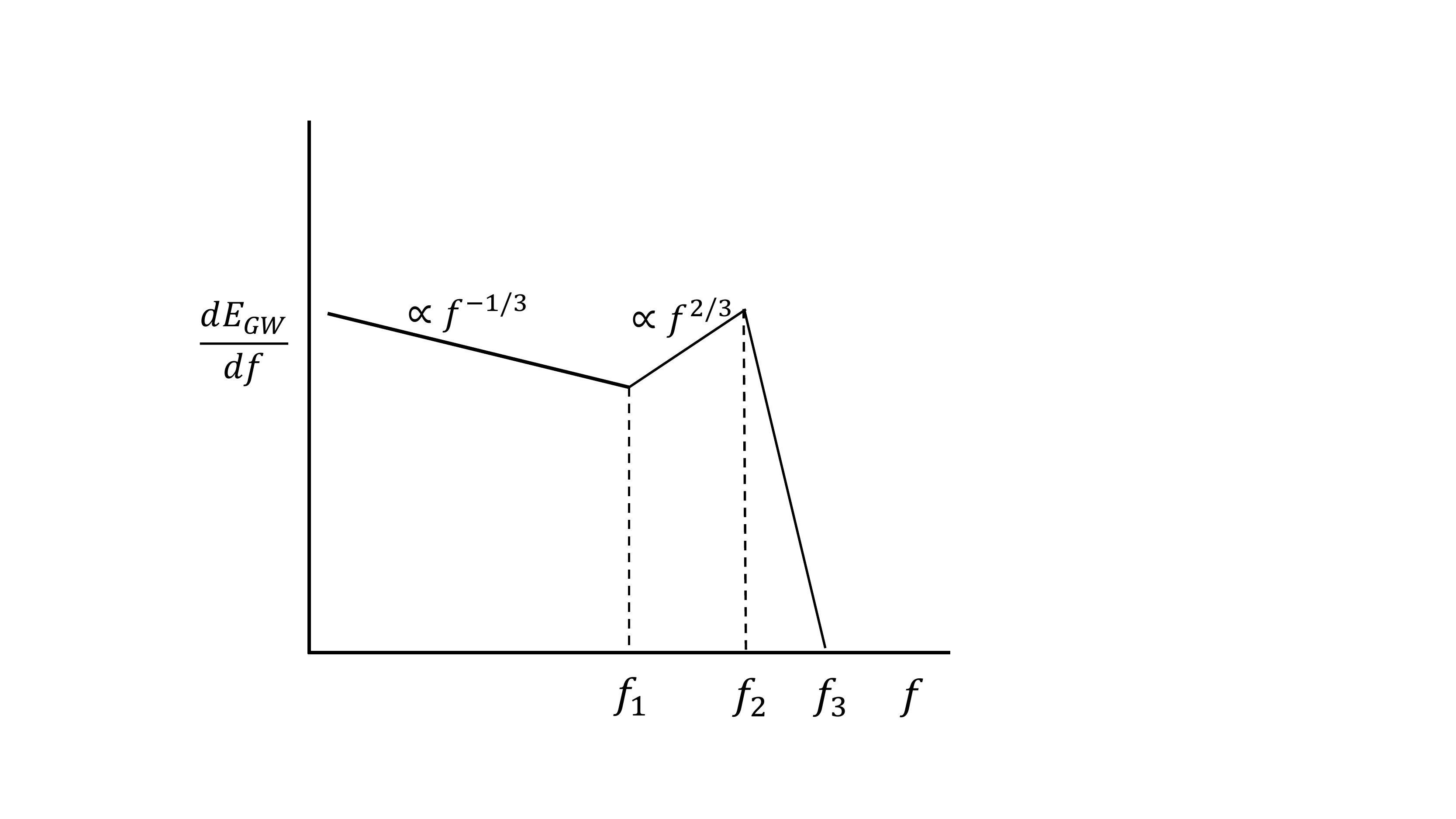}
\end{center}
\caption{Schematic shape of the spectral energy density given by Eq.~(\ref{sed}).}
\label{spectral-energy}
\end{figure}

It is customary to represent the strength of the stochastic GWs in terms of the GW energy density $\rho_{\rm GW}$
per logarithmic frequency bin normalized by the critical density $\rho_c$ (for instance \cite{Maggiore:1999vm});
\be
\Omega_{\rm GW}(f)=\frac{1}{\rho_c} \frac{d \rho_{\rm GW}}{d \ln f}.
\ee
For GWs that are emitted by the mergers of compact objects in binaries, $\Omega_{\rm GW}$ can be written as
\cite{TheLIGOScientific:2016wyq}
\be
\Omega_{\rm GW}(f)=\frac{f}{\rho_c H_0}
\int_0^\infty dz~\frac{{\cal R}(z)}{(1+z) \sqrt{\Omega_m {(1+z)}^3+\Omega_\Lambda}}
\frac{dE_{\rm GW}(f')}{df'}\Bigg|_{f'= (1+z)f}, \label{pbhb-omega-gw}
\ee
where ${\cal R}(z)$ is the merger rate of the source we are interested in,
and $dE_{\rm GW}/df$ is the spectral energy density of a source,
which describes how much energy is released from the source in the form of GWs with frequency $f$.
The spectral energy density can be determined by general relativity.
Based on the phenomenological waveforms in the Fourier domain for the
inspiral, merger, and ringdown phases for non-spinning BH binaries \cite{Ajith:2007kx},
Zhu {\it et al.} \cite{Zhu:2011bd} converted it to the spectral energy density as
\be
\label{sed}
\frac{dE_{\rm GW}}{df}=\frac{ {(G\pi)}^{2/3} M_c^{5/3}}{3}
\begin{cases}
f^{-1/3},~~~~~(f < f_1,~{\rm inspiral~phase}), \\
f_1^{-1} f^{2/3},~~~~~(f_1<f< f_2,~{\rm merger~phase}), \\
f_1^{-1} f_2^{-4/3} \frac{f}{1+4{\left( \frac{f-f_2}{\sigma} \right)}^2},~~~~~(f_2<f< f_3,~{\rm ringdown~phase}),
\end{cases}
\ee
where $M_c$ is the chirp mass and $f_1,~f_2,~f_3,~\sigma$ are fitting parameters.
Schematic shape of this function is shown in Fig.~\ref{spectral-energy}.
Modified waveforms for BH-BH binaries, generalized to non-precessing spins, are given in Ajith {\it et al.} \cite{Ajith:2009bn}.

\begin{figure}[t]
\begin{center}
\includegraphics[width=140mm]{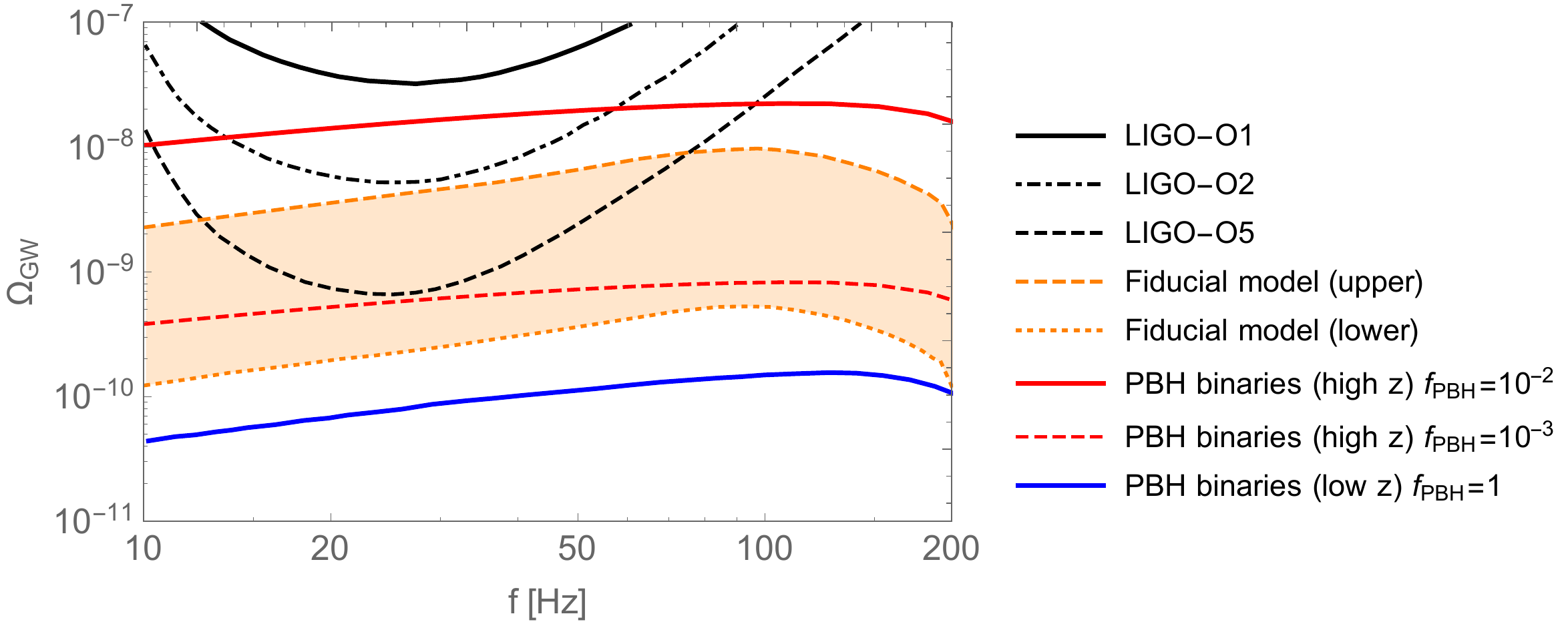}
\end{center}
\caption{Predicted $\Omega_{\rm GW}$ in the PBH scenarios.
Two red curves obtained in \cite{Raidal:2017mfl} are the prediced $\Omega_{\rm GW}$ from PBH binaries that are formed in the
radiation dominated epoch for two different values of $f_{\rm PBH}$.
Blue curve obtained in \cite{Mandic:2016lcn} is the predicted $\Omega_{\rm BH}$ from PBH binaries that are formed in the present 
Universe with $f_{\rm PBH}=1$.
Orange band computed in \cite{Mandic:2016lcn} shows the expected $\Omega_{\rm GW}$ from the astrophysical BHs within uncertainties.
Black curves given in \cite{TheLIGOScientific:2016wyq} represent sensitivities of LIGO at different observation stages.}
\label{PBH-binary-omegagw}
\end{figure}

For the evaluation of $\Omega_{\rm GW}$ from the mergers of PBH binaries,
we use either Eq.~(\ref{eu-merger-rate}) or (\ref{present-er}) for ${\cal R}(z)$ in Eq.~(\ref{pbhb-omega-gw}).
For the PBH binaries formed in the radiation dominated epoch, the expected $\Omega_{\rm GW}$ was computed
for the first time in \cite{Ioka:1998gf} for MACHO mass range $\sim 0.5~M_\odot$. 
In light of the detection of BH-BH binaries of $\sim 30~M_\odot$ by LIGO,
the computation of $\Omega_{\rm GW}$ for $\sim 30~M_\odot$ was done in \cite{Wang:2016ana}.
The analysis was generalized to the case where the PBH mass function is not monochromatic in \cite{Raidal:2017mfl}.
For the PBH binaries formed in the present Universe, the expected $\Omega_{\rm GW}$ was computed
in \cite{Mandic:2016lcn} (assuming monochromatic PBH mass function).
Fig.~\ref{PBH-binary-omegagw} shows curves of $\Omega_{\rm GW}(f)$ predicted in the aforementioned PBH scenarios,
red curves for PBH binaries formed in the radiation dominated epoch ($M_{\rm PBH}=30~M_\odot$, $f_{\rm PBH}=10^{-2}$ and $f_{\rm PBH}=10^{-3}$)
and blue one for PBH binaries formed in the present Universe ($M_{\rm PBH}=30~M_\odot$ and $f_{\rm PBH}=1$).
As is anticipated from Fig.~\ref{merger-rate}, PBH binaries formed in the radiation dominated epoch produce
larger stochastic GWs than the ones formed in the present Universe for the same $f_{\rm PBH}$.
The band colored by orange represents the contribution from the astrophysical BH binaries in the fiducial model
defined in \cite{TheLIGOScientific:2016wyq}. 
The band width originates from the statistical uncertainty of the inferred merger rate at local Universe.
As more merger events are accumulated in the future, the width will shrink for fixed model. 
We also show the sensitivity curves of LIGO-O1, O2, and O5 provided in \cite{TheLIGOScientific:2016wyq}.
Quite interestingly, Fig.~\ref{PBH-binary-omegagw} shows that the stochastic GWs from PBHs can be potentially
detected by LIGO for the interesting range of $f_{\rm PBH}$ for the PBH binaries formed in the early Universe.
Thus, there is a good motivation to search and use the stochastic GWs to test the PBH scenario.

While different scenarios predict different curves of $\Omega_{\rm GW}$ in Fig.~\ref{PBH-binary-omegagw},
their shapes look similar.
This is understandable, given that the shape is essentially determined by the shape of the spectral
energy density $\frac{dE_{\rm GW}}{df}$ and the source in different scenarios is physically identical, namely, 
BH-BH binaries. 
Since the physics that determines the spectral energy density is well understood (just general relativity),
the qualitative shape of $\Omega_{\rm GW}$-curves is robust.
But this is two-edged sword for those who try to test the PBH scenarios by using the stochastic GWs.
When $\Omega_{\rm GW}$ is measured, 
it becomes a big challenge to determine if the observed stochastic GWs
originate from the PBHs or from the astrophysical BH binaries \cite{Mandic:2016lcn, Wang:2016ana}.
Furthermore, if PBHs exist,
the real $\Omega_{\rm GW}$ would a superposition of the one from PBHs
and the one from the astrophysical BHs.
In order to extract the PBH signal from $\Omega_{\rm GW}$,
it is indispensable to reduce the theoretical uncertainties about how much the astrophysical
BHs produce $\Omega_{\rm GW}$.
More studies are needed to figure out how far we can go.

Finally, we have to mention that $\Omega_{\rm GW}$-curves in the PBH scenarios in 
Fig.\ref{merger-rate}
contain contributions from merger events that can be identified as a single event.
How the individual events and the residual are decomposed depends on the detector's sensitivity and the
implementation of data analysis.
Thus, we have to apply the similar operation to the predicted $\Omega_{\rm GW}$ when one compares
the PBH scenario with the real data.

\subsubsection{Cosmic evolution of the merger rate}
Second observable that can be potentially used for distinguishing the PBH scenario from the
astrophysical ones is the time evolution of the BH-BH merger rate.
While the PBHs exist from almost the beginning of the Universe, the BHs resulting from
the death of stars appear at low redshift Universe.
Thus, the redshift evolution of the number density of PBHs differs from the one of astrophysical BHs.
From this simple fact, we expect that the redshift dependence of the merger rate of the BH-BH binaries
should also exhibit difference between the different scenarios.
 
\begin{figure}[t]
\begin{center}
\includegraphics[width=140mm]{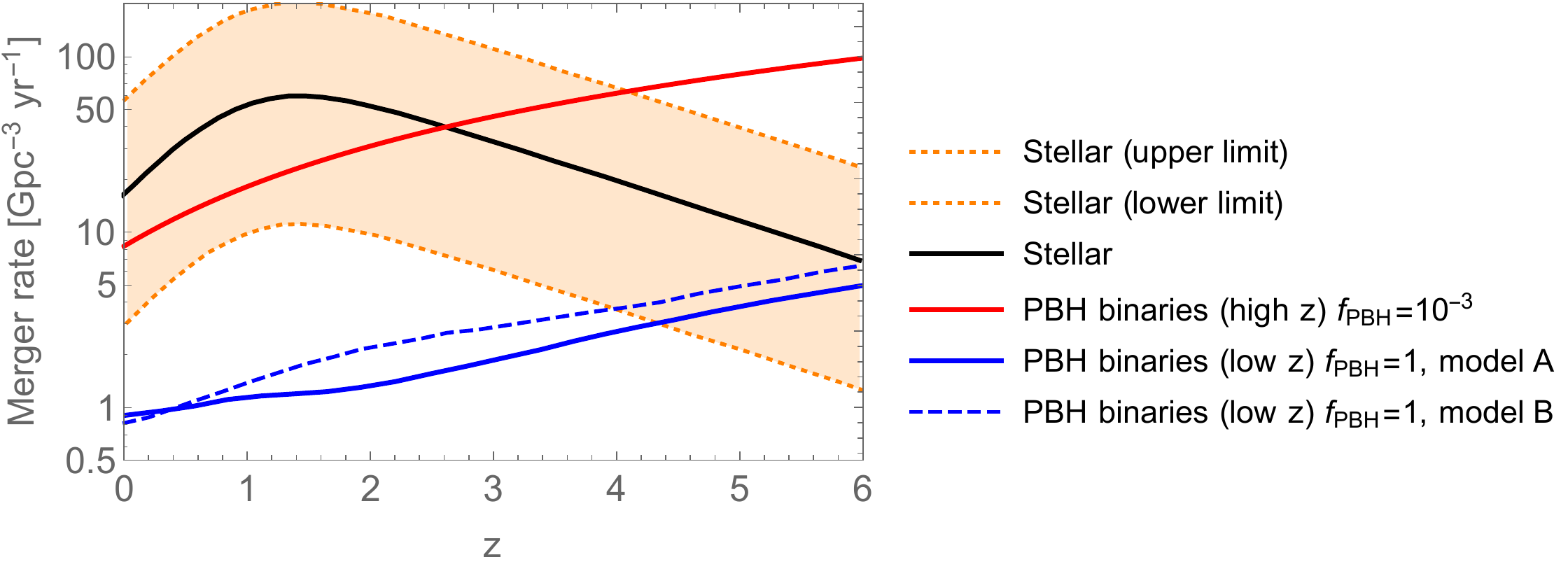}
\end{center}
\caption{Redshift dependence of the merger rate of the BH-BH binaries per unit source time and unit comoving volume.
Unit is ${\rm Gpc}^{-3} {\rm yr}^{-1}$.
Black and orange curves, taken from \cite{Mandic:2016lcn}, show the uncertainties of the expected merger rate from astrophysical BHs.
Red curve is the merger rate from the PBH binaries formed in the radiation dominated epoch with $f_{\rm PBH}=10^{-3}$.
Blue curves, taken from \cite{Mandic:2016lcn}, are the merger rate from the PBH binaries
formed in the low-redshift Universe.}
\label{Merger-rate-z-dependence}
\end{figure}

Fig.~\ref{Merger-rate-z-dependence} shows the redshift evolution of the merger rate
per unit source time and unit comoving volume (${\rm Gpc}^{-3} {\rm yr}^{-1}$) for different scenarios.
The black curve represents the merger rate of the astrophysical BHs in the fiducial model mentioned in \ref{SGWB-PBH}.
The orange band is the uncertainty of the model.
These curves are taken from \cite{Mandic:2016lcn}.
The merger rate has a peak at $z=1\sim 2$.
This is caused by the assumption that the formation rate of the BH binaries is proportional to the
star formation rate below the half of the solar metallicity \cite{TheLIGOScientific:2016wyq}.
Although there is a large uncertainty, 
there is a clear tendency that the merger rate drops sharply beyond the redshift $1\sim 2$.

The red curve is the merger rate Eq.~(\ref{eu-merger-rate}) of the PBH binaries formed in the radiation dominated epoch
for $f_{\rm PBH}=10^{-3}$ and $M_{\rm PBH}=30~M_\odot$.
Interestingly, the merger rate continuously increases for higher redshift even beyond $z=1\sim 2$.
For the chosen parameters, the merger rate is comparable to the merger rate of the stellar BHs at low redshift,
but exceeds the astrophysical prediction at higher redshift $z \gtrsim 5$. 
Thus, the increasing feature of the merger rate at higher redshift is the smoking gun of the PBH scenario, and
searching for such events definitely constitutes one of the routes we have to tread in the future \cite{Nakamura:2016hna}.
For the sake of completeness, 
we also show the merger rate of the PBH binaries formed in the low redshift Universe
for $f_{\rm PBH}=1$ and the chirp mass $30M_\odot$ as blue curves, which is taken from \cite{Mandic:2016lcn}.
The Press-Schechter formalism is used to obtain the halo mass function,
and the thick and dashed curves correspond to different choices of 
the halo concentration models (for more details, see \cite{Mandic:2016lcn}).
We find that similarly to the PBH binaries formed in the early Universe
the merger rate in this case also increases monotonically for higher redshift.
Yet, the predicted merger rate is significantly lower than the one of the stellar origin up
to $z\sim 6$ and than the red curve at any redshift for the chosen parameter.

\begin{figure}[t]
\begin{center}
\includegraphics[width=100mm]{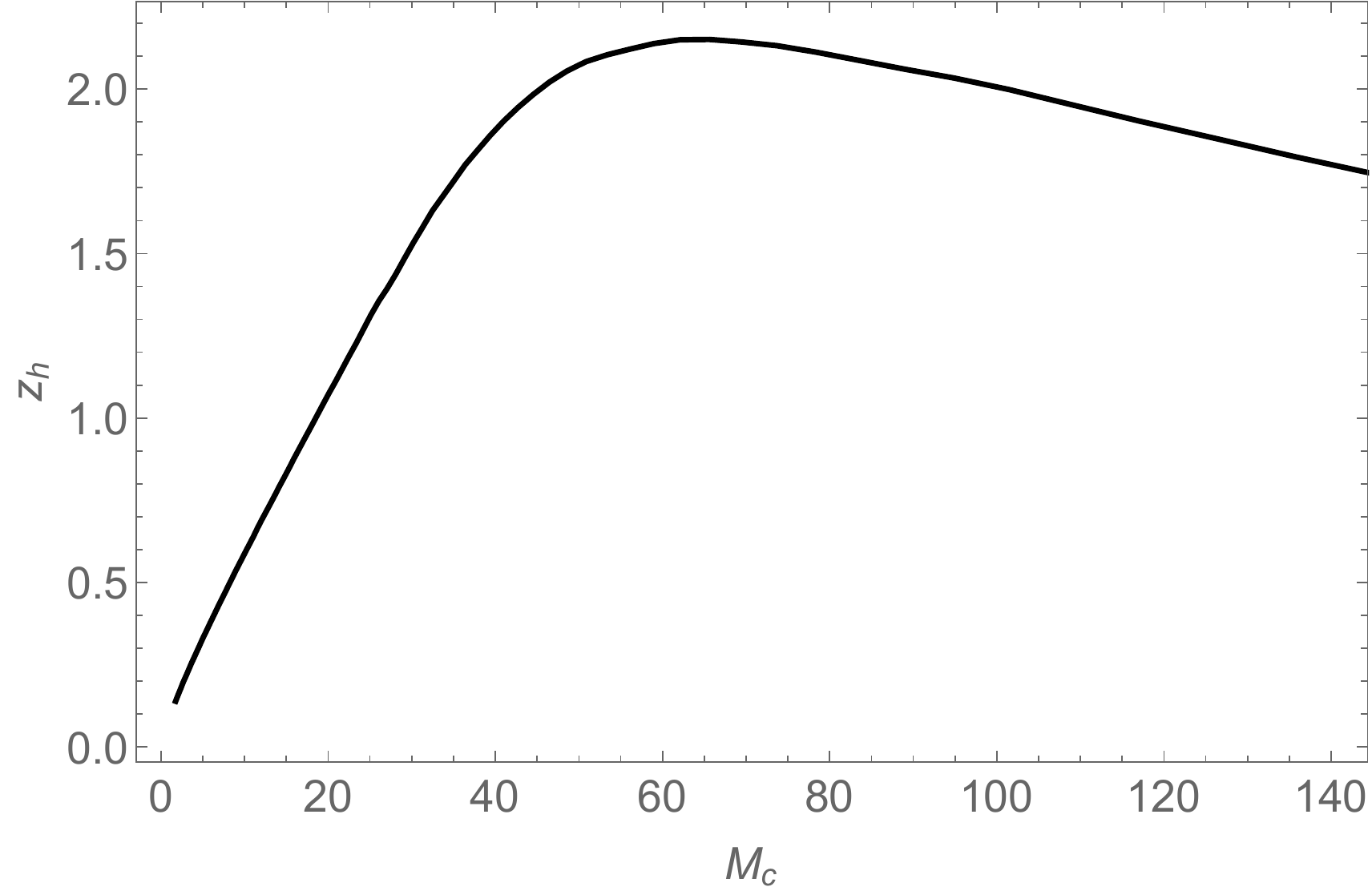}
\end{center}
\caption{Horizon redshift as a function of the (source-frame) chirp mass for the advanced LIGO with the design sensitivity constructed
from \cite{Martynov:2016fzi}.
}
\label{LIGO-hdf}
\end{figure}

Detecting the merger events at high redshift $z \gg 1$ is challenging.
Figure \ref{LIGO-hdf} shows the maximum redshift that the advanced LIGO with the design
sensitivity can detect the merger event of the binaries with a signal-to-noise ratio 8 as a function of the (source-frame) chirp mass,
which is constructed from \cite{Martynov:2016fzi}.
We find that the advanced LIGO at the design sensitivity can detect the merger events
of BH-BH binaries up to $z \sim 1.5$ for $M_c =30M_\odot$.
This shows that the LIGO may not be powerful enough to distinguish the PBH scenario from the astrophysical
scenarios in the context of looking into the high redshift merger events.

Beyond LIGO, there are several proposed GW detectors both on ground and in space.
They include Einstein Telescope, Cosmic Explorer, eLISA, and DECIGO.
These detectors will be able to detect the GWs coming from much more distant places than LIGO can.
For instance, it was shown in \cite{Nakamura:2016hna} that pre-DECIGO (DECihertz laser Interferometer
Gravitational wave Observatory),
which consists of three spacecrafts cruising around the Earth in a triangle with its arm length $100$ km
and is planned to be launched in the late 2020s,
can detect the merger events of $30M_\odot$ BHs up to $z \sim 10$ with a signal-to-ratio 8.
At such high redshift, the merger events of astrophysical origin are rare,
and we can perform a clear test of the PBH scenario.

\subsubsection{Mass distribution}
\label{PBH-mass-distribution}
Masses of the individual BHs before the merger are $(m_1,m_2)=(35,30)$ for GW150914, 
$(m_1,m_2)=(14,8)$ for GW151226, $(m_1,m_2)=(31,19)$ for GW170104, $(m_1,m_2)=(12,7)$ for GW170608, 
and $(m_1,m_2)=(30,25)$ for GW170814 in units of solar mass.
Obviously, there is some spread in the mass distribution. 
It is natural to think that the event rate distribution in the 2-dimensional mass plane should reflect to a certain degree the formation
mechanism of the BH binaries and its statistical nature can be used to discriminate different formation scenarios.
Although merger events that have been discovered are currently countable by hand, 
it is almost sure that much more merger events will be detected in the coming decades.
In such era, we should have a plenty of information about the statistical properties of the merger event distribution
in the 2-dimensional plane $(m_1,m_2)$. 
It is a purpose of this subsection to review some studies that aim to test the PBH scenario by using the
event rate distribution that will become available in the future.

Generically speaking, if the PBHs exist in the Universe, the merger events of the BH binaries 
we will observe are mixture of the PBH binaries and those formed by the astrophysical mechanisms.
As we have seen in the previous sections, because of our ignorance of the PBH abundance and
mass function and the large uncertainties about the astrophysical processes forming the BH binaries,
there is not a definite conclusion on which one (primordial or astrophysical) dominates the merger events.
It is important to keep in mind that the following discussions are based on some simplified assumptions
reflecting the aforementioned unknown factors and their conclusions may need to be modified as we become able
to reduce the uncertainties in the future.

In \cite{Kovetz:2016kpi}, considering a possibility that some BH merger events involve PBHs,
the case was investigated where the PBH mass function has a sharp spike 
around a certain mass $M_0$ with $f_{\rm PBH}=1$ and the PBH binaries are formed in the low redshift Universe.
The resultant merger rate, obtained as a function of the heavier BH in binaries, 
was superposed on the one predicted in the fiducial astrophysical model.
It was found that merger-event histogram exhibits a visible bump at around $M_0$ on top of the smooth 
astrophysical background if the accumulated number of events reaches a few thousands,
which is feasible by the advanced LIGO with the design sensitivity.
According to the analysis in \cite{Kovetz:2017rvv}, the above PBH scenario with $f_{\rm PBH}>0.5$ will be excluded
at $99.9\%$ confidence level by the advanced LIGO.
It is also shown that use of the data in the two-dimensional mass plane is more powerful than using only the
heavier mass in constraining $f_{\rm PBH}$.

In \cite{Kocsis:2017yty}, the case was investigated where the PBHs with extended mass function form binaries in the radiation
dominated era, aiming at finding the unique feature of the PBH scenario in the merger-event distribution.
For simplicity, the astrophysical contribution was not included in the merger rate.
Furthermore, PBHs are assumed to distribute randomly in space with no correlation between different mass.
As explained in \ref{PBH-binary-EU}, the major-axis and eccentricity of the PBH binary is determined by the 
gravitational force between two BHs that form the binary and torque exerted by the outer BHs not involved in the binary, respectively.
Denoting by $m_1$ and $m_2$ the mass of the individual PBHs that form the binary and by
$m_t=m_1+m_2$ the total mass,
$a$ and $e$ are given by \cite{Kocsis:2017yty}
\begin{align}
\label{a-and-e}
&a=\frac{1}{1+z_{\rm eq}} \frac{\rho_{c,0} \Omega_m}{m_t} x^4, \nonumber \\
&1-e^2= \frac{9}{4} \zeta^2,~~~~~~~~~~\zeta=|{\vec \zeta}|,~~~~~~~~~ 
{\vec \zeta}=\sum_{i=1}^N \frac{x^3}{y_i^3} \frac{M_i}{m_t}
\sin (2\theta_i ) \frac{({\vec e_z} \times {\vec e_i})}{ |{\vec e_z} \times {\vec e_i}| }.
\end{align}
Here, contrary to the analysis in \ref{PBH-binary-EU}, not only the closest outer BH 
but also other more distant BHs have been included to evaluate the eccentricity for completeness.
The upper limit $N$, which can be taken to be infinity practically, is the number of PBHs contained in the
Hubble horizon at the binary formation time.
Furthermore, $y_i, M_i, \theta_i, {\vec e}_i$ are distance to, mass of, direction to, and angle measured from the major axis
of the $i$-th outer BH.
Then, the merger probability of the PBH binary consisting of BHs with mass $m_1$
and $m_2$ at cosmic time $t$ becomes
\be
\mathcal{R}_{\rm intr}(m_1,m_2,t) = \int_0^{e_m} de ~F(x(a), \zeta(e)) \frac{dx}{da} \frac{d\zeta}{de} \frac{\partial a}{\partial t}, \label{formal-Pint}
\ee
where $\frac{dx}{da}$ and $\frac{d\zeta}{de}$ can be computed from Eqs.~(\ref{a-and-e}) 
and $\frac{\partial a}{\partial t}$ from Eq.~(\ref{binary-life-time}).
The function $F(x,\zeta)$, which gives the probability density of $x$ and $\zeta$, is formally written as
\begin{align}
F (x(a),\zeta(e))=& \Theta (a_{\rm max}-a) \frac{4\pi x^2 (a)}{n_{\rm BH}^{-1}}
\int \lim_{N\to \infty}
\prod_{i=1}^N \frac{dV_i}{n_{\rm BH}^{-1}} \frac{f(M_i)dM_i}{n_{\rm BH}}
\frac{\sin \theta_i d\theta_i d\phi_i}{4\pi}~\Theta (y_i-y_{i-1} ) \nonumber \\
&\times e^{ -\frac{4\pi}{3} n_{\rm BH} y_N^3}
\delta \left(  \zeta-g (x,y_i,M_i,\theta_i,\phi_i ) \right), \label{formal-F}
\end{align}
where 
$\Theta(\cdot)$ is the Heaviside step function and $\delta (\cdot)$ is the Dirac's delta function,
and $g$ is given by
\be
g (x,y_i,M_i,\theta_i,\phi_i) \equiv \bigg| \sum_{i=1}^N \frac{x^3}{y_i^3} \frac{M_i}{m_t}
\sin (2\theta_i ) \frac{({\vec e_z} \times {\vec e_i})}{| {\vec e_z} \times {\vec e_i} |} \bigg|. \label{function-g}
\ee
Finally, the observable merger rate ${\cal R}$ per unit time and unit comoving volume
is given by 
\be
\mathcal{R} (m_1,m_2,t)=\frac{1}{2 n_{\rm PBH}} {\cal R}_{\rm intr} (m_1,m_2,t) f(m_1) f(m_2),
\ee
where $f(m)$ is the PBH mass function.

In \cite{Kocsis:2017yty}, since exact computation of the integral (\ref{formal-F}) 
for arbitrary shape of the mass function is impossible,
${\cal R}_{\rm intr}$ was computed under two different approximations.
The first is to include only the closest outer BH ($N=1$), just as we discussed in \ref{PBH-binary-EU},
without specifying the specific shape of $f(m)$.
The second is to consider the case with $N \gg 1$. For this case, analytic form of $F(x,\zeta)$ given by
\be
F(x,\zeta)=6\sqrt{3} \gamma^{1/3} n_{\rm BH} {\tilde \sigma}^2 \zeta x^2 {\left( \frac{m_t}{m_{\rm max}} \right)}^2 
{\bigg[ {\left( \frac{m_t}{m_{\rm max}} \right)}^3 \zeta^3+\gamma {\tilde \sigma}^6 \bigg]}^{-1}, \label{flat-F-ansatz}
\ee
was adopted. 
This function was confirmed to correctly reproduce the numerically computed $F(x,\zeta)$ by the Monte-Carlo calculations
for the flat PBH mass function.
Quite interestingly, in both cases, it was found that the merger rate ${\cal R}$ depends on
the total mass $m_t=m_1+m_2$ in a specific way almost independent of the mass function. 
In order to extract this information, a dimensionless quantity $\alpha$ defined by
\be
\alpha (m_1,m_2,t)= -m_t^2 \frac{\partial^2}{\partial m_1 \partial m_2} \ln {\cal R} (m_1,m_2,t),
\ee
was introduced.
Both cases considered above predict this quantity to be
\be
\frac{36}{37} \le 
\alpha (m_1,m_2,t)
\le \frac{22}{21}, \label{PBH-consistency}
\ee
which sharply concentrates around unity.
Notice that ${\cal R}$ itself strongly depends on the mass function like $\propto {\tilde f}(m_1) {\tilde f}(m_2) m_t^\alpha$,
where ${\tilde f}(m)$ is related to $f(m)$ in a non-trivial way,
but such part decouples from ${\cal R}$ by taking its logarithm and differentiation twice.

The result (\ref{PBH-consistency}) shows that measuring the quantity $\alpha$ by observations
is a powerful method to test the PBH scenario, independently of the PBH mass function.
Raw merger rate distribution in the $m_1-m_2$ plane directly determined by observations will not coincide 
with ${\cal R}$ computed above since the former is affected by the detector bias.
Thus, extracting the collect value of $\alpha$ from observations is a challenging task.
When statistically enough merger events have been accumulated in the $m_1-m_2$ plane in the future,
it should be in principle possible to compare the observation with the prediction (\ref{PBH-consistency}).
How to achieve this is an important topic, but we will not discuss it here.

The quantity $\alpha$ can be used not only to test the PBH scenario, but also to discriminate different 
scenarios of the BH binary formation.
For instance, the quantity $\alpha$ for the PBH binaries that formed in the low redshift Universe was also derived
and found to take a unique value given by
\be
\alpha (m_1,m_2,t)=\frac{10}{7},
\ee
which is different from (\ref{PBH-consistency}) that PBH binaries that formed in the radiation dominated era predict.
The dynamical formation scenario, which is one of the strong astrophysical candidates to form BH binaries,
predicts $ \alpha \approx 4$ \cite{OLeary:2016ayz}.
As these examples vividly demonstrate, different scenarios predict different values of $\alpha$.
When the observational data reaches a sensitivity $\Delta \alpha \lesssim 0.1$,
$\alpha$ will be used as a powerful discriminator.

\subsubsection{Spin distribution}
Distribution of the BH spin is also useful to constrain the formation scenarios
of the BH-BH binaries.
Magnitude of the BH spin is commonly expressed in terms of a dimensionless quantity
$\chi$ defined by
\be
\chi=\frac{|\vec S|}{GM_{\rm BH}^2},
\ee
where ${\vec S}$ is the BH spin.
Physical requirement that no naked-singularity appears restricts the range of $\chi$
as $0 \le \chi \le 1$,
and BHs with $\chi=1$ correspond to the maximally rotating BHs.

Spins of the individual BHs in the binary affects the GW waveform primarily through
a particular combination given by
\be
\chi_{\rm eff}=\frac{m_1 \chi_1 \cos \theta_1+m_2 \chi_2 \cos \theta_2}{m_t},
\ee
where $\theta_i~(i=1,2)$ is the angle between the spin of the $i$-th BH and
the orbital angular momentum.
Thus, the allowed range of $\chi_{\rm eff}$ is $-1 \le \chi_{\rm eff} \le 1$.
At the time of writing this article, five merger events have been robustly detected.
The measured values of $\chi_{\rm eff}$ are
\be
\chi_{\rm eff}=-0.06^{+0.14}_{-0.14},~~~
0.21^{+0.20}_{-0.10},~~~
-0.12^{+0.21}_{-0.30},~~~
0.07^{+0.23}_{-0.09},~~~
0.06^{+0.12}_{-0.12},   \label{value-chieff}
\ee
for GW150914, GW151226, GW170104, GW170608, and GW170814 respectively (with $90\%$ credible intervals).
Detection of much more merger events in the future will
allow us to construct the distribution of $\chi_{\rm eff}$.

There are several astrophysical scenarios for the origin of the BH-BH binaries.
For instance, the isolated field binary scenario, in which stars are formed as a binary and
later individual stars collapse to BHs (heavier one first),
BH spins are likely to be aligned with the orbital angular momentum,
and $\chi_{\rm eff} >0$ is a natural consequence in this scenario \footnote{
Yet, we have to keep in mind that natal kick velocity of the BH by the
supernova explosion can alter this conclusion.}.
Since there has not been robust prediction of the magnitude of $\chi_{\rm eff}$ yet,
low values of $\chi_{\rm eff}$ of the detected events (\ref{value-chieff})
are used to constrain the properties of the field binaries \cite{Kushnir:2016zee,Hotokezaka:2017esv,Postnov:2017nfw,Hotokezaka:2017dun}.
Tidal lock of the progenitor of the lighter BH from the heavier BH will result in
the high spin of the lighter BH.
Thus the progenitor of the lighter BH must be sufficiently compact, which
favors the Wolfe-Rayet star as the progenitor of the lighter BH \cite{Hotokezaka:2017esv}.
Dynamical formation scenario, in which BH binaries are formed
by dynamical interactions among BHs in dense stellar environments such as
globular clusters, predicts isotropic distribution of the individual BH spins.
Thus, positive and negative $\chi_{\rm eff}$ are equally probable in this scenario.

Despite of small number of detections of merger events,
the comparison between the alignment hypothesis of the BH spins and the isotropic one
has been already made against the observed values of $\chi_{\rm eff}$ in \cite{Farr:2017uvj}.
Considering some simple distribution functions of spins extending to high values,
the odds ratio against the alignment hypothesis compared to the
isotropic one was found to be $0.015$, that corresponds to about $2.4~\sigma$. 
Furthermore, among different models of the isotropic distribution,
the most favored one has low spin magnitude distribution,
although the statistical significance is not as strong as the above odds ratio.
However, see \cite{Farr:2017gtv} in which slight preference to the alignment hypothesis was suggested. 
Definitely, more observational data is needed to establish a robust conclusion about this issue.

As the discussions in the previous sections demonstrate, 
the PBH binaries, whether they are formed in the radiation dominated era
or in the low redshift Universe, have isotropic spin distribution.
What remains an open issue is the distribution of spin magnitude $a$
of the individual PBHs.
In \cite{Chiba:2017rvs}, the PBH spin distribution has been derived under
several assumptions in the following way.
In the direct collapse scenario of the PBH formation in the radiation dominated era,
in which the nearly spherical overdense region undergoes the gravitational collapse
upon the horizon re-entry,
PBHs form when the density contrast exceeds a threshold $\delta_{\rm th}$.
For the PBH that formed out of the density contrast slightly above the threshold,
which will be the case in the realistic scenario in which the probability of realizing much
higher density contrast is significantly suppressed,
its mass and spin may have simple dependence on $\delta_{\rm th}$ as
\be
M_{\rm PBH} \approx C_M {| \delta-\delta_{\rm th} (q)|}^{\gamma_M},~~~~
S_{\rm PBH} \approx C_S {| \delta-\delta_{\rm th} (q)|}^{\gamma_J} q,~~~~
\delta_{\rm th}(q)=\delta_{\rm th,0}+K q^2,
\ee  
where $q$ is a parameter that characterizes the amount of the PBH spin (see \cite{Baumgarte:2016xjw}),
and $C_M,~C_S$,$~\gamma_M$, $\gamma_J$, $\delta_{\rm th,0}$, and $K$ are all constants.
In particular, $K (\approx 5.7\times 10^{-3})$ is positive.
This is natural since stronger gravity, thus larger amplitude of the density contrast, 
is needed to defeat the centrifugal force and to form spinning PBHs.

These scalings were found numerically in the asymptotically flat spacetime \cite{Baumgarte:2016xjw}.
In \cite{Chiba:2017rvs}, these relations were adopted.
Formally, one can write the above relations as $M_{\rm PBH}=M_{\rm PBH}(\delta,q)$ and 
$S_{\rm PBH}=S_{\rm PBH}(\delta,q)$.
Then, denoting the probability density of $(\delta,q)$ as $P(\delta,q)$,
we can convert the probability in the $\delta-q$ plane to the one in the $M_{\rm PBH}-\chi$ plane as
\be
dP=P(\delta, q) d\delta dq=F(M_{\rm PBH},\chi)dM_{\rm PBH} d\chi,
\ee
where $\chi=S_{\rm PBH}/(G M_{\rm PBH}^2)$.
In \cite{Chiba:2017rvs}, it was assumed that there is no correlation between
$\delta$ and $q$ and probability density for $\delta$ and $q$ are Gaussian and flat, respectively. 
This latter assumption may be oversimplification given the vectorial nature of the angular momentum
in three dimensional space.
In more realistic situation, the probability density at small $q$ would be suppressed as $\sim q^2$,
and the flatness assumption could be understood as a limiting case in which the intial
overdensity can easily have a large angular momentum.

Then, integration of the above probability over $M_{\rm PBH}$ was found to be Gaussian,
\be
dP \propto \exp \bigg[ -{\left( \frac{\chi}{\chi_*} \right)}^2 \bigg]d\chi,~~~~~\chi_* \approx 0.46. \label{PBH-spin-dis}
\ee
Thus, PBHs with higher spins $(\chi \approx 1)$ are less likely to be realized than those with lower spins,
which can be traced back to the fact that the threshold amplitude of the PBH formation increases as we increase
the PBH spin.
The value of $\chi_*$ depends on $\delta_{\rm th,0}$ and variance $\sigma^2$ of the density perturbation.
In the above equation, $\delta_{\rm th,0}=1/3,~~\sigma=0.15~\delta_{\rm th,0}$ were adopted as fiducial values.

\begin{figure}[t]
\begin{center}
\includegraphics[width=100mm]{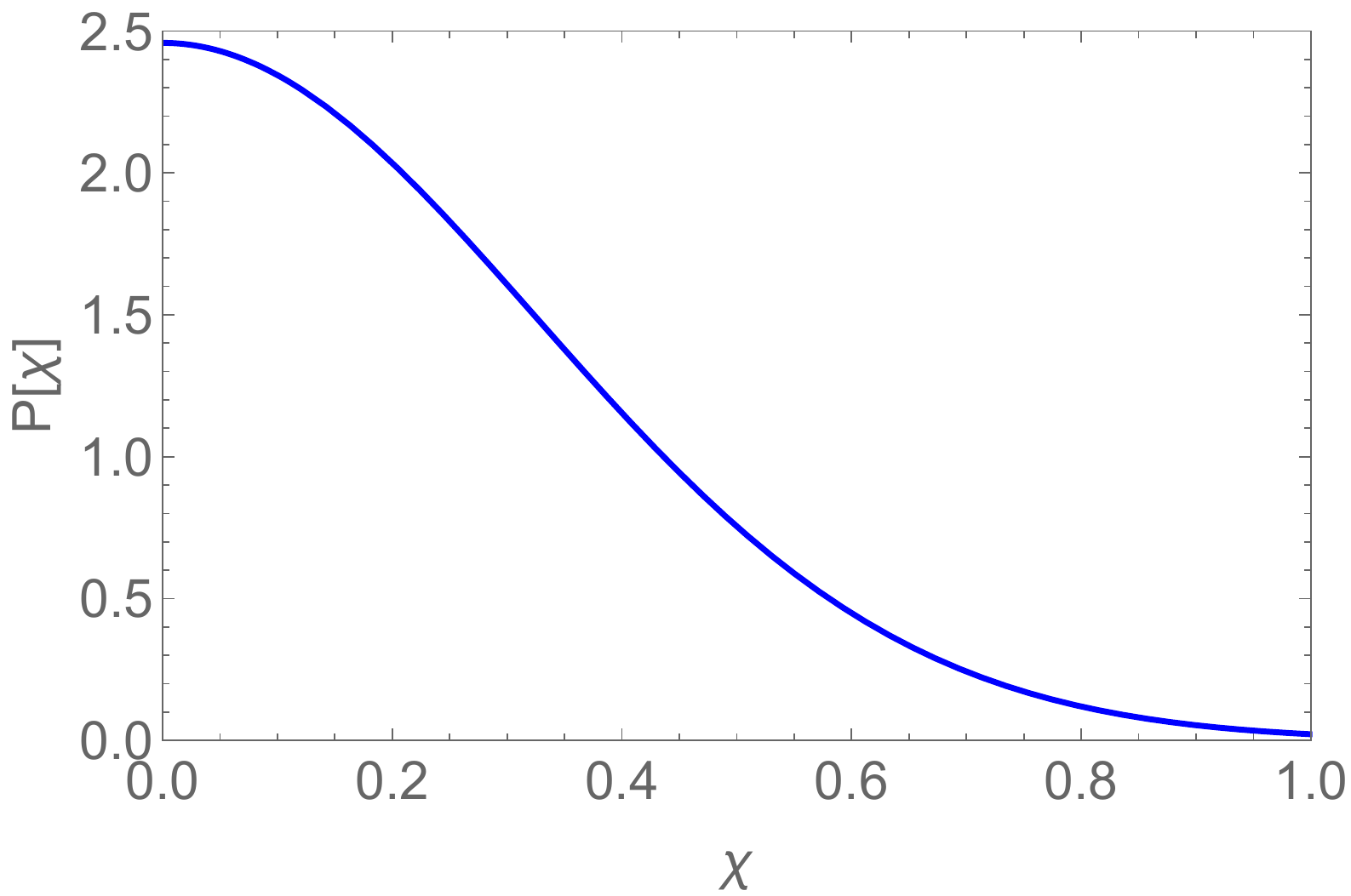}
\end{center}
\caption{PBH spin distribution (\ref{PBH-spin-dis}).
}
\label{PBH-spin}
\end{figure}

Fig.~\ref{PBH-spin} shows the PBH spin distribution given by Eq.~(\ref{PBH-spin-dis}).
As mentioned above, the distribution at low $\chi$ could be significantly modified 
in more realistic situation. 
Nevertheless, it seems robust to conclude that rapidly spinning PBHs $(\chi \approx 1)$ are very unlikely. 
If spin distribution turns out to have a peak at a high value $\chi \simeq 1$ in the future,
the PBH scenario as the origin of the BH-BH binaries will then be strongly disfavored.

Yet, we have to keep in mind that even the observational confirmation of the spin
distribution (\ref{PBH-spin-dis}) does not necessarily prove the PBH scenario.
The above result has been derived under a couple of non-trivial assumptions.
Neither the Gaussian shape nor the value of $\chi_*$ should be taken seriously as unique prediction of the PBH scenario.
For instance, the flatness assumption on $q$ may be overestimation of the probability for higher value of $q$
and the typical spin magnitude $\chi_*$ could be much smaller than the value in Eq.~(\ref{PBH-spin-dis}).
Obviously, more studies are needed to clarify the spin distribution of PBHs.
At this moment, a general lesson we can learn from this result is that PBHs with lower spin 
are more favored than the higher spin, which is reasonable from a physical point of view\footnote{Spins of PBHs formed in the matter-dominated era was discussed in Ref.~\cite{Harada:2017fjm}. This work predicted that in the matter-dominated era PBH formation would be much suppressed due to the angular momentum and formed PBHs could have higher spin.}.

\subsubsection{Cross-correlation with galaxies}
Cross-correlation between the spatial distribution of the BH-BH merger events and that of galaxies
offers another method to test the PBH scenario.
This method will become powerful when a sufficiently large number of the merger events enough to discuss the statistical
distribution in space has accumulated ($\gtrsim 100$).

On large scales, much longer than the typical distance between galaxies ($\sim {\rm Mpc}$),
we can define number density of galaxies and its density contrast $\delta_{\rm g}({\vec x})$.
Since galaxies form by the gravitational collapse of the baryonic gas where the dark matter density is higher than average,
there are more galaxies in sites where the dark matter density contrast is higher.
It is known that $\delta_{\rm g}({\vec x})$ does not coincide with the density contrast of matter 
$\delta_{\rm m}({\vec x})$ due to the fact that the galaxies are formed at the high density peaks \cite{Kaiser:1984sw}. 
The mismatch is commonly represented as a bias parameter $b_{\rm g}$ by
\be
\delta_{\rm g}=b_{\rm g} \delta_{\rm m}.
\ee
The bias parameter can evolve on cosmological time scales.
The observations of the galaxy clustering and weak lensing
suggest $b_{\rm g}=1.3 \sim 1.7$ for $z \lesssim 0.5$ and increase of $b_{\rm g}$ for higher redshift \cite{Abbott:2017wau}.

The fact that the bias parameter $b_{\rm g}$ is different from unity is a good news for the purpose
of testing the origin of the BH binaries.
The reason for this is simple.
If the BH binaries are of astrophysical origin, then $\delta_{\rm BH}$, the number density contrast of the merger events
of the BH binaries, should coincide with $\delta_{\rm g}$.
If on the other hand BH binaries consist of PBHs, then $\delta_{\rm BH}$ should trace $\delta_{\rm m}$ unless the BH binary evolution is not significantly affected by the interaction with baryons.
When taken together, PBHs and astrophysical BHs predict different magnitude of 
cross-correlation between $\delta_{\rm BH}$ and $\delta_{\rm g}$.

The idea of taking the cross-correlation between $\delta_{\rm BH}$ and $\delta_{\rm g}$ (as well as weak lensing)
to test the spatial clustering of BH binaries was proposed and investigated in \cite{Namikawa:2016edr}.
The similar methodology was used for the purpose of testing the PBH scenario in \cite{Raccanelli:2016cud}
in which the close encounter mechanism for the binary formation (see \ref{bphbfpu}) was assumed.
According to the notation in \cite{Raccanelli:2016cud}, 
the cross-correlation between $\delta_{\rm BH}$ and $\delta_{\rm g}$ in multi-pole decomposition is given by 
\be
C_\ell^{\rm BH, g}=r \int \frac{4\pi dk}{k}~\Delta^2 (k) W_\ell^{\rm BH} (k) W_\ell^{\rm g} (k), \label{cross-correlation-bh-g}
\ee
where $r$ is the cross-correlation coefficient, which represents how much galaxy distribution actually traces 
the matter distribution, $\Delta^2$ is the dimensionless power spectrum, and $W_\ell^{\rm BH}$
and $W_\ell^{\rm g}$ are defined by
\be
W_\ell^{\rm X}=\int \frac{dN_{\rm X}}{dz} b_X (z) j_\ell (k\chi (z)) dz,
\ee
where $X=\{ {\rm PBH, g} \}$, $dN_{\rm X}/dz$ is the source redshift distribution, 
$b_X$ is the bias parameter, and $\chi (z)$ is the comoving distance.
In \cite{Raccanelli:2016cud}, the constant redshift distribution of galaxies
and the constant galaxy bias $b_{\rm g}=1.4$ were assumed,
and the redshift distribution of the PBH binaries given in \cite{Bird:2016dcv} and
the constant BH bias $b_{\rm PBH}=0.5$ were adopted.
The reason why $b_{\rm PBH}$ is smaller than unity is that BH merger events in the scenario in \cite{Bird:2016dcv}
occur dominantly inside the small halos whose spatial distribution are more spread than the heavier halos
that host galaxies.

The theoretical prediction of the cross-correlation (\ref{cross-correlation-bh-g}) is the one that can be compared 
with observational data in the future.
The right-hand side of Eq.~(\ref{cross-correlation-bh-g}) is proportional to $b_{\rm g}$ if the BH binaries 
are formed by the astrophysical processes and to $b_{\rm PBH}$ if the BH binaries are PBHs.
Thus, the target sensitivity of the future observations to discriminate the PBH scenario from the astrophysical ones
is $\delta b \lesssim b_{\rm g}-b_{\rm PBH} =0.9$.
What then matters is whether the future observations are powerful enough to achieve this sensitivity.
According to the analysis in \cite{Raccanelli:2016cud},
such achievement is possible by the Einstein Telescope and could be even possible
by the long-term observations by aLIGO, depending on the BH merger rate (see also \cite{Raccanelli:2016fmc}).

\subsubsection{Eccentricity}
Eccentricity of the binary orbit has also a potential to discriminate the PBH scenario from
the astrophysical scenarios.

\begin{figure}[t]
\begin{center}
\includegraphics[width=100mm]{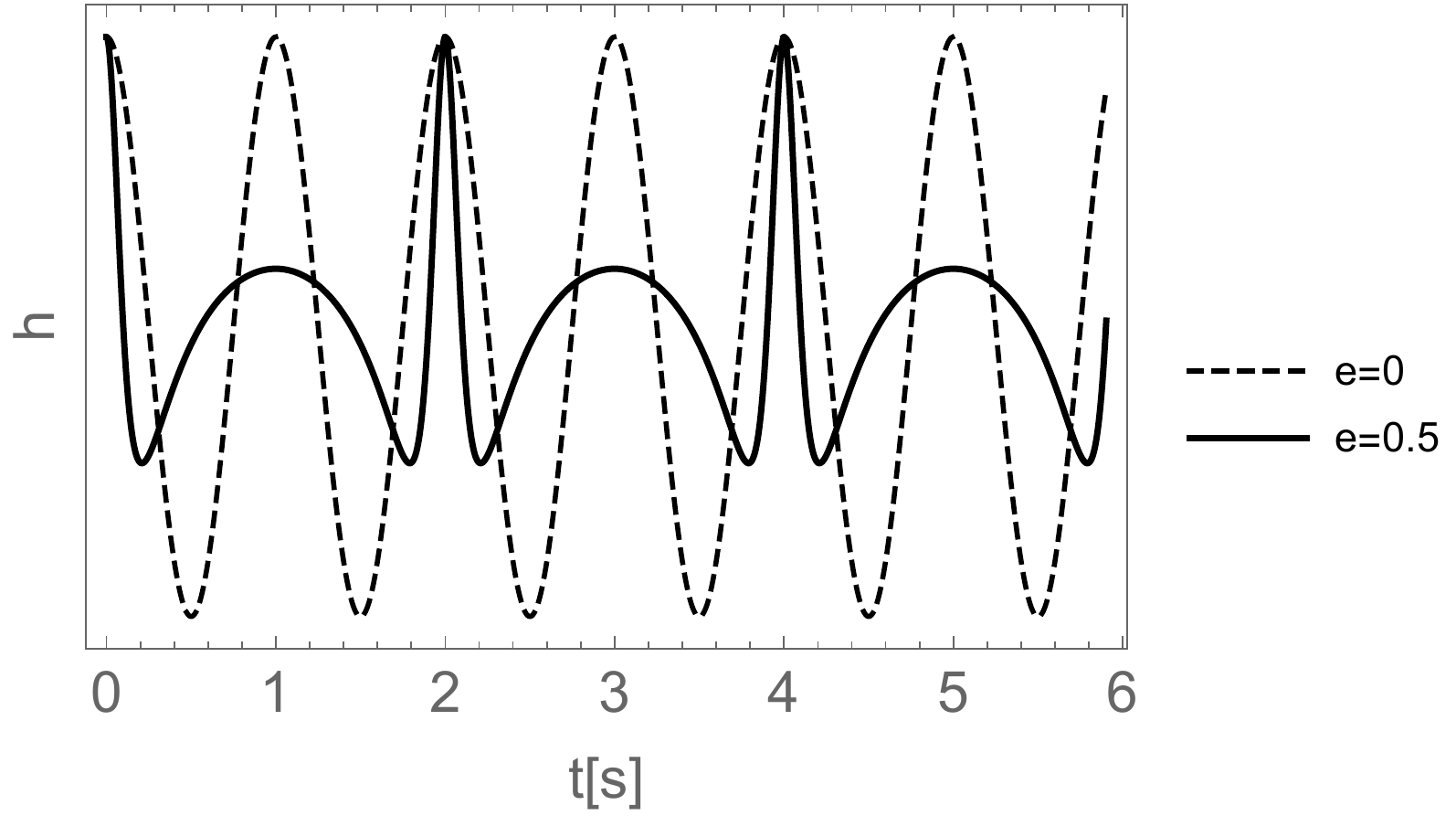}
\end{center}
\caption{Gravitational waveforms from BH-BH binaries with equal masses ($30~M_\odot$) with the
circular orbit and the eccentric one ($e=0.5$) when the orbital period is $2~s$. 
Waveforms are normalized appropriately.
}
\label{eccentricity-waveform}
\end{figure}

First of all, circular binary and eccentric binary emit GWs whose waveforms look differently.
Fig.~\ref{eccentricity-waveform} shows two GW waveforms from BH binaries consisting of the equal mass 
($m_{\rm BH}=30~M_\odot$)
with different eccentricities ($e=0$ and $e=0.5$) \cite{Martel:1999tm}. 
Amplitude of the waveforms are normalized.
We find that the waveform for the eccentric orbit is modulated and show two
phases, gentle hill and tall spike.
The spike originates at the periastron where the acceleration of BHs as well as velocities are higher.
This figure demonstrates that it is in principle possible to know the eccentricity of the binary
from measurement of the time dependence of the waveform.

While the individual binary formation scenarios predict (in principle) the distribution of the
eccentricity at the binary formation time,
GW observations do not measure the initial eccentricities, but those in the inspiral phase corresponding to 
the frequency band which the detectors are sensitive to.
This fact is important since the eccentricity changes as the binary shrinks by the GW emission.
According to Peters \cite{Peters:1964zz}, the orbital eccentricity changes due to the gravitational radiation reaction as
\be
\frac{de}{dt}=-\frac{304}{15} e \frac{G^3 m_1 m_2 (m_1+m_2)}{a^4 {(1-e^2)}^{5/2}}
\left( 1+\frac{121}{304}e^2 \right).
\ee
Clearly, $de/dt \le 0$ for any $e$.
Thus, the eccentric orbit is always circularized by the GW emission.

Generally speaking, binaries born with larger separation have smaller eccentricities in the final inspiral phase
than those with initially shorter separation since the formers have more time to circularize the orbit.
For the stellar-mass BH binaries, frequency band around $100~{\rm Hz}$
to which ground-based detectors such as LIGO are the most sensitive corresponds to
the last several revolutions before the merger.
BH binaries from the isolated field binaries and those dynamically formed in the globular clusters 
are expected to have negligible eccentricities in the LIGO band \cite{Kowalska:2010qg, Rodriguez:2016kxx}.
On the other hand, BH binaries formed in the vicinity of the supermassive BHs in galactic nuclei 
are mostly highly eccentric $(e \simeq 1)$ even at the LIGO band \cite{OLeary:2008myb}.

In \cite{Cholis:2016kqi}, the eccentricity distribution of PBH binaries that are formed
in the low redshift Universe in the LIGO frequency band was investigated.
A point is that since those binaries are formed with high eccentricity and 
have much shorter lifetime than the Hubble time,
fraction of them retain some eccentricity even when they enter the LIGO frequency band.
According to the analysis in \cite{Cholis:2016kqi}, ${\cal O}(1)$ merger events with non-zero eccentricity 
are expected to be detected by a few years observation by LIGO and ${\cal O}(10)$ events by
ten years observation by the Einstein Telescope.
On the other hand, the eccentricity distribution for the LIGO band for the PBH binaries 
formed in the radiation dominated era is expected to have a strong spike at $e=0$ 
since their lifetime is about the Hubble time.

Space-based laser interferometers such as eLISA and DECIGO are sensitive to lower frequency 
band than the ground-based detectors. 
They can observe stellar-mass BH binaries in the inspiral phase much before the mergers.
Thus, the orbits in such low-frequency phase are more eccentric than the nearly merging phase,
if binaries have eccentricities initially.
The analysis in \cite{Nishizawa:2016jji, Nishizawa:2016eza} shows that several years observations
by eLISA have a potential to distinguish between the field and cluster formation scenarios.
Yet, there is no similar study for the distinguishability of the PBH scenario.

\newpage
 
\section{Summary}
\begin{quote}
``I think over the coming decades we will see enormous numbers of things. 
Just as electromagnetic astronomy was begun in essence, at least modern astronomy, 
by Galileo pointing his telescope in the sky and discovering Jupiter's moons. 
This is the same thing but for gravitational waves...'' \hspace{4cm}
{\it Kip S. Thorne} \cite{Thorne-interview}
\end{quote}

LIGO's first observation of the GWs finally opened an era of GW astronomy. 
{\it Direct} observations of the BHs vividly demonstrated that GWs bring us information 
of the Universe which can never be obtained by the observations of the electromagnetic signals. 
PBH is no exception regarding this point. 
Since the original proposal around 1970, electromagnetic searches for the PBHs have been performed over decades. 
Until now, none of these searches detected solid evidence for the existence of PBHs and 
tight constraints on PBH with various masses have been obtained. 
The situation has drastically changed by the LIGO's event which
suddenly raised an interesting possibility; 
LIGO might have detected the PBHs for the first time!
The purpose of this article is to deliver recent proposals of the PBHs as the source
of the LIGO events and give a review of various ideas to test the PBHs using the future GW observations,
simultaneously covering the basics of the PBH formation as well as the existing constraints on the non-evaporating PBHs
from the electromagnetic observations.

After having introduced the basics of the PBH formation and the relevant inflation models in Sec.~2, 
we reviewed various existing constraints on the PBH abundance of 
the non-evaporating PBHs in Sec.~\ref{Observational-constraint} as well as the constraints that
will be imposed by the future observations. 
For the sake of completeness, we not only addressed the stellar-size PBHs but also PBHs in
wider mass range from $\sim 10^{-16}~M_\odot$ to $\sim 10^{10}~M_\odot$. 
In Sec.~\ref{GW-PBH}, we reviewed the PBH scenario as an explanation of the LIGO events
and various proposals to test it by the future GW observations.
 
GW astronomy has just begun.
As the GW astronomy progresses, it will continuously bring us new findings and
also stimulate related theoretical studies.
Definitely, we will gain more knowledge about the PBHs and the early Universe.
PBHs are dark but the future of the PBH research is bright.

\section*{Acknowledgments}
It is a pleasure to thank Yacine Ali-Haimoud, Christian Byrnes, Bernard Carr, Anne Green, Tomohiro Harada,
Kazunori Kohri, and Masamune Oguri for a careful reading of the manuscript and their enlightening comments.
This work was supported in part by JSPS KEKENHI Grant Numbers
JP17H06358 (T.T.), JP17H06359 (T.S.), 
JP15H05888 (M.S, T.S and S.Y.), JP15H02087 (T.T.), JP16H01103 (S.Y.)
JSPS Grant-in-Aid for Young Scientists (B) JP15K17632 (T.S.) and JP15K17659 (S.Y.), 
and the Grant-in-Aid for Scientific Research JP26287044 (T.T.).

\section*{Permission of the reuse}
Following are permissions of reusing the figures/data in the literature to produce some
figures in this article.
We thank all the persons concerned for having kindly allowed us to reuse their materials.

\begin{enumerate}
\item{Fig.~\ref{Constraints}} \newline
  Ref.\cite{Tisserand:2006zx}: P.~Tisserand et al. Astron.Astrophys. 469, 387 (2007), reproduced with permission \copyright ~ESO.\newline
  Ref.\cite{Wyrzykowski:2011tr}: Reproduced with permission from L. Wyrzykowski et al. The OGLE view of microlensing towards the Magellanic Clouds – IV. OGLE-III SMC data and final conclusions on MACHOs. MNRAS (2011) 416 (4): 2949-2961. Published by Oxford University Press on behalf of The Royal Astronomical Society. All rights reserved. 
  The figure is not covered by the Open-Access licence of this publication. 
  For permissions contact Journals.Permissions@OUP.com \newline
  Ref.\cite{Griest:2013aaa}:  K.~Griest et al.~ApJ 786, 158 (2014), reproduced with permission from the author and \copyright ~AAS. \newline 
  Ref.\cite{Quinn:2009zg}: Reproduced with permission from D.P.Quinn etal. On the reported death of the MACHO era. MNRAS: Letters (2009) 396 (1):L11-L15. Published by Oxford University Press on behalf of The Royal Astronomical Society. All rights reserved. 
The figure is not covered by the Open-Access licence of this publication. 
For permissions contact Journals.Permissions@OUP.com\newline
  Ref.\cite{Brandt:2016aco}: T.~D.~Brandt, ApJ 824, L31 (2016), reproduced with permission from the author and \copyright ~AAS. \newline
  Ref.\cite{Niikura:2017zjd,Oguri:2017ock}: Reproduced with permission from the authors. \newline 
  Ref. \cite{Wilkinson:2001vv,Barnacka:2012bm,Graham:2015apa,Capela:2013yf,Ali-Haimoud:2016mbv,Gaggero:2016dpq,Poulin:2017bwe}: Reproduced with permission from the authors and \copyright ~APS. \newline
  Ref.~\cite{Inoue:2017csr}: Reproduced with permission \copyright~IOP.\newline

\item{Fig.~\ref{merger-constraint}} \newline
 Ref.~\cite{Ali-Haimoud:2017rtz}: Reproduced with permission from the author and \copyright ~APS.\newline

\item{Fig.~\ref{PBH-binary-omegagw}} \newline
 Ref.~\cite{Mandic:2016lcn, TheLIGOScientific:2016wyq}: Reproduced with permission from the authors and \copyright ~APS. \newline
 Ref.~\cite{Raidal:2017mfl}: Reproduced with permission \copyright~IOP.\newline
 
\item{Fig.~\ref{Merger-rate-z-dependence}} \newline
 Ref.~\cite{Mandic:2016lcn}: Reproduced with permission from the author and \copyright ~APS. \newline
 
\item{Fig.~\ref{LIGO-hdf}}\newline
 Ref.~\cite{Martynov:2016fzi}: Reproduced with permission from the author and \copyright ~APS. \newline
\end{enumerate}

\bibliographystyle{h-physrev}
\bibliography{draft}

\end{document}